\documentclass[10pt,journal,compsoc]{IEEEtran}

\usepackage{booktabs} % For formal tables
\usepackage{tabu}

\usepackage{bookmark}
\usepackage{multirow}
\usepackage{array}
\usepackage[normalem]{ulem}
\usepackage{ulem}

\makeatletter  
\newif\if@restonecol  
\makeatother

\usepackage[linesnumbered,ruled,vlined]{algorithm2e} 
\usepackage{algpseudocode}
\usepackage{amsmath}  
\usepackage{cite}
\usepackage[pdftex]{graphicx}
  
\graphicspath{IMAGES/}
\DeclareGraphicsExtensions{.pdf,.jpeg,.png}
\usepackage{threeparttable}
\usepackage{microtype}

\usepackage{dblfloatfix}
\usepackage{diagbox}

\usepackage{arydshln}

\usepackage{ragged2e}
\usepackage{url}

\usepackage{url}

\usepackage{lipsum}
\usepackage{xcolor}
\usepackage{listings}

\begin{document}

\title{RobustECD: Enhancement of Network Structure for Robust Community Detection}

\author{Jiajun Zhou,
        Zhi~Chen,
        Min~Du,
        Lihong Chen,
        Shanqing~Yu,\\
        % Feifei~Li,
        Guanrong~Chen,~\IEEEmembership{Fellow,~IEEE},
        and~Qi~Xuan,~\IEEEmembership{Member,~IEEE}% <-this % stops a space
\IEEEcompsocitemizethanks{
\IEEEcompsocthanksitem J. Zhou, L. Chen, S. Yu, and Q. Xuan are with the Institute of Cyberspace Security, College of Information Engineering, Zhejiang University of Technology, Hangzhou 310023, China. E-mail: jjzhou012@163.com, \{yushanqing, 2111803032, xuanqi\}@zjut.edu.cn).
\IEEEcompsocthanksitem Z. Chen and M. Du are with the Department of Electrical Engineering and Computer Sciences, University of California, Berkeley, CA 94720, USA. E-mail: \{zhichen98, min.du\}@berkeley.edu.
% \IEEEcompsocthanksitem F. Li is with the Department of Computer Science, University of Utah, Salt Lake City, UT 8411, USA. E-mail: lifeifei@cs.utah.edu.
\IEEEcompsocthanksitem G. Chen is with the Department of Electrical Engineering, City University of Hong Kong, Hong Kong SAR, China. E-mail: eegchen@cityu.edu.hk.
\IEEEcompsocthanksitem J. Zhou and Z. Chen make equal contribution.
\IEEEcompsocthanksitem Corresponding author: Qi Xuan.
}

%\IEEEcompsocitemizethanks{\IEEEcompsocthanksitem M. Shell was with the Department
%of Electrical and Computer Engineering, Georgia Institute of Technology, Atlanta,
%GA, 30332.\protect\\
%E-mail: see http://www.michaelshell.org/contact.html
%\IEEEcompsocthanksitem J. Doe and J. Doe are with Anonymous University.}% <-this % stops an unwanted space
\thanks{
%  \\
%  \\
%  \\
%  \\
%  \\
%  \\
}
}

\markboth{IEEE TRANSACTIONS ON KNOWLEDGE AND DATA ENGINEERING}%
{Shell \MakeLowercase{\textit{et al.}}: Bare Demo of IEEEtran.cls for Computer Society Journals}


\IEEEtitleabstractindextext{%
\begin{abstract}
    \justifying
	Community detection, which focuses on clustering vertex interactions, plays a significant role in network analysis. However, it also faces numerous challenges like missing data and adversarial attack. How to further improve the performance and robustness of community detection for real-world networks has raised great concerns. 
    In this paper, we explore \emph{robust community detection} by enhancing network structure, with two generic algorithms presented: one is named \emph{robust community detection via genetic algorithm} (\emph{RobustECD-GA}), in which the modularity and the number of clusters are combined in a fitness function to find the optimal structure enhancement scheme; the other is called \emph{robust community detection via similarity ensemble} (\emph{RobustECD-SE}), integrating multiple information of community structures captured by various vertex similarities, which scales well on large-scale networks.
    Comprehensive experiments on real-world networks demonstrate, by comparing with two traditional enhancement strategies, that the new methods help six representative community detection algorithms achieve more significant performance improvement. 
    Moreover, experiments on the corresponding adversarial networks indicate that the new methods could also optimize the network structure to a certain extent, achieving stronger robustness against adversarial attack. 
    The source code of this paper is released on \url{https://github.com/jjzhou012/RobustECD}.
\end{abstract}

}

\maketitle

\IEEEdisplaynontitleabstractindextext

\IEEEpeerreviewmaketitle

\IEEEraisesectionheading{\section{Introduction}\label{sec:introduction}}

% \IEEEPARstart{R}{ecently,} as an interdiscipline, network science has been widely applied to model complex systems in different fields like sociology, biology, transportation and computer science~\cite{durugbo2011modelling,strogatz2001exploring,newman2003structure}. Real-world networks share various common properties such as power-law degree distribution~\cite{barabasi2009scale,barabasi1999emergence}, 
% small-world features~\cite{watts1998collective}, and community structures~\cite{girvan2002community}. 

\IEEEPARstart{C}{ommunity} detection, or network clustering, which aims to identify groups of interacting vertices in a network in term of their structural properties, has recently attracted considerable attention from different fields like sociology, biology and computer science~\cite{strogatz2001exploring,newman2003structure}.
Community structure is of ultra importance in network analysis. Typically in networks, vertices are organized into groups, called {\em {communities, clusters}} or {\em {modules}}, with dense connections within groups and sparse connections between them. 
For instance, in co-author networks, communities are formed by scientists with similar research interests in close fields; in social networks like Facebook, they can represent people focusing on similar topics.
Many recent researches suggest that network properties at the community level are quite different 
from those at the global level, and thus ignoring community structure may miss many interesting features~\cite{newman2006finding}. 
In fact, identifying communities in networks has played a significant role in exploiting essential network structures.

Until now, a large number of techniques have been developed to detect community structures in networks.
However, despite the advance of various community detection methods, from spectral method, label propagation to deep learning, their capability to discover the true community structure faces numerous challenges, since these approaches strongly rely on the topological structure of the underlying network, which is vulnerable in real-world scenarios.

First, missing data and adversarial noise seriously affect the performance of community detection algorithms.
Real-world networks are often flawed in integrity and suffer from missing data, since not all real-world relationships are reflected in a single network. 
For instance, users in social networks like Twitter seldom follow all their friends in activities. Moreover, missing data also occurs when crawling datasets from online networks with privacy restrictions.
On the other hand, the accuracy of a network is very likely to be questioned when the information encoded in the network topology is perturbed by artificial noise, especially when the network suffers from adversarial attacks, which leads to the degradation of the performance of many network analysis methods. In particular, adversarial attacks against community detection aim to hide target communities or sensitive edges~\cite{chen2019ga,fionda2017community}, and finally generate specific adversarial networks, which can strongly impact the performance of detection algorithms. 
Existing community detection methods rarely consider missing data and adversarial noise in networks, increasing the risk to obtain wrong community structures. 

Another challenge is the lack of a consensus on the formal definition of a network community structure~\cite{ciglan2013community}. 
Currently, there are no universal standards for the definition of community, and a large number of  detection algorithms based on different technologies and ideas have been proposed, which led to a quality discrepancy among different results.
Moreover, modularity optimization in community detection has a resolution limit~\cite{fortunato2007resolution}. 
Clusters consisting of a number of vertices smaller than a threshold would not be detected because these clusters tend to merge into larger ones by modularity optimization. 
Large, but locally sparse communities probably tend to be subdivided into smaller ones during community partition.

It is believed that such challenges are mostly from unstable network structures.
Networks with sparse community structures are vulnerable to adversarial attacks which can destroy network structures, leading to community detection deception. 
Generally, communities with weak structures could be absorbed from the outside or disintegrated from the inside of the network.
% , and are easily disturbed by imperceptible noise. 
Enhancing the network structure and improving the robustness of the network could be an effective way to address these challenges.
In this paper, we explore \emph{robust community detection} by enhancing network structures, and develop two algorithms.
A heuristic idea comes from the fact that community structures show a high connection density of intra-communities and a sparse one of inter-communities. Agglomerating the intra-communities by adding edges between internal vertices and dividing the inter-communities by removing edges between communities, therefore, can strengthen the community structure in a network.
It's a natural reversal of the studies about community detection deception in \cite{chen2019ga,fionda2017community} where they are proposed to weaken community structures via intra-community edge deletion and inter-community edge addition.
Another idea for robust community detection is to enhance network structure with edge prediction, of which the task is to complement missing edges or predict future edges between pairwise vertices based on the current network structure.
The vertex similarity indices can be used to guide network structure optimization, according to the following two assumptions: 1) vertices in the same community are aggregated based on their high similarity; 2) a larger similarity of pairwise vertices leads to a higher likelihood of edges between them~\cite{lu2011link}.
% contribution
The main contributions of our work are summarized as follows:
\begin{itemize}
    \item 
    First, we study the \emph{\uline{\textbf{Robust}} \uline{\textbf{C}}ommunity \uline{\textbf{D}}etection} via network structure \uline{\textbf{E}}nhancement (\emph{RobustECD}), which can improve the performance of existing detection algorithms. 
    To the best of our knowledge, our work is the first for enhancing community detection in both real-world networks and adversarial networks. 
    \item 
    Second, we develop two generic enhancement algorithms, namely \emph{robust community detection via genetic algorithm} (\emph{RobustECD-GA}) and \emph{robust community detection via similarity ensemble} (\emph{RobustECD-SE}).
    Experimental results in six real-world networks demonstrate the superiority of our methods in helping six community detection algorithms to achieve significant improvement of performances.
    \item 
    Third, we test our enhancement algorithms on four adversarial networks, the results show that both \emph{RobustECD-GA} and \emph{RobustECD-SE} can optimize the network structure to a certain extent, and achieve robust community detection against adversarial attack.
    %  destroyed by missing data or artificial noise, to a certain extent, 
    %  and achieve stronger robustness against missing data and artificial noise.
    \item
    Finally, our methods could alleviate the resolution limit in modularity optimization, and help various community detection algorithms to achieve consensus, i.e., getting more consistent partitions.
    % Finally, since our methods are designed to resolve the resolution limit in modularity optimization, they can help various community detection algorithms to achieve consensus, i.e., getting consistent partition.
\end{itemize}
The rest of the paper is organized as follows. First, in Sec.~\ref{sec:related-work}, we review the related works. Then, in Sec.~\ref{sec:method}, we describe our approaches in detail. Thereafter, we present extensive experiments in Sec.~\ref{sec:experiment}, with a series of discussions. Finally, we conclude the paper and outline future work in Sec.~\ref{sec:conclusion}.

\section{Related Work} \label{sec:related-work}
\subsection{Community Detection}
Community detection strives to identify groups of interacting vertices by maximizing cluster quality measures such as modularity~\cite{newman2004finding} and normalized mutual information~\cite{danon2005comparing}. 
The literature~\cite{mohamed2019comprehensive} has provided comprehensive reviews on community detection.
For community detection in undirected networks, widely used methods concentrate on agglomerative~\cite{blondel2008fast}, divisive~\cite{girvan2002community, newman2006finding}, hierarchical~\cite{girvan2002community, clauset2004finding}, spectral~\cite{von2007tutorial, fortunato2010community}, random walk~\cite{rosvall2008maps, zlatic2010topologically}, label propagation~\cite{raghavan2007near, li2020community}, high-order~\cite{huang2020mumod} and deep learning~\cite{bo2020structural, fan2020one2multi} methods.
\vspace{-10pt}
\subsection{Traditional Enhancement of Community Detection}
Due to the deficiency of many detection methods, how to improve their performance in complicated real applications has become an important issue.
In this paper, we focus on the problem of enhancing existing community detection methods.
Existing traditional enhancement approaches suggest preprocessing networks via weighting or rewiring.
% Most strategies suggest first preprocessing networks and then feeding them into community detection algorithms so as to improve their performances. 
% For instance, it was found that the resolution limits of modularity optimization can be alleviated by weighting network edges in different ways, which make it more suitable for community detection. 
Meo et al.~\cite{de2013enhancing} introduced a measure of $\kappa$-path edge centrality and proposed a weighting algorithm called WERW-Kpath to effectively compute the centrality as edge weight, which is better for community detection. 
Sun~\cite{sun2014weighting} weighted networks via a series of edge centrality indices and detected communities in the weighted network using a function that considers both links and link weights. 
Lai et al.~\cite{lai2010enhanced} considered random walk for simulation on dynamic processes, and applied it to enhance modularity optimization, based on the intuition that pairwise vertices in the same community have similar dynamic patterns.
% Interestingly, Li et al.~\cite{li2019edmot} proposed an edge enhancement approach for motif-aware community detection, called EdMot, which not only can leverage higher-order 
% connections of the network, but also can resolve the hypergraph fragmentation issue.
Interestingly, Li et al.~\cite{li2019edmot} considered motif-aware community detection which achieves community partition using motif information in networks, and proposed an edge enhancement approach called Edmot.
Their method transfers the network into motif-based hypergraph and partitions it into modules, and then a new edge set is constructed to enhance the connectivity structure of the original network by fully connecting all modules.
Lancichinetti et al.~\cite{lancichinetti2012consensus} proposed consensus clustering algorithm, which combines the information of different outputs to obtain a more representative partition, to analyze the time evolution of clusters in dynamics networks. 
Dahlin et al.~\cite{1997Ensemble} proposed the ensemble cluster that combines the ensemble method with clustering, and improve community detection by aggregating multiple runs of algorithms.
On the other hand, model-based methods tend to integrate the enhancement into the whole community detection procedure. For example, He et al.~\cite{he2016model} provided a framework to enhance the ability of non-negative matrix factorization (NMF) models to detect communities, which uses the NMF method to train a stochastic model constrained by vertex similarity.
\vspace{-10pt}
\subsection{Adversarial Attack on Community Detection}
In this paper, since some experiments are conducted on networks with adversarial noise which are generated via adversarial attack~\cite{759851e20d2e47aaad2a560211f6a126}, we briefly review the research on adversarial attack for community detection.
Waniek et al.~\cite{waniek2018hiding} proposed a simple heuristic method deployed by intra-community edge deletion and inter-community edge addition, and introduced a measure of concealment to express how well a community is hidden. 
Fionda et al.~\cite{fionda2017community} introduced and formalized the community deception problem, and proposed a community deception algorithm based on safeness, which achieves a success in hiding a target community. 
Chen et al.~\cite{chen2019ga} proposed an effective evolutionary computation strategy, namely genetic algorithm (GA)-based $\mathcal{Q}$-Attack, to achieve deception by negligibly rewiring networks.
Li et al.~\cite{li2020adversarial} proposed an end-to-end graph neural framework that combines graph generator and graph partitioner, and achieved the generation of adversarial examples of high quality and generalization.
\vspace{-5pt}
\section{Methodology} \label{sec:method}
In this section, we first formulate the problem of community detection, and then present two enhancement strategies. The main notations used in this paper are listed in TABLE~\ref{tb:notation}.
% table1
\begin{table}[!t]
    \renewcommand\arraystretch{1.2}
    \centering
    \setlength{\abovecaptionskip}{-3pt}
    \caption{Main notations used in this paper.}
    \label{tb:notation}
    \resizebox{\linewidth}{!}{%
    \begin{tabular}{l r} 
    \hline\hline
    Symbol & Description                                    \\ 
    \hline
    $\mathcal{G}$                    & The target network                                \\
    $\mathcal{V,E,M}$                & Sets of vertices, edges, communities in $\mathcal{G}$  \\
    $n,m$                            & Numbers of vertices, edges in $\mathcal{G}$         \\
    $v,e$                            & Vertex, edge in $\mathcal{G}$               \\
    $\mathcal{S}$                    & The studied community detection algorithm            \\
    $\mathcal{M}_\textit{S},\phi_\textit{S}$& Set/Number of communities found by $\textit{S}$ in $\mathcal{G}$\\
    $\mathcal{M}_\textit{real},\phi_\textit{real}$& Set/Number of the ground-truth communities in $\mathcal{G}$\\
    % $\mathcal{E}_\textit{mod}$       & The robust enhancement scheme                \\
    $\mathcal{E}_\textit{add}, \mathcal{E}_\textit{del}$  & The schemes of edge addition/deletion     \\
    % $\mathcal{G}^{\ast}$             & The rewired network                               \\
    % $\mathcal{E}^{\ast}, \mathcal{M}^{\ast}$    & Sets of edges, communities in graph $\mathcal{G}^{\ast}$     \\  
    $\beta_\textit{a}, \beta_\textit{d}$     & budget of edge addition/deletion \\
    $\mathcal{Q}$                    & Modularity \\
    \hline
    % $\mathcal{P}$                 & Population \\
    % $\mathcal{F}$                 & Set of fitness in population $\mathcal{P}$  \\
    $\phi_\textit{p}$             & Size of population  \\
    % $\mathcal{P}_\textit{c}$             & Crossover rate                        \\
    % $\mathcal{P}_\textit{m}$             & Mutation rate                         \\
    % $\mathcal{P}_\textit{e}$             & Elitist pres rate                     \\
    $\mathcal{T}_\textit{ga}$            & Number of iterations                   \\
    \hline
    $\mathcal{G}_\textit{co}, \mathcal{A}_\textit{co}$  & Co-occurrence network and its adjacency matrix   \\      
    $\mathcal{T}$                 & Threshold of prune in $\mathcal{G}_\textit{co}$   \\
    $\mathcal{H}$                 & Similarity metric (or similarity matrix)  \\
    $\mathcal{G}_\textit{co}^{\mathcal{T}}$& Co-occurrence graph pruned with threshold $\mathcal{T}$\\
    $\mathcal{M^T}$     & Community partition in pruned graph $\mathcal{G}_\textit{co}^{\mathcal{T}}$\\
    $c,C$     & Consensus of a cluster/partition \\
    % $C$     & Consensus score of cluster partition   \\
    % $z$     & Number of partitions in \emph{RobustECD-SE}-E        \\
    % $\mathcal{A}_{\mathsf{cn}} , \ldots, \mathcal{A}_{\mathsf{rwr}}$    & Similarity score matrices for graph $\mathcal{G}$   \\
    % \hline
    % $\mathcal{K}$             & Number of selected connected components (EdMot).                             \\
    % \midrule[1pt]
    % $\kappa$           &  Length of walk path (WERW-$\kappa$Path)  \\
    % $\rho$             &  Iterations of walk (WERW-$\kappa$Path)   \\
    % \midrule[1pt]
    % $\beta$      & Attack cost in adversarial attack algorithms   \\
    \hline\hline
    \end{tabular}
    }
    \vspace{-8pt} 
\end{table}
\vspace{-5pt}  
\subsection{Problem Formulation}
Assume that an undirected and unweighted network is represented by a graph $\mathcal{G = (V,E)}$, 
which consists of a vertex set $\mathcal{V}=\left\{v_{i} \mid i=1, \ldots, n\right\}$ and an edge set $\mathcal{E}=\left\{e_{i} \mid i=1, \ldots, m\right\}$.
The topological structure of graph $\mathcal{G}$ is represented by an $n \times n$ adjacency matrix $\mathcal{A}$ with $\mathcal{A}_{ij}=1$ if $(v_i, v_j) \in \mathcal{E}$ and $\mathcal{A}_{ij}=0$ otherwise.
The nonexistent edge set $\bar{\mathcal{E}}$ is represented by $\left\{\left(v_{i}, v_{j}\right) \mid \mathcal{A}_{i j}=0 ; i \neq j\right\}$.
The task of community detection in a network is to find a vertex partition $\mathcal{M}=\left\{\mathcal{M}_{i} \mid i=1, \ldots, k \right\}$, with 
$\bigcup \mathcal{M}_{i}=\mathcal{V}$ and $\mathcal{M}_{i} \bigcap \mathcal{M}_{j}=\emptyset$ for $i \neq j$, where set $\mathcal{M}_{i}$ is called a {\em{community}}. 
The ground-truth community partition of the network is denoted as $\mathcal{M}_\textit{real}$. Note that the community overlapping problem will not be considered in this paper.

We further explore the network structure enhancement from heuristic and optimized approaches. In the enhancement scenario, a network will be rewired via edge modification, during which the edges removed from network are sampled from the candidate set $\mathcal{E}_\textit{del}^\textit{c}$, while the edges added to the network are sampled from the candidate pairwise vertices set $\mathcal{E}_\textit{add}^\textit{c}$. The construction of candidate sets varies for different methods, as further discussed below.

For a network $\mathcal{G}$, one can get the set of edges added/removed from $\mathcal{G}$ via sampling from the candidate sets:
\begin{equation}
    \setlength{\abovedisplayskip}{3pt}
    \setlength{\belowdisplayskip}{3pt}
    \begin{aligned}
    &\mathcal{E}_{\textit{del}}=\left\{\tilde{e}_{i} \mid i=1, \ldots,\left\lceil m \cdot\beta_\textit{d}\right\rceil\right\} \subset \mathcal{E}_{\textit{del}}^\textit{c} \ , \\
    &\mathcal{E}_{\textit{add}}=\left\{\tilde{e}_{i} \mid i=1, \ldots,\left\lceil m \cdot\beta_\textit{a}\right\rceil\right\} \subset \mathcal{E}_{\textit{add}}^\textit{c} \ ,    
    \end{aligned}
\end{equation}
where $\beta_\textit{a}$, $\beta_\textit{d}$ are the budget of edge addition/deletion and $\lceil x\rceil=\mathbf{ceil}(x)$.
Then, based on the modification scheme $\mathcal{E}_\textit{mod} = (\mathcal{E}_\textit{add}, \mathcal{E}_\textit{del})$, the connectivity structure of the original network is optimized to generate a rewired network:
\begin{equation} \label{eq:rewire}
    \setlength{\abovedisplayskip}{3pt}
    \setlength{\belowdisplayskip}{3pt}
    \mathcal{G}^{\ast}=(\mathcal{V}, \mathcal{E}^{\ast} ) \quad {\rm with} \  \mathcal{E}^{\ast} = \mathcal{E} \cup \mathcal{E}_\textit{add} \backslash \mathcal{E}_\textit{del}. 
\end{equation}
% In this way, we can find the solution $\mathcal{E}_\textit{mod}$ to optimize the network structure by a rewiring process.
%without changing the vertices and get the rewired network $\mathcal{G}^{\ast}$.
For the rewired networks obtained via enhancement, we expect that the community detection methods perform significantly better and the new partition $\mathcal{M}^{\ast}$ is closer to the ground-truth communities, i.e., there is a significant improvement in evaluation metrics after assigning $\mathcal{M}^{\ast}$ to $\mathcal{G}$.
% \vspace{-9pt}
\subsection{Modularity-Based Structure Enhancement}
\begin{figure*} 
    \centering
    \includegraphics[width=\textwidth]{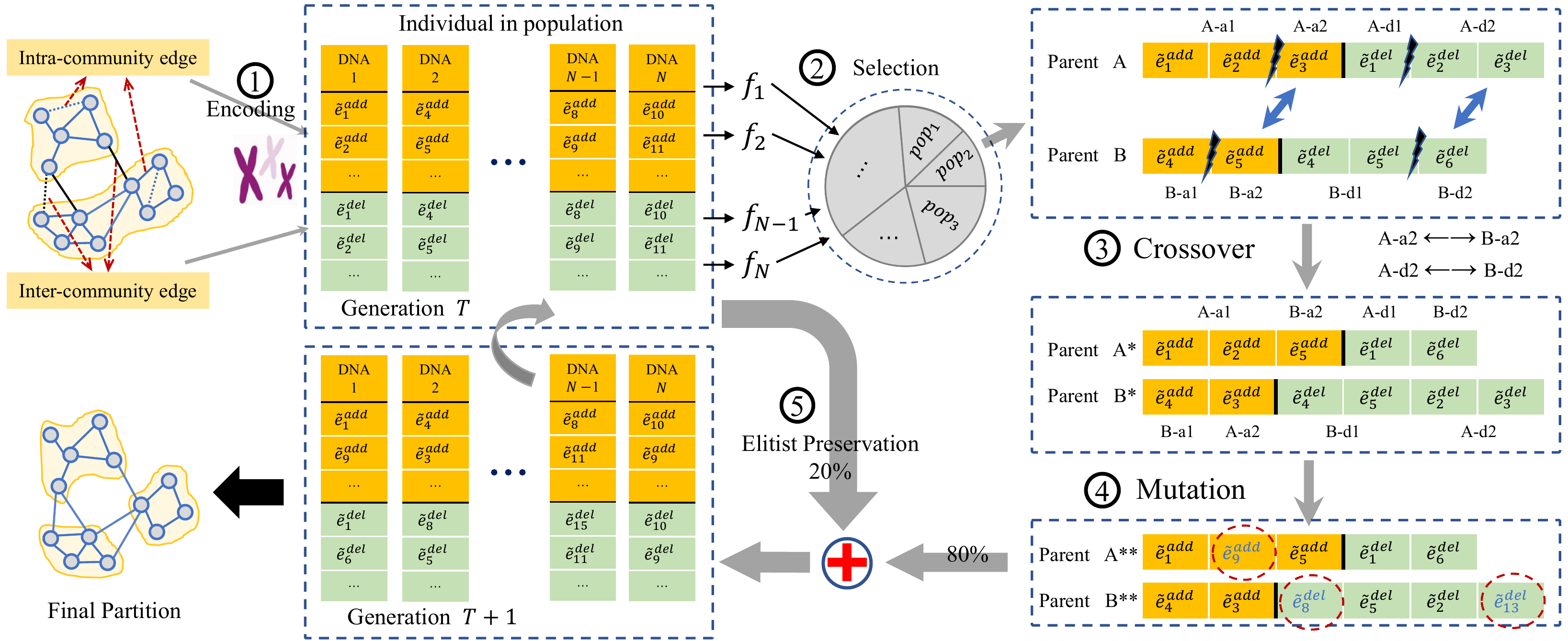}
    \setlength{\abovecaptionskip}{-10pt}
    \caption{Schematic depiction of \emph{RobustECD-GA}. 
             The workflow of evolution iteration proceeds as follows: 
             1) chromosome encoding for population initialization;
             2) fitness calculation and individual selection;
             3) chromosome crossover;
             4) chromosome mutation;
             5) elitist preservation.}
    \label{fig:framework-RobustECD-GA}
    \vspace{-13pt}
\end{figure*}

% Many real-world complex networks have relatively distinct community structure, showing a high 
% connection density of intra-community and a sparse one of inter-community. Agglomerating the 
% intra-community by adding edges between internal vertices and divisiving inter-community 
% by deleting edges between communities, thus, can strengthen the community structures in original % networks and repair the broken structures in adversarial ones.
% Several related works \cite{fionda2017community, waniek2018hiding} have verified that intra-community 
% edge deletion and inter-community edge addition can effectively deploy community deception attack, 
% which in turn also shows that intra-community edge addition and inter-community edge deletion can
% stabilize the community structure.
% Based on the fact that community structures show a high connection density of intra-communities and a sparse one of inter-communities, 
Previous works~\cite{chen2019ga,fionda2017community, waniek2018hiding} have shown that intra-community edge deletion and inter-community edge addition can facilitate the deployment of community deception attacks.
By contrast, it is natural that agglomerating the intra-communities by adding edges between internal vertices and dividing the inter-communities by removing edges between communities, i.e. intra-community edge addition and inter-community edge deletion, can strengthen the community structures in a network.
Meanwhile, the resolution limitation problem that cannot be neglected requires the proposed approach to be capable of combining these four basic community edge modifications organically.
Therefore, based on modularity, we propose the first method, named \emph{robust community detection via genetic algorithm} (\emph{RobustECD-GA}), which aims to enhance community structure via adaptable community edge rewiring.
The schematic depiction of \emph{RobustECD-GA} is shown in Fig.~\ref{fig:framework-RobustECD-GA}.
% \vspace{-6pt}
\subsubsection{Network Rewiring}%\label{aemga-netrewire}
% Therefore, by contrast, edge modification schemes in \emph{RobustECD-GA} are designed with two basic operations, i.e., intra-community edge addition and inter-community edge deletion, which can stabilize the community structure. 
%Therefore, having a general prior knowledge of community structure may facilitate the deployment of our enhancement strategies.
%In \emph{RobustECD-GA}, edge modification schemes consist of two basic operations, i.e., intra-community edge addition and inter-community edge deletion.
% Given a network $\mathcal{G}$, 
% the neighbors of a target vertex $v_{i}$ is denoted as $\mathcal{V}^{-}_{i} = \left\{v_{j} \mid v_{j} \neq v_{i}; (v_{i}, v_{j}) \in \mathcal{E} \right\}$,
% while its nonneighbor set is denoted as $\mathcal{V}^{+}_{i} = \mathcal{V} \backslash (\mathcal{V}^{-}_{i} \cup \{v_{i}\})$.
Given a network $\mathcal{G}$, a community edge optimization strategy requires knowledge of the community structure to pick the optimal edge modification schemes and thus depends on the prior community detection algorithm $\mathcal{S}$ that generates the estimated partition 
$\mathcal{M}_\textit{S}=\left\{\mathcal{M}_{i} \mid i=1, \ldots, k \right\}$.
For arbitrary pairwise vertices $(v_i,v_j)$, the candidate sets of four basic community edge modifications are represented as follows:
\begin{equation} \label{eq:comm-edge-op}
    \setlength{\abovedisplayskip}{4pt}
    \setlength{\belowdisplayskip}{4pt}
    \begin{aligned}
        &\mathcal{E}_{\textit{intra-add}}^{\textit{c}}=\{(v_{i},v_{j}) \mid v_{i},v_{j}\in\mathcal{M}_{i}, \mathcal{A}_{ij}=0 \} \ , \\
        &\mathcal{E}_{\textit{intra-del}}^{\textit{c}}=\{(v_{i},v_{j}) \mid v_{i},v_{j}\in\mathcal{M}_{i}, \mathcal{A}_{ij}=1 \} \ , \\
        &\mathcal{E}_{\textit{inter-add}}^{\textit{c}}=\{(v_{i},v_{j}) \mid v_{i}\in\mathcal{M}_{i}, v_{j}\in\mathcal{M}_{j}, \mathcal{A}_{ij}=0 \} \ , \\
        &\mathcal{E}_{\textit{inter-del}}^{\textit{c}}=\{(v_{i},v_{j}) \mid v_{i}\in\mathcal{M}_{i}, v_{j}\in\mathcal{M}_{j}, \mathcal{A}_{ij}=1 \} \ , 
    \end{aligned}
\end{equation}
where $\mathcal{M}_{i},\mathcal{M}_{j} \in \mathcal{M}_\textit{S}$, $\mathcal{E}_{\textit{intra-del}}^{\textit{c}} \cup \mathcal{E}_{\textit{inter-del}}^{\textit{c}} =\mathcal{E}$ and  $\mathcal{E}_{\textit{intra-add}}^{\textit{c}} \cup \mathcal{E}_{\textit{inter-add}}^{\textit{c}}=\bar{\mathcal{E}}$.

The adaptable community edge rewiring consists of two parts, one is the required intra-community edge addition and inter-community edge deletion, and the other is an optional part that depends on the comparison between the number of communities in the estimated partition and the one in the ground truth:
\begin{equation}\label{eq:adaptable-mod}
    \setlength{\abovedisplayskip}{4pt}
    \setlength{\belowdisplayskip}{4pt}
    \mathcal{E}_{\textit{mod}}=
    \begin{cases}
        \left(\mathcal{E}_{\textit{add}} \subset \bar{\mathcal{E}}, \mathcal{E}_{\textit{del}} \subset \mathcal{E}_{\textit{inter-del }}^{\textit{c}}\right)& {\phi_{\textit{S}}>\phi_{\textit{real}}}\\
        \left(\mathcal{E}_{\textit{add}} \subset \mathcal{E}_{\textit{intra-add }}^{\textit{c}}, \mathcal{E}_{\textit{del}} \subset \mathcal{E} \right)& {\phi_{\textit{S}}<\phi_{\textit{real}}}\\
        \left(\mathcal{E}_{\textit{add}} \subset \mathcal{E}_{\textit{intra-add }}^{\textit{c}}, \mathcal{E}_{\textit{del}} \subset \mathcal{E}_{\textit{inter-del }}^{\textit{c}}\right)& {\phi_{\textit{S}}=\phi_{\textit{real}}}
    \end{cases}
\end{equation}
where $\phi_{\textit{real}}$ is the number of the ground-truth communities and $\phi_{\textit{S}}$ is the number of communities in the estimated partition $\mathcal{M}_\textit{S}$. 
Eq~(\ref{eq:adaptable-mod}) lists three scenarios caused by resolution limit during initialization of edge modification:
% \vspace{-1.5pt} 
\begin{itemize}
    \item {}
    When $\phi_{\textit{S}}>\phi_{\textit{real}}$, there is a relatively high resolution, and large but locally sparse communities tend to be subdivided into smaller fragments. Ideally, extra inter-community edge addition is conducive to aggregate those fragments into integrate communities;
    \item {}
    When $\phi_{\textit{S}}<\phi_{\textit{real}}$, there is a relatively low resolution, and clusters consisting of a number of vertices smaller than a threshold tend to merge into larger ones. Ideally, extra intra-community edge deletion is conducive to subdivide large clusters;
    \item {}
    When $\phi_{\textit{S}}=\phi_{\textit{real}}$, there is a relatively suitable resolution. Both the inter-community edge addition and intra-community edge deletion are inoperative.
\end{itemize}
Notably, the aforementioned adaptable community edge rewiring requires knowledge of the ground-truth community (i.e., $\mathcal{M}_\textit{real}$ and $\phi_{\textit{real}}$). 
When it comes to the dilemma that ground-truth communities are not available, the rewiring mechanism preserves only the required part, i.e., intra-community edge addition and inter-community edge deletion.
\begin{figure} 
    \centering
    \includegraphics[width=\linewidth]{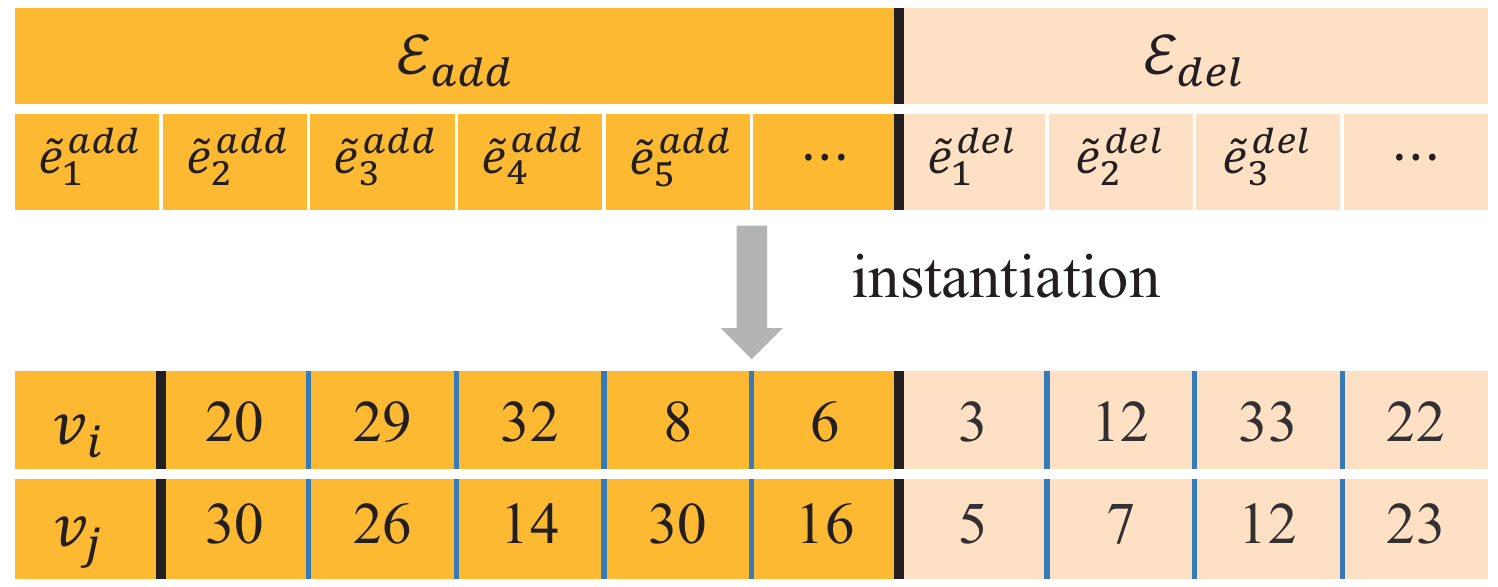}
    \setlength{\abovecaptionskip}{-10pt}
    \caption{The diagram of chromosome in \emph{RobustECD-GA}. It consists of two parts including edge addition segment $\mathcal{E}_\textit{add}$ and edge deletion segment $\mathcal{E}_\textit{del}$. The instance of chromosome is initialized in the experiment for Karate dataset, with an edge addition segment of length 5 and an edge deletion segment of length 4.}
    \label{Fig:chromosome}
    \vspace{-13pt}
\end{figure}
% \vspace{-5pt}
\subsubsection{Evolutionary Optimization}
We use the genetic algorithm (GA) due to its good performance in solving combinatorial optimization problems. Specifically, we design the encoding scheme of chromosome and the function of fitness as follows.
% \vspace{-5pt}
\begin{itemize}
    \item {\textbf{Chromosome}.}
    Chromosome represents an edge modification scheme $\mathcal{E}_\textit{mod}$, consisting of two parts: $\mathcal{E}_\textit{add}$ and $\mathcal{E}_\textit{del}$, where a gene denotes an edge modification operation, including edge addition or deletion. The diagram of chromosome is shown in Fig.~\ref{Fig:chromosome}. 
    \item {\textbf{Fitness}.}    
    Modularity~\cite{newman2004finding} has been widely applied in community detection task and is an effective evaluation metric to assess the quality of network partition (more details please refer to Sec.~\ref{sec:evaluation-measures}). 
    Here, the fitness is defined as:
    \begin{equation}
        \setlength{\abovedisplayskip}{3pt}
        \setlength{\belowdisplayskip}{3pt}
        % f =  \frac{|\mathcal{Q}|}{e^{|\phi_\textit{S} - \phi_\textit{real}|}} \ ,   
        f =  |\mathcal{Q}| / e^{|\phi_\textit{S} - \phi_\textit{real}|} \ ,
    \end{equation}
    where $\mathcal{Q}$ is the modularity of the partition for the target network and $e$ is Euler's number.
    Denominator $e^{|\phi_\textit{S} - \phi_\textit{real}|}$ is typically chosen to impose a penalty on the size of resolution. 
    Modularity is divided by a large penalty term which is no less than $e$ when $\phi_\textit{S} \neq \phi_\textit{real}$, and the fitness function degenerates to modularity when $\phi_\textit{S} = \phi_\textit{real}$ or no access to the ground-truth community.
    Individuals with larger modularity and more accurate partition generally have larger fitness.
\end{itemize}

% Note that 
% algorithm 1
\begin{algorithm}[!th]
    \caption{\emph{RobustECD-GA}}  
    \LinesNumbered  
    \label{alg:RobustECD-GA} 
	\KwIn{Target network $\mathcal{G}$, community detection algorithm $\mathcal{S}$, parameter for GA($\phi_\textit{p}$, $\mathcal{P}_{c}$,  $\mathcal{P}_\textit{m}$, $\mathcal{P}_\textit{e}$,  $\mathcal{T}_\textit{ga}$), budget $\beta_\textit{a}$, $\beta_\textit{d}$ .}  
    \KwOut{New community partition $\mathcal{M}^{\ast}$}  
    $\mathcal{M}_\textit{S} \leftarrow \mathsf{detectCommunity} (\mathcal{S,G}) $ \;
    $\mathcal{P, F} \leftarrow \mathsf{initializePop} (\mathcal{G},\mathcal{M}_\textit{S}, \phi_\textit{p},\beta_\textit{a},\beta_\textit{d})$ \;
    Initialize current generation $i=0$ \;
    \While{$i < \mathcal{T}_\textit{ga}$ }  
    {  
        $\mathcal{P}_\textit{elitist} \leftarrow \mathsf {retainElitist}(\mathcal{F}, \mathcal{P}, \mathcal{P}_{e})$\;
        $\mathcal{P}_\textit{select} \leftarrow \mathsf {selection}(\mathcal{F}, \mathcal{P})$ \;
        $\mathcal{P}_\textit{crossover} \leftarrow \mathsf {crossover}(\mathcal{P}_\textit{select}, \mathcal{P}_\textit{c})$ \;
        $\mathcal{P}_\textit{mutate} \leftarrow \mathsf {mutation}(\mathcal{P}_\textit{crossover}, \mathcal{P}_\textit{m}, \mathcal{M}_\textit{S})$ \;
        $\mathcal{F} \leftarrow \mathsf {getFitness}(\mathcal{G,S}, \mathcal{P}_\textit{mutate})$ \;
        $\mathcal{P} \leftarrow \mathsf {getNextGeneration}(\mathcal{P}_\textit{mutate}, \mathcal{P}_\textit{elitist}) $    
    }   
    Get the individual with highest fitness from the last population: $\mathcal{E}_\textit{mod} \leftarrow \mathsf {getBestIndividual}(\mathcal{F,P}) $ \;
    Rewire the original network to obtain $\mathcal{G}^{\ast}$ via Eq.~(\ref{eq:rewire}) \;
    Feed $\mathcal{G}^{\ast}$ into $\mathcal{S}$ to generate new community partition: $\mathcal{M}^{\ast} \leftarrow \mathsf{detectCommunity} (\mathcal{S},\mathcal{G}^{\ast})$. \; 
    \textbf{end} \;
    \textbf{return} $\mathcal{M}^{\ast}$;  
\end{algorithm} 
% algorithm 2
\begin{algorithm}[!th]
    \caption{\emph{RobustECD-SE}}  
    \LinesNumbered  
    \label{alg:RobustECD-SE} 
    \KwIn{Target network $\mathcal{G}$, community detection algorithm $\mathcal{S}$, budget $\beta_\textit{a}$.}  
    \KwOut{New community structure $M^{\ast}$}  
    Compute similarity indices listed in TABLE~\ref{tb:similarity}: $\{\mathcal{H}_{\mathsf{CN}}, \ldots, \mathcal{H}_{\mathsf{RWR}}\} \leftarrow \mathsf{computeSimilarity}(\mathcal{G})$  \;
    Obtain rewire schemes via sampling:  $\{\mathcal{E}_\textit{mod}^{1}, \ldots \} \leftarrow \mathsf{sample}(\mathcal{G}, \beta_\textit{a}, \{\mathcal{H}_{\mathsf{CN}}, \ldots, \mathcal{H}_{\mathsf{RWR}}\})$ \;

    Update graph via network rewiring : $\{\mathcal{G}^{\ast}_{1}, \ldots\} \leftarrow \mathsf{rewire}(\mathcal{G}, \{\mathcal{E}_\textit{mod}^{1}, \ldots,  \}) $\;
    Obtain multiple partitions via community detection:   $\{\mathcal{M}^{\ast}_{1}, \ldots\}  \leftarrow \mathsf{detectCommunity}(\mathcal{S}, \{\mathcal{G}^{\ast}_{1}, \ldots\})$ \;
    Get co-occurrence network from multiple partitions:  $\mathcal{G}_\textit{co}, \mathcal{A}_\textit{co} \leftarrow \mathsf{getCoNetwork}(\{\mathcal{M}^{\ast}_{1}, \ldots\})$  \;
    Threshold selection:    $\mathcal{T}, {\mathcal{M}^{\mathcal{T}}_\textit{core}} \leftarrow \mathsf{getOptimalThreshold}(\mathcal{G}_\textit{co}) $   \;
    
    Get the final partition by assigning isolated vertices to core communities:  $\mathcal{M}^{\ast} \leftarrow \mathsf{getFinalPartition}( {\mathcal{M}^{\mathcal{T}}_\textit{core}}, \{\mathcal{H}_{\mathsf{CN}}, \ldots, \mathcal{H}_{\mathsf{RWR}}\})$ \;
    \textbf{end} \;
    \textbf{return} $\mathcal{M}^{\ast}$;  
\end{algorithm} 
% \vspace{-3pt}
\begin{table}[!t]
    \renewcommand\arraystretch{1.2}
    \centering
    \caption{Summary of similarity indices used in this paper. $\mathcal{H}(i,j)$ is the similarity score of pairwise vertices $(v_i,v_j)$, $\Gamma(i)$ is the 1-hop neighbors of $v_i$, $g(\cdot)$ is the nonnegative function under the given network, $\gamma$ is a decaying factor between 0 and 1, $\eta$ is a positive constant/function of $\gamma$.}
    \label{tb:similarity}
    \resizebox{\linewidth}{!}{%
    \begin{tabular}{ccll} 
        \hline\hline
        Order  & \multicolumn{1}{c}{Category}                                   & \multicolumn{1}{c}{Form}                                          & \multicolumn{1}{c}{Used in paper}  \\ 
        \hline
        First  & Local                                                          & $\mathcal{H}(i,j)=|\Gamma(i) \cap \Gamma(j)| \cdot g(i, j)$                & \begin{tabular}[c]{@{}l@{}}Common Neighbors (CN)~\cite{lu2011link}\\Salton~\cite{salton1983introduction}, Jaccard~\cite{jaccard1901etude}\\Hub Promoted Index (HPI)~\cite{ravasz2002hierarchical}\end{tabular}             \\ 
        \hline
        Second & Local                                                          & $\mathcal{H}(i,j)=\sum\limits_{z \in \Gamma(i) \cap(j)} g(z)$            & \begin{tabular}[c]{@{}l@{}}Adamic–Adar Index (AA)~\cite{adamic2003friends}\\Resource Allocation Index (RA)~\cite{zhou2009predicting}\end{tabular}              \\ 
        \hline
        High   & \begin{tabular}[c]{@{}c@{}}Quasi-local,\\Global\end{tabular}   & $\mathcal{H}(i,j)=\eta \sum\limits_{l=1}^{\infty} \gamma^{l} g(i, j, l)$~\cite{zhang2018link} & \begin{tabular}[c]{@{}l@{}}Local Path Index (LP)~\cite{lu2009similarity}\\Random Walk with Restart (RWR)~\cite{brin1998anatomy}\end{tabular}              \\
        \hline\hline
    \end{tabular}
    }
    \vspace{-12pt}
\end{table}
The procedure of \emph{RobustECD-GA} is shown in Algorithm~\ref{alg:RobustECD-GA}. As mentioned above, \emph{RobustECD-GA} requires the knowledge of the community structure, which guides the edge modification. 
We feed the target network $\mathcal{G}$ into community detection algorithm $\mathcal{S}$ to generate a general community partition $\mathcal{M}_\textit{S}$ and then construct the candidate edge sets (line 1). %The core of AE-M-GA is evolutionary optimization.

During \textbf{initialization}, a parental generation 
$\mathcal{P} = \left\{\mathcal{E}_\textit{mod}^i \mid i=1, \ldots ,\phi_\textit{p} \right\}$ 
is randomly generated with a population size $\phi_\textit{p}$ and each individual $\mathcal{E}_\textit{mod}^i$ in the population has an unfixed size, i.e., the quantity of modified edges is not fixed for each initial modification scheme (line 2). 
During \textbf{selection}, the operation is conducted on \emph{roulette}, which means that the probability for an individual to be selected is proportional to its fitness (line 6). 
\textbf{Crossover} is the process of combining the parental generation to generate new schemes and we apply \emph{multi-point crossover} to swap gene segments between two parental chromosomes with a crossover rate $\mathcal{P}_\textit{c}$ (line 7). 
\textbf{Mutation} prevents the algorithm from falling into local optimization. We traverse each gene in the chromosome and conduct the
mutation operation with a mutation rate $\mathcal{P}_\textit{m}$ (line 8). In so doing, we randomly replace the edge modification operation $\tilde e^\textit{add}$ or $\tilde e^\textit{del}$ with another one.
Finally, \textbf{elitist preservation} is applied to retain excellent individuals, which refer to modification schemes with higher fitness. In particular, we retain excellent individuals by replacing the worst 20\% of the offspring with the best 20\% of the parents (line 5). 
Evolution is a process of iteration and we set the number of iterations $\mathcal{T}_\textit{ga}$ as the evolutionary generation. The evolutionary optimization stops when it is convergent or this condition is satisfied.

\subsection{Similarity-Based Structure Enhancement}
\begin{figure*}[!t]
    \centering
    \includegraphics[width=\textwidth]{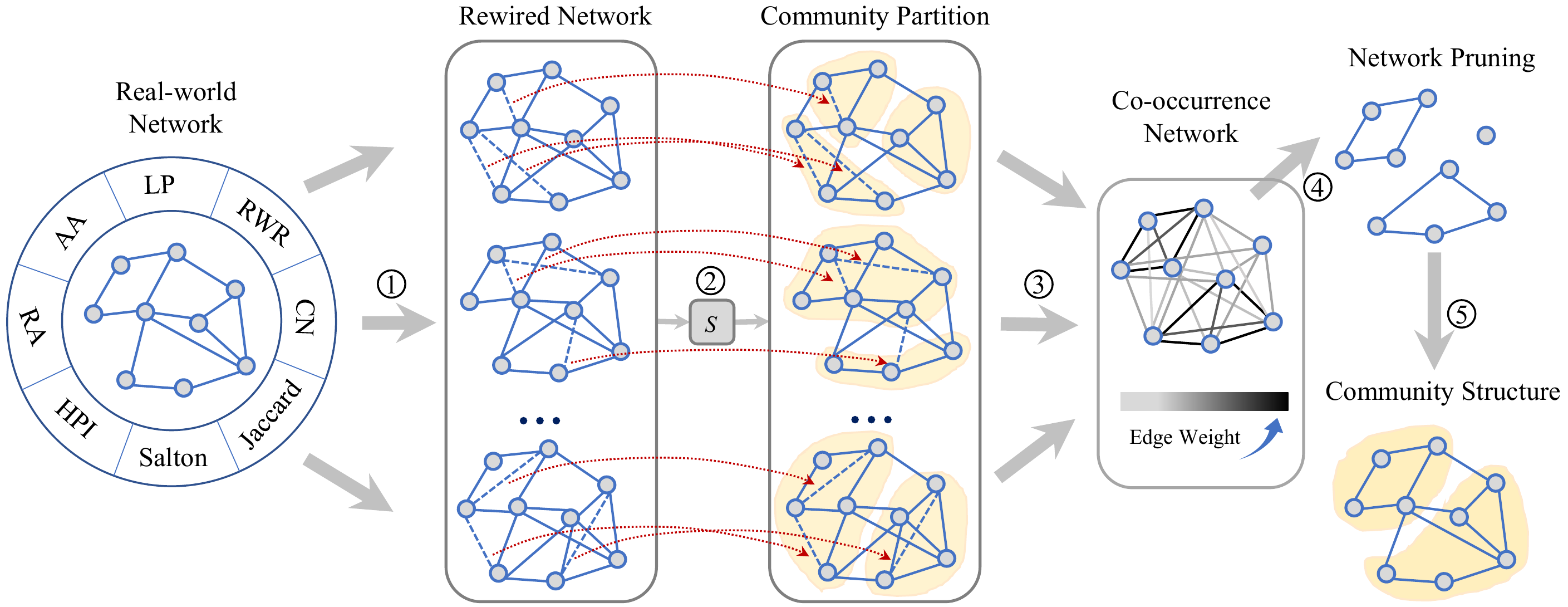}
    \setlength{\abovecaptionskip}{-10pt}
    \caption{Schematic depiction of \emph{RobustECD-SE}. 
             The complete workflow proceeds as follows:
             1) similarity rewiring to generate rewired networks;
             2) community detection to generate community partitions;
             3) partition ensemble to generate co-occurrence network;
             4) network pruning to identify core communities;
             5) isolated vertices reassignment to get final community structure.}
    \label{fig:framework-RobustECD-SE}
    \vspace{-12pt}
\end{figure*}

Empirically, vertices in the same community is aggregated due to their high similarity. 
Vertex similarity can be defined as the number of common features that a pair of vertices share~\cite{lin1998information}. 
Previous works~\cite{lu2011link, zhang2018link} have shown that local, global and random-walk-based similarity indices perform excellently in capturing network structure features, and further unified them into three general forms of heuristics according to the subgraph involved in the similarity calculation, as summarized in TABLE~\ref{tb:similarity} and Appendix~\ref{sec: detail-si}.
% For instance, in social networks, communities are formed by users with similar interests; in citation networks, they represent papers on related topics, etc. A typical application of vertex similarity analysis is link prediction, which is based on current network structure and attempts to complement missing links or predict future links between pairwise vertices. A basic assumption of link prediction via similarity is that if the similarity 
% of pairwise vertices is greater, the likelihood of the existence of links between them could be higher~\cite{lu2011link}. 
Therefore, we adopt the heuristics to aggregate those vertices of high similarity, i.e., considering the vertex similarity indices as the guidance of edge modification. 
Based on this, we propose the second method, named \emph{robust community detection via similarity ensemble} (\emph{RobustECD-SE}), which rewires a network via multiple similarity indices and aggregates corresponding community partitions to generate more accurate community structures.
The schematic depiction of \emph{RobustECD-SE} is shown in Fig.~\ref{fig:framework-RobustECD-SE}, and the procedure of \emph{RobustECD-SE} is shown in Algorithm~\ref{alg:RobustECD-SE}.
% \vspace{-3pt}
\begin{figure*} 
    \centering
    \includegraphics[width=\linewidth]{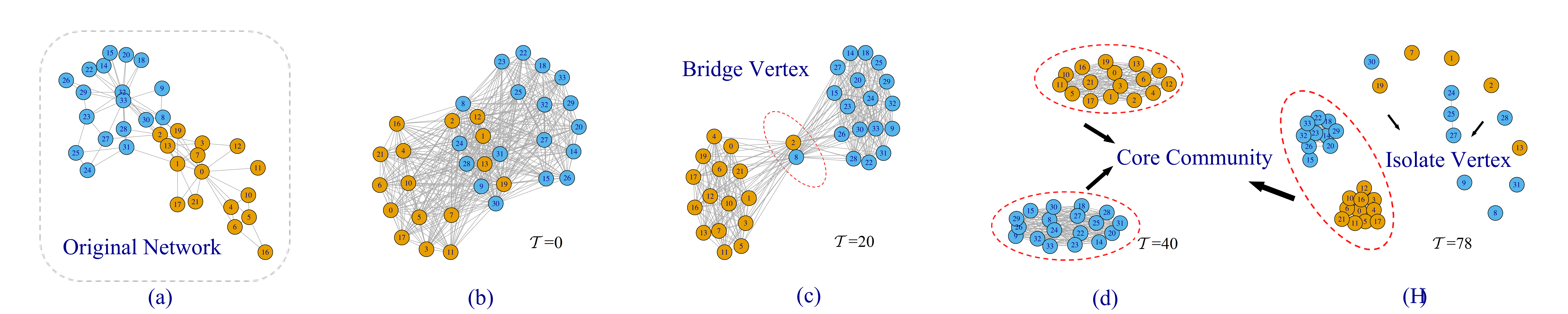}
    \setlength{\abovecaptionskip}{-10pt}
    \caption{Visualization of network pruning in the co-occurrence network of Karate dataset. The last four represent the co-occurrence network pruned with various values of threshold $\mathcal{T}$. Note that vertices with the same color share the same ground truth community label.}
    \label{fig:vis-puring}
    \vspace{-12pt}
\end{figure*}
% \vspace{-10pt}
\subsubsection{Network Rewiring}\label{sec:RobustECD-SE-nr}
In \emph{RobustECD-SE}, the similarity rewiring is a modification of network structure at the global level, so that the candidate set of edge modification is defined as $\mathcal{E}_\textit{add}^\textit{c}=\bar{\mathcal{E}}$.
Note that only edge addition is considered in \emph{RobustECD-SE} and edge deletion is neglected for several reasons: 1) slight improvement of performance; 2) extra time consumption, as further discussed in Appendix~\ref{app:1}.

Given a target network $\mathcal{G}$, the similarity matrix $\mathcal{H}$, which consists of similarity scores of arbitrary pairwise vertices, can be directly calculated (line 1).
During \textbf{similarity rewiring}, we first assign all entries in $\mathcal{E}_\textit{add}^\textit{c}$ with relative weights that are associate with the vertex similarity scores. And we get the $\mathcal{E}_\textit{add}$, the set of edges added to $\mathcal{G}$, via weighted random sampling from $\mathcal{E}_\textit{add}^\textit{c}$ with a budget of $\beta_\textit{a}$, which means that the probability for an entry in $\mathcal{E}_\textit{add}^\textit{c}$ to be selected is proportional to its similarity score $\mathcal{H}_{ij}$ (line 2).
After edge sampling, we update network via Eq.(~\ref{eq:rewire}) (line 3). 

For each similarity index listed in TABLE~\ref{tb:similarity}, we conduct similarity rewiring for certain times, and finally obtain a series of rewired networks.

\subsubsection{Ensemble Optimization}
After network rewiring, we feed all rewired networks into the \textbf{community detection} algorithm $\mathcal{S}$ to generate a series of community partitions (line 4).
Due to the diversity of similarity indices and rewiring schemes, these partitions are likely to be non-unique and not necessarily better than the original partition $\mathcal{M}$. Ensemble learning, which achieves better classification or prediction performance by integrating multiple weak models, has been used for clustering tasks. Previous works on consensus and ensemble clustering~\cite{lancichinetti2012consensus, 1997Ensemble} have
shown that these techniques can be combined with existing clustering methods and improve the stability and accuracy of community partitions.

During \textbf{partition ensemble}, we aggregate multiple partitions using a consensus matrix $\mathcal{A}_\textit{co} = \{a_{ij}\}_{n \times n}$, in which element $a_{ij}$ indicates the frequency of two vertices $v_i$ and $v_j$ assigned to the same community. 
A weighted co-occurrence network $\mathcal{G}_\textit{co}$ can be generated by using
$\mathcal{A}_\textit{co}$ as the adjacency matrix (line 5).
Once pairwise vertices appear in the same community in some partitions,
$\mathcal{G}_\textit{co}$ links them and assigns weights that correspond to the frequency of co-occurrence. A larger/smaller weight means a higher/lower likelihood that the pairwise vertices belong to the same community. %, while a smaller weight means a low confidencethat the algorithms to cluster these vertices. 

For the co-occurrence network $\mathcal{G}_\textit{co}$, a natural idea is to consider those edges with larger/lower weights as intra/inter-community edges in the original network.
Then, we can deploy community edge rewiring in $\mathcal{G}_\textit{co}$ to optimize network structure.
For inter-community edge deletion, we can prune $\mathcal{G}_\textit{co}$ by setting a weight threshold $\mathcal{T}$. During \textbf{network pruning}, all edges with weights less than $\mathcal{T}$ are considered as inter-community edges and will be removed from $\mathcal{G}_\textit{co}$ (line 6).
% We neglect intra-community edge addition for the following reasons:
% 1) all partitions generated during community detection are used to construct $\mathcal{G}_\textit{co}$;
% 2) the addition of new edges in $\mathcal{G}_\textit{co}$ depends on the co-occurrence of pairwise vertices, but there is no access to more new partitions.
We neglect intra-community edge addition since that the addition of new edges in $\mathcal{G}_\textit{co}$ depends on the co-occurrence of pairwise vertices, but there is no access to more new partitions at this stage.

A visualization of network pruning in the co-occurrence network of Karate dataset is shown in Fig.~\ref{fig:vis-puring}. In this paper, we use eight similarity indices, and for each index, ten samplings are performed, to generate a total of eighty partitions, which
determine the range of threshold $\mathcal{T} \in [0, 80]$.
The original Karate network is shown in Fig.~\ref{fig:vis-puring} (a), and there are two communities, with the vertices of the same color sharing the same ground-truth community label.
Fig.~\ref{fig:vis-puring} (b) shows the co-occurrence network, which aggregates the information of eighty partitions and has dense connections.
The last three subgraphs show the different pruned co-occurrence networks with various thresholds. 
When $\mathcal{T} = 20$, the pruned co-occurrence network still has only one connected component but two bridge vertices emerge, as shown in Fig.~\ref{fig:vis-puring} (c). 
With the increase of the threshold, $\mathcal{G}_\textit{co}$ is pruned to two connected components, matching exactly with the two clusters in the original network, as shown in Fig.~\ref{fig:vis-puring} (d).
%and an interesting finding is that such pruning achieves a completely correction in community partition. 
When the threshold approaches the upper limit, generally, we'll get several small connected components that contain few vertices, or even isolated vertices, as shown in Fig.~\ref{fig:vis-puring} (e).
%The impact of threshold $\mathcal{T}$ on network pruning is intuitively shown in Fig.~\ref{fig:vis-puring}.
This phenomenon indicates that the selection of threshold actually influences the result of community partition, which is similar to the resolution limit problem in community detection. 
% The effect of threshold $\mathcal{T}$ on network pruning is intuitively shown in Fig.~, and we optimize the threshold via traversal (line 6). The domain of threshold is $\{1,2, \ldots z \}$, i.e., $\mathcal{T}$ will always be smaller than the number of partitions $z$. We prune $\mathcal{G}_{co}$ with a threshold $\mathcal{T}$ to yield a pruned network $\mathcal{G}_{co}^{\mathcal{T}}$, and evaluate the cluster 
% partition of $\mathcal{G}_{co}^{\mathcal{T}}$ via {\em{cluster consensus}} metric, which can be used to quantify the stability of clusters~\cite{monti2003consensus}.
% For a pruned co-occurrence network $\mathcal{G}^{\mathcal{T}}_{co}$ with  
% cluster partition $\mathcal{M}^{\mathcal{T}} = \{\mathcal{M}_{k} \mid  k=1,\ldots,M^{\mathcal{T}} \}$, the consensus of cluster $\mathcal{M}_{k}$ is defined as
% \begin{equation}
%     c(\mathcal{M}_{k})=\frac{1}{M_{k}\left(M_{k}-1\right) / 2} \sum_{i, j \in \mathcal{M}_{k} \atop i<j} \mathcal{A}_{co}(i, j),
% \end{equation}
% where $M_{k}$ is the size of $\mathcal{M}_{k}$. The optimal threshold corresponds to the maximum partition score, which can be computed
% via a weighted sum of cluster consensus, as follows:
% \begin{equation}
%     C(\mathcal{M^T}) = \sum_{k=1}^{M^{\mathcal{T}}} \frac{M_{k}}{n}  c(\mathcal{M}_{k}),
% \end{equation}
% \begin{equation}
%     \mathcal{T} =  \mathop{\arg\max}_{\mathcal{T}}  C(\mathcal{M^T}).
% \end{equation}
\begin{table*}
    \renewcommand\arraystretch{1.2}
    \centering
    \setlength{\abovecaptionskip}{0pt}
    \caption{Real-world networks. $\phi_\textit{S}$ is the number of communities found by the specific community detection method $\mathcal{S}$.}
    \label{tb:dataset}
    \resizebox{\textwidth}{!}{%
    \begin{tabular}{lccclcccccc} 
        \hline\hline
        \multicolumn{1}{l}{\multirow{2}{*}{Network}} & \multirow{2}{*}{$m$} & \multirow{2}{*}{$n$} & \multirow{2}{*}{$\phi_\textit{real}$} & \multicolumn{1}{c}{\multirow{2}{*}{Description}} & \multicolumn{6}{c}{Number of communities ($\phi_\textit{S}$)}    \\ 
        \cline{6-11}
        \multicolumn{1}{c}{}  &     &     &     & \multicolumn{1}{c}{}                                & INF & FG  & WT  & LOU & LP  & N2VKM                                       \\ 
        \hline
        Karate                & 34  & 78  & 2   & Zachary Karate club~\cite{zachary1977information}   & 3   & 3   & 4   & 4   & 2   & 2      \\
        Polbooks              & 105 & 441 & 3   & Books about US politics~\cite{newman2006modularity} & 6   & 4   & 5   & 4   & 4   & 3      \\
        Football              & 115 & 613 & 12  & American College football~\cite{girvan2002community}& 12  & 6   & 10  & 10  & 9   & 12     \\
        Polblogs              & 1490& 19090 & 2 & Political blogs~\cite{lada2005political}            & 306 & 277 & 416 & 276 & 272 & 2      \\ 
        \hline
        \multicolumn{1}{c}{Amazon-sub} & 10077& 24205& 3251& Amazon product co-purchasing~\cite{yang2015defining}& 651 & 97  & 509 & 69  & 638 & \multirow{2}{*}{\diagbox{}{}} \\
        DBLP-sub                       & 26183& 137529     & 5051 & DBLP collaboration~\cite{yang2015defining}         & 1143& 120 & 4361& 49  & 787  &                       \\
        \hline\hline
        \end{tabular}
    }
    \vspace{-12pt}
\end{table*}
After pruning, the co-occurrence network $\mathcal{G}_\textit{co}$ is divided into several connected components, and those with large sizes will be treated as core communities $\mathcal{M}^{\mathcal{T}}_\textit{core}$ (line 6). 
Generally, there exists several small connected components that contain few vertices, or even isolated vertices, when pruning $\mathcal{G}_\textit{co}$ with a relatively large threshold. In order to get a final partition, these vertices in small connected components will be treated as isolated ones and assigned to the core community, to which it has the maximum average similarity (line 7). 
The ID of the core community that an isolated vertex $v_i$ is assigned to, is defined as:
\begin{equation}
    \setlength{\abovedisplayskip}{3pt}
    \setlength{\belowdisplayskip}{3pt}
    \textit{ID}_{\mathcal{H}} = \mathop{\arg\max}_{k} \frac{1}{\phi_\textit{core}^{k}} \sum_{j \in \mathcal{M}_\textit{core}^{k} } \mathcal{H}(i,j) \ ,
\end{equation}
where $\mathcal{M}_\textit{core}^{k}$ is the $k$-th core community and $\phi_\textit{core}^{k}$ is the number of vertices in $\mathcal{M}_\textit{core}^{k}$. 
The final ID is determined by a \emph{relative majority vote} of all similarity indices used in this paper:
\begin{equation}
    \setlength{\abovedisplayskip}{3pt}
    \setlength{\belowdisplayskip}{3pt}
    \textit{ID} = \mathsf{realtiveMajorityVote}\{\textit{ID}_{\mathcal{H}}\ |\ {\mathcal{H}} =\textsf{CN, \ldots, RWR} \} \ .
\end{equation}

\subsubsection{Threshold Selection}
In order to address the resolution limit problem, we search the optimal threshold via a traversal procedure.
Actually, the range of threshold $\mathcal{T}\in [0,80]$ can be narrowed down to $\mathcal{T}\in \{a_{ij} \  | \  \forall a_{ij} \in \mathcal{A}_\textit{co}\}$, which reduce the access times during traversal.
% In order to address the resolution limit problem, we optimize the threshold via a traversal procedure (line 6). The domain of threshold is $\{1,2, \ldots z \}$, i.e., $\mathcal{T}$ will always be smaller than the number of partitions $z$. 
We prune $\mathcal{G}_\textit{co}$ with an accessed threshold $\mathcal{T}$ to yield a pruned network $\mathcal{G}_\textit{co}^{\mathcal{T}}$, and evaluate the cluster partition of $\mathcal{G}_\textit{co}^{\mathcal{T}}$ using \emph{cluster consensus} metric, which can quantify the stability of clusters~\cite{monti2003consensus}.
For a pruned co-occurrence network $\mathcal{G}^{\mathcal{T}}_\textit{co}$ with  
cluster partition $\mathcal{M}^{\mathcal{T}} = \{\mathcal{M}_{k} \mid  k=1,\ldots,\phi^\mathcal{T} \}$, the consensus of cluster $\mathcal{M}_{k}$ is defined as
\begin{equation}
    \setlength{\abovedisplayskip}{3pt}
    \setlength{\belowdisplayskip}{3pt}
    c(\mathcal{M}_{k})=\frac{1}{\phi_{k}\left(\phi_{k}-1\right) / 2} \sum_{i, j \in \mathcal{M}_{k} \atop i<j} a_{ij} \ ,
\end{equation}
where $\phi_{k}$ is the number of vertices in $\mathcal{M}_{k}$. The optimal threshold corresponds to that yield the $\mathcal{G}_\textit{co}^{\mathcal{T}}$ with the maximum partition score, which is computed via a weighted sum of cluster consensus, as follows:
\begin{equation}
    \setlength{\abovedisplayskip}{2pt}
    \setlength{\belowdisplayskip}{2pt}
    C(\mathcal{M^T}) = \sum_{k=1}^{\phi^\mathcal{T}} \frac{\phi_{k}}{n}  c(\mathcal{M}_{k}) \ ,
\end{equation}
\begin{equation}\label{eq:threshold-optimization}
    \setlength{\abovedisplayskip}{2pt}
    \setlength{\belowdisplayskip}{0pt}
    \mathcal{T} =  \mathop{\arg\max}_{\mathcal{T}}  \ C(\mathcal{M^T}) \ .
\end{equation}

Note that Eq~(\ref{eq:threshold-optimization}) is the heuristic definition of the optimal threshold, which can alleviate the resolution limit problem to a certain extent.
Considering the complexity of the calculation during threshold selection, we can further simplify this process by the following approximation: 
\begin{equation}
    \mathcal{T}' = \mathop{\arg\min}_{\mathcal{T}} \ |\phi^\mathcal{T}_\textit{core} - \phi_\textit{real}| \ ,
\end{equation}
where $\phi^\mathcal{T}_\textit{core}$ is the number of core communities yielded during network pruning. $\mathcal{T}'$ is the approximate optimal threshold, and we can obtain the core communities, of which the number is closest to that of the ground-truth. Note that this approximation depends on the knowledge of the ground-truth community.

\section{Experiments} \label{sec:experiment}
\subsection{Datasets} \label{Datasets}
We evaluate the proposed approaches against six real-world networks and four adversarial networks. 
For all networks, the ground-truth community labels are available. 
TABLE~\ref{tb:dataset} provides an overview of the networks considered, including the number of communities found by each community detection method.
Specifically, the six real-world networks consist of four small benchmark networks and two large-scale networks with missing data.
The networks with missing data are sub-networks extracted from the Amazon product co-purchasing network and DBLP collaboration network~\cite{yang2015defining}, respectively.
The four adversarial networks are generated via adversarial attack on four benchmark networks.
Refer to Appendix~\ref{app:2} for more details about datasets.
% \begin{itemize}
%     \item {\textbf {Zachary Karate club (Karate)} \cite{zachary1977information}.} 
%     The network is about the pattern of relationship among the members of a Karate club at an American university, which splits into two groups after a dispute.
%     % \item {\textbf {Dolphin social network (Dolphins)} \cite{lusseau2003bottlenose}.} 
%     % The network created by Lusseau is about the frequent associations between 62 dolphins living off Doubtful Sound, New Zealand.
%     \item {\textbf {Books about US politics (Polbooks)} \cite{newman2006modularity}.} 
%     It is a network of books about US politics published in 2004 for presidential election. Links between books represent their frequent purchasing by the same buyers.
%     \item {\textbf {American College football (Football)} \cite{girvan2002community}.} 
%     The network is about the schedule of games between American college football teams in regular season Fall 2000.
%     \item {\textbf {Political blogs (Polblogs)} \cite{lada2005political}.}
%     The network consists of hyperlinks between weblogs on US politics recorded in 2005.
% \end{itemize}
% \vspace{-5pt}
\subsection{Evaluation Metrics} \label{sec:evaluation-measures}
Benefiting from the availability of the ground-truth community labels, we evaluate the community partitions using supervised metrics like normalized mutual information~\cite{danon2005comparing} and adjusted rand index. Note that we design the fitness in \emph{RobustECD-GA} using modularity $\mathcal{Q}$, thus it is not suitable as the evaluation metric. 
%which is an unsupervised metric and evaluate the cluster partition without the ground-truth community labels. 
% The two evaluation metrics are briefly introduced in the following.
\begin{itemize}
    \item {\textbf {Modularity ($\mathcal{Q}$)} \cite{newman2004finding}.} 
    Modularity is commonly used to measure the quality of community partition for a network with unknown community structure. 
    The basic idea is to compare the network with the corresponding \emph{null models}, which refer to random graph models that have some of the same properties as the network but are completely random in other aspects. 
    For a given network and a specific community partition $\mathcal{M}$, modularity is defined as:
    \begin{equation}
        \setlength{\abovedisplayskip}{3pt}
        \setlength{\belowdisplayskip}{3pt}
        \mathcal{Q}=\frac{1}{2 m} \sum_{i j}\left(\mathcal{A}_{ij}-\frac{k_{i} k_{j}}{2 m}\right) \delta (l_{i}, l_{j}) \ ,
    \end{equation}
    where $m$ is the number of edges, $\mathcal{A}_{ij}$ is the element of adjacency matrix $\mathcal{A}$, $k_{i},k_{j}$ are the degree of vertices $v_{i},v_{j}$, respectively. 
    $l_{i},l_{j}$ are the community labels of vertices $v_{i},v_{j}$ in $\mathcal{M}$, respectively. 
    % $\delta$ is the Kronecker function. 
    $\delta (l_{i}, l_{j}) = 1$ if $l_{i} = l_{j}$ and $\delta (l_{i}, l_{j}) = 0$ otherwise.
    \item {\textbf {Normalized Mutual Information (NMI)} \cite{danon2005comparing}.} 
    NMI is a commonly used criterion to evaluate the similarity of two clustering results. It quantifies how much information the estimated partition
    contains in the real partition.
    For two clustering results $X$ and $Y$, the NMI is defined as:
    \begin{equation}
        \setlength{\abovedisplayskip}{3pt}
        \setlength{\belowdisplayskip}{3pt}
        I_{\text { norm }}(X, Y)=\frac{2 I(X, Y)}{H(X)+H(Y)},
    \end{equation}
    where $I(X,Y) = H(Y) - H(X|Y)$ is the mutual information of $X$ and $Y$, $H(Y)$ is the Shannon entropy of $Y$, 
    and $H(X|Y)$ is the conditional entropy of $X$ given $Y$. 
    \item {\textbf {Adjusted Rand Index (ARI)} \cite{hubert1985comparing}.}
    ARI is the corrected-for-chance version of the Rand index (RI), which measures the degree of agreement between an estimated partition and a real partition. It is defined as 
    \begin{equation}
        \setlength{\abovedisplayskip}{3pt}
        \setlength{\belowdisplayskip}{3pt}
        ARI=\frac{RI-E[RI]}{\max (RI)-E[RI]}.
    \end{equation}
\end{itemize}
Both NMI and ARI require the ground-truth community labels for evaluation purpose and the values are generally in the range between 0 to 1. 
% Note that ARI can yield negative values if the index is less than the expected index~\cite{hubert1985comparing}. 
For both metrics, a larger value indicates a better partition.
\vspace{-5pt}
\subsection{Community Detection Methods}
We consider the following six community detection algorithms in our experiments. 
% The first five are available in the Python version of the \texttt{igraph} \footnote{https://igraph.org/python/} library.
% The implementation of \texttt{Node2vec} is available online \footnote{https://github.com/eliorc/node2vec}.
\begin{itemize}
    \item {\textbf {Infomap (INF)} \cite{rosvall2008maps}.}
    Infomap decomposes a network into modules by compressing the description of the information flow, i.e., it detects communities by minimizing the encoding length for a random walk. 
    % This algorithm runs in time $\mathcal{O} (|E|)$.
    \item {\textbf {Fast Greedy (FG)} \cite{clauset2004finding}.}
    This is a bottom-up hierarchical agglomeration algorithm. It merges individual vertices into communities based on a greedy modularity maximization strategy.
    % This algorithm runs in time $\mathcal{O}\left(|V| \log ^{2}|V|\right)$.
    \item {\textbf {WalkTrap (WT)} \cite{pons2005computing}.}
    It detects communities based on the idea that short random walks tend to stay in the same community.
    % This algorithm runs in time $\mathcal{O}\left(|V|^{2} \log |V|\right)$.
    \item {\textbf {Louvain (LOU)} \cite{blondel2008fast}.}
    This is a multi-level modularity optimization algorithm. It initializes each vertex with a separate community, and moves vertices between communities iteratively in a way that maximizes the vertices' local contributions to the overall modularity.
    % This algorithm runs in time $\mathcal{O}(|V| \log |V|)$.
    \item {\textbf {Label Propagation (LP)} \cite{raghavan2007near}.}
    This method detects communities by initializing each vertex with a unique label and re-assigning each vertex the dominant label in its neighbourhood in each iteration.
    % This algorithm runs in time $\mathcal{O} (|E|)$.
    \item {\textbf {Node2vec + Kmeans (N2VKM)} \cite{grover2016node2vec}.}
    This is a network embedding method, which learns lower-dimensional representations for vertices by biased random walk and skip-Gram. The $K$-means algorithm is then used to detect communities by clustering the embedded vectors of vertices in an Euclidean space. 
    % % This algorithm runs in time $\mathcal{O}(|V|(|E|+|V|))$.
\end{itemize} 
% Then, we obtain the same number of communities as the ground truth after feeding the network into N2VKM according to TABLE~\ref{tb:dataset}.
% \subsection{Adversarial Attack Methods}
% \textcolor{blue}{We consider the following two community deception methods to generate a series of adversarial networks:}
% \begin{itemize}
%     \item {\textbf {$\mathcal{Q}$-Attack}~\cite{chen2019ga}.} 
%     It is an evolutionary attack strategy based on genetic algorithm, in which the modularity is used to design the fitness function. This strategy deploys attack via negligible network rewiring, which doesn't change the degree of vertices, and achieves the state-of-the-art attack effect.
%     \item {\textbf {$\mathcal{D}_{m}$-Deception via Modularity}~\cite{fionda2017community}.} 
%     $\mathcal{D}_{m}$ is a community deception algorithm based on modularity, which can hide a target community via intra-community edge deletion and inter-community edge addition.
% \end{itemize}
% baseline
% !!! add more comparative method
\vspace{-5pt}
\subsection{Baseline Enhancement Methods}
We compare our methods with the following two typical traditional enhancement methods, one is based on network rewiring and the other utilizes network weighting.
\begin{itemize}
    \item {\textbf {EdMot} \cite{li2019edmot}.} 
    It is an edge enhancement approach for motif-aware community detection via network rewiring and is proposed to address the hypergraph fragmentation issue. This strategy is recently proposed and achieves the state-of-the-art enhancement effect
    in several community detection methods.
    \item {\textbf {WERW-KPath} \cite{de2013enhancing}.} 
    It is an enhancement approach for community detection via network weighting. It exploits random walks to compute the $\kappa$-path edge centrality, which is then used to weight the edges. The strategy shows better interpretability and effectiveness among a series of weighted methods.
\end{itemize}
% We have the following consideration for selecting the prior community detection method:
% \begin{itemize}
%     \item {\textbf{Significant randomness.} Note that since operations such as random walk are conducted randomly, there is no guarantee that 
%     a community detection algorithm contain these operations returns the same partition result after each run. After experimental observation, 
%     the algorithm LP and N2VKM have significant randomness.} 
% \end{itemize}
\vspace{-5pt}
\subsection{Experiment Setup}
% paratemeter set
For all datasets, edges in networks are treated as undirected and self-loops will be removed. The main parameter settings for the proposed algorithms are shown in TABLE~\ref{tb:para}.  

In \emph{RobustECD-GA}, the general parameters of GA are set as empirical values.
Note that the budget $\beta_\textit{a}$ and $\beta_\textit{d}$ controls the upper limit of the chromosome size during Initialization, i.e., each chromosome will be initialized with an unfixed size not larger than $\lceil m \cdot (\beta_\textit{a} + \beta_\textit{d})\rceil$. 
We vary $\beta_\textit{a}$ and $\beta_\textit{d}$ in \{0.01, 0.02, ..., 2.9\} and \{0.01, 0.02, ..., 0.29\}, respectively.
% The sample rate of edge addition $\beta_\textit{a}$ is fixed to 3.0 for all experiments while the sample rate of edge deletion $\beta_\textit{d}$ is unfixed (0 for original networks and 0.2 for adversarial networks). 
% In \emph{RobustECD-SE}, the selection of sample rate varies across datasets, and we will study in Sec.~\ref{analysis} the impact of the sample rate on the performance of \emph{RobustECD-SE}. Specifically, the sample rate $\beta_\textit{a}$ is set to 1.5, 2.7, 0.2 and 2.7 for Karate, Polbooks, Football and Polblogs, respectively.
In \emph{RobustECD-SE}, we vary $\beta_\textit{a}$ in \{0.1, 0.2, ..., 2.9\}.

In addition, for the community detection algorithm N2VKM, 
we take the number of clusters $K$ in $K$-means the same as that of the ground-truth communities, and use the default setting for parameters in \emph{Node2vec}. Specifically, the walk length is 80, the number of walks per node is 10, the embedding dimension is 128, and both return hyper parameter $p$ and in-out parameter $q$ are equal to 1. 
% The number of ground-truth communities is set as the input to $K$-means. 
We repeat all experiments for 50 times and report the average metrics and their standard deviations of community detection.
% Moreover, because of the randomness of LP and N2VKM, we repeat experiments for 20 times and report the average results of community detection.
% We generate noise networks for all benchmark datasets via adversarial attacks.
% Details of noise networks are shown in TABLE~\ref{tb:adv-detail}.
% For two large-scale networks, we don't deploy the attack since that the noise has been introduced during sub-network extraction.
% We generate adversarial networks for the two smaller datasets, Karate and Polbooks, via $\mathcal{Q}$-Attack, and use $\mathcal{D}_{m}$ to attack the other two larger datasets. 
% Details of the adversarial networks are shown in TABLE~\ref{adv-detail}.
% For two large-scale networks, we don't deploy the attack since that the noise has been introduced during sub-network extraction.
% We generate adversarial networks for the two smaller datasets, Karate and Polbooks, via $\mathcal{Q}$-Attack, and use $\mathcal{D}_{m}$ to attack the other two larger datasets. Details of the adversarial networks are shown in TABLE~\ref{adv-detail}.
% \usepackage{multirow}
\begin{table}
    \renewcommand\arraystretch{1.2}
    % \large
    \centering
    \setlength{\abovecaptionskip}{0pt}
    \caption{Main parameters setting for the proposed algorithms.}
    \label{tb:para}
    \resizebox{\linewidth}{!}{%
    \begin{tabular}{c|c|c} 
        \hline\hline
        Method                        & Parameter & Value  \\ 
        \hline
        \multirow{3}{*}{RobustECD-GA} & GA($\phi_\textit{p}$, $\mathcal{P}_{c}$,  $\mathcal{P}_\textit{m}$, $\mathcal{P}_\textit{e}$,  $\mathcal{T}_\textit{ga}$)    & \{120, 0.8, 0.02, 0.2, 1000\}      \\
                                      & $\beta_\textit{a}$   & \{0.01, 0.02, ..., 2.9\}      \\
                                      & $\beta_\textit{d}$   & \{0.01, 0.02, ..., 0.29\}       \\ 
        \hline
        RobustECD-SE                  & $\beta_\textit{a}$   & \{0.1, 0.2, ..., 2.9\}       \\
        \hline\hline
        \end{tabular}
    }
    \vspace{-12pt}
\end{table}
\begin{table*}
    \renewcommand\arraystretch{1.4}
    \large
    \centering
    \caption{Community detection results in the real networks.}
    \label{tb:ori-effect}
    \resizebox{\textwidth}{!}{%
    \begin{tabular}{l|l|cccccc|c|cccccc|c} 
    \hline\hline
    \multicolumn{1}{c|}{\multirow{3}{*}{Dataset}}                         & \multicolumn{1}{c|}{\multirow{3}{*}{Method}} & \multicolumn{14}{c}{Community Detection}                  \\ 
    \cline{3-16}
    \multicolumn{1}{c|}{}                                                 & \multicolumn{1}{c|}{}                          & \multicolumn{7}{c|}{NMI}                                                                                                                                                                  & \multicolumn{7}{c}{ARI}                                                                                                                                                                   \\ 
    \cline{3-16}
    \multicolumn{1}{c|}{}                                                 & \multicolumn{1}{c|}{}                          & \multicolumn{1}{c}{INF} & \multicolumn{1}{c}{FG} & \multicolumn{1}{c}{WT} & \multicolumn{1}{c}{LOU} & \multicolumn{1}{c}{LP} & \multicolumn{1}{c|}{N2VKM} & \multicolumn{1}{c|}{Avg RIMP} & \multicolumn{1}{c}{INF} & \multicolumn{1}{c}{FG} & \multicolumn{1}{c}{WT} & \multicolumn{1}{c}{LOU} & \multicolumn{1}{c}{LP} & \multicolumn{1}{c|}{N2VKM} & \multicolumn{1}{c}{Avg RIMP}  \\ 
    \hline
    \multirow{5}{*}{Karate}                                               & original             & 0.699\footnotesize{$\pm$0.000}             & 0.598\footnotesize{$\pm$0.000}            & 0.600\footnotesize{$\pm$0.000}            & 0.587\footnotesize{$\pm$0.000}             & 0.689\footnotesize{$\pm$0.283}            & 0.705\footnotesize{$\pm$0.175}                & ----                          
                                                                                                 & 0.702\footnotesize{$\pm$0.000}             & 0.491\footnotesize{$\pm$0.000}            & 0.513\footnotesize{$\pm$0.000}            & 0.462\footnotesize{$\pm$0.000}             & 0.687\footnotesize{$\pm$0.320}            & 0.716\footnotesize{$\pm$0.212}                & ----                          \\
                                                                          & WERW-Kpath           & 0.618\footnotesize{$\pm$0.071}             & 0.607\footnotesize{$\pm$0.040}            & 0.528\footnotesize{$\pm$0.059}            & 0.518\footnotesize{$\pm$0.043}             & 0.622\footnotesize{$\pm$0.185}            & 0.644\footnotesize{$\pm$0.186}                & -8.70\%                       
                                                                                                 & 0.557\footnotesize{$\pm$0.124}             & 0.593\footnotesize{$\pm$0.046}            & 0.398\footnotesize{$\pm$0.107}            & 0.407\footnotesize{$\pm$0.042}             & 0.604\footnotesize{$\pm$0.229}            & 0.652\footnotesize{$\pm$0.227}                & -9.20\%                       \\
                                                                          & Edmot                & 0.699\footnotesize{$\pm$0.000}             & 0.598\footnotesize{$\pm$0.000}            & 0.600\footnotesize{$\pm$0.000}            & 0.587\footnotesize{$\pm$0.000}             & 0.685\footnotesize{$\pm$0.203}            & 0.713\footnotesize{$\pm$0.177}                & 0.09\%                        
                                                                                                 & 0.702\footnotesize{$\pm$0.000}             & 0.491\footnotesize{$\pm$0.000}            & 0.513\footnotesize{$\pm$0.000}            & 0.462\footnotesize{$\pm$0.000}             & 0.686\footnotesize{$\pm$0.237}            & 0.728\footnotesize{$\pm$0.200}                & 0.26\%                        \\
                                                                          & \emph{RobustECD-GA}~ & \textbf{0.912\footnotesize{$\pm$0.176}}    & 0.878\footnotesize{$\pm$0.071}            & \textbf{0.984\footnotesize{$\pm$0.049}}   & 0.867\footnotesize{$\pm$0.090}             & 0.705\footnotesize{$\pm$0.180}            & \textbf{0.838\footnotesize{$\pm$0.019}}       & 35.03\%                       
                                                                                                 & \textbf{0.923\footnotesize{$\pm$0.165}}    & 0.912\footnotesize{$\pm$0.051}            & \textbf{0.988\footnotesize{$\pm$0.035}}   & 0.898\footnotesize{$\pm$0.083}             & 0.714\footnotesize{$\pm$0.217}            & \textbf{0.872\footnotesize{$\pm$0.024}}       & 54.98\%                       \\
                                                                          & \emph{RobustECD-SE}  & 0.825\footnotesize{$\pm$0.220}             & \textbf{1.000\footnotesize{$\pm$0.000}}   & 0.821\footnotesize{$\pm$0.088}            & \textbf{1.000\footnotesize{$\pm$0.000}}    & \textbf{0.847\footnotesize{$\pm$0.136}}   & 0.834\footnotesize{$\pm$0.114}                & \textbf{38.95\%}              
                                                                                                 & 0.797\footnotesize{$\pm$0.240}             & \textbf{1.000\footnotesize{$\pm$0.000}}   & 0.852\footnotesize{$\pm$0.086}            & \textbf{1.000\footnotesize{$\pm$0.000}}    & \textbf{0.871\footnotesize{$\pm$0.124}}   & 0.869\footnotesize{$\pm$0.100}                & \textbf{57.98\%}              \\ 
    \hline
    \multirow{5}{*}{Polbooks}                                             & original             & 0.493\footnotesize{$\pm$0.000}             & 0.531\footnotesize{$\pm$0.000}            & 0.559\footnotesize{$\pm$0.000}            & 0.512\footnotesize{$\pm$0.000}             & 0.554\footnotesize{$\pm$0.025}            & 0.556\footnotesize{$\pm$0.017}                & ----                          
                                                                                                 & 0.536\footnotesize{$\pm$0.000}             & 0.638\footnotesize{$\pm$0.000}            & 0.681\footnotesize{$\pm$0.000}            & 0.558\footnotesize{$\pm$0.000}             & 0.647\footnotesize{$\pm$0.041}            & 0.662\footnotesize{$\pm$0.009}                & ----                          \\
                                                                          & WERW-Kpath           & 0.462\footnotesize{$\pm$0.002}             & 0.546\footnotesize{$\pm$0.020}            & 0.531\footnotesize{$\pm$0.031}            & 0.509\footnotesize{$\pm$0.020}             & 0.552\footnotesize{$\pm$0.034}            & 0.563\footnotesize{$\pm$0.016}                & -1.36\%                       
                                                                                                 & 0.435\footnotesize{$\pm$0.000}             & 0.658\footnotesize{$\pm$0.026}            & 0.591\footnotesize{$\pm$0.069}            & 0.571\footnotesize{$\pm$0.034}             & 0.654\footnotesize{$\pm$0.054}            & 0.660\footnotesize{$\pm$0.015}                & -4.30\%                       \\
                                                                          & Edmot                & 0.493\footnotesize{$\pm$0.000}             & 0.531\footnotesize{$\pm$0.000}            & 0.559\footnotesize{$\pm$0.000}            & 0.512\footnotesize{$\pm$0.000}             & 0.561\footnotesize{$\pm$0.026}            & 0.564\footnotesize{$\pm$0.016}                & 0.45\%                        
                                                                                                 & 0.536\footnotesize{$\pm$0.000}             & 0.638\footnotesize{$\pm$0.000}            & 0.681\footnotesize{$\pm$0.000}            & 0.558\footnotesize{$\pm$0.000}             & 0.661\footnotesize{$\pm$0.035}            & 0.667\footnotesize{$\pm$0.018}                & 0.49\%                        \\
                                                                          & \emph{RobustECD-GA}~ & 0.526\footnotesize{$\pm$0.121}             & 0.554\footnotesize{$\pm$0.000}            & 0.554\footnotesize{$\pm$0.000}            & 0.554\footnotesize{$\pm$0.000}             & 0.554\footnotesize{$\pm$0.014}            & 0.589\footnotesize{$\pm$0.017}                & 4.04\%                       
                                                                                                 & 0.621\footnotesize{$\pm$0.143}             & \textbf{0.652\footnotesize{$\pm$0.000}}   & 0.652\footnotesize{$\pm$0.000}            & 0.652\footnotesize{$\pm$0.000}             & \textbf{0.670\footnotesize{$\pm$0.007}}   & \textbf{0.684\footnotesize{$\pm$0.013}}       & 6.25\%                        \\
                                                                          & \emph{RobustECD-SE}  & \textbf{0.574\footnotesize{$\pm$0.014}}    & \textbf{0.569\footnotesize{$\pm$0.001}}   & \textbf{0.586\footnotesize{$\pm$0.017}}   & \textbf{0.560\footnotesize{$\pm$0.011}}    & \textbf{0.598\footnotesize{$\pm$0.009}}   & \textbf{0.589\footnotesize{$\pm$0.009}}       & \textbf{8.61\%}               
                                                                                                 & \textbf{0.677\footnotesize{$\pm$0.014}}    & 0.636\footnotesize{$\pm$0.000}            & \textbf{0.687\footnotesize{$\pm$0.015}}   & \textbf{0.669\footnotesize{$\pm$0.006}}    & 0.665\footnotesize{$\pm$0.010}            & 0.677\footnotesize{$\pm$0.011}                & \textbf{8.64\%}               \\ 
    \hline
    \multirow{5}{*}{Football}                                             & original             & 0.924\footnotesize{$\pm$0.000}             & 0.698\footnotesize{$\pm$0.000}            & 0.887\footnotesize{$\pm$0.000}            & 0.890\footnotesize{$\pm$0.000}             & 0.888\footnotesize{$\pm$0.037}            & 0.912\footnotesize{$\pm$0.012}                & ----                          
                                                                                                 & 0.897\footnotesize{$\pm$0.000}             & 0.474\footnotesize{$\pm$0.000}            & 0.815\footnotesize{$\pm$0.000}            & 0.807\footnotesize{$\pm$0.000}             & 0.784\footnotesize{$\pm$0.103}            & 0.872\footnotesize{$\pm$0.025}                & ----                          \\
                                                                          & WERW-Kpath           & 0.924\footnotesize{$\pm$0.000}             & 0.698\footnotesize{$\pm$0.000}            & 0.887\footnotesize{$\pm$0.000}            & 0.890\footnotesize{$\pm$0.000}             & 0.885\footnotesize{$\pm$0.030}            & 0.915\footnotesize{$\pm$0.011}                & 0.00\%                        
                                                                                                 & 0.897\footnotesize{$\pm$0.000}             & 0.474\footnotesize{$\pm$0.000}            & 0.815\footnotesize{$\pm$0.000}            & 0.807\footnotesize{$\pm$0.000}             & 0.779\footnotesize{$\pm$0.083}            & 0.876\footnotesize{$\pm$0.020}                & -0.03\%                       \\
                                                                          & Edmot                & 0.924\footnotesize{$\pm$0.000}             & 0.698\footnotesize{$\pm$0.000}            & 0.887\footnotesize{$\pm$0.000}            & 0.890\footnotesize{$\pm$0.000}             & 0.885\footnotesize{$\pm$0.030}            & 0.915\footnotesize{$\pm$0.011}                & 0.00\%                       
                                                                                                 & 0.897\footnotesize{$\pm$0.000}             & 0.474\footnotesize{$\pm$0.000}            & 0.815\footnotesize{$\pm$0.000}            & 0.807\footnotesize{$\pm$0.000}             & 0.779\footnotesize{$\pm$0.083}            & 0.876\footnotesize{$\pm$0.020}                & -0.03\%                       \\
                                                                          & \emph{RobustECD-GA}~ & \textbf{0.927\footnotesize{$\pm$0.000}}    & 0.862\footnotesize{$\pm$0.018}            & \textbf{0.927\footnotesize{$\pm$0.000}}   & \textbf{0.909\footnotesize{$\pm$0.000}}    & 0.896\footnotesize{$\pm$0.023}            & \textbf{0.927\footnotesize{$\pm$0.000}}       & \textbf{5.50\%}               
                                                                                                 & 0.889\footnotesize{$\pm$0.000}             & 0.746\footnotesize{$\pm$0.042}            & \textbf{0.889\footnotesize{$\pm$0.000}}   & 0.847\footnotesize{$\pm$0.000}             & 0.793\footnotesize{$\pm$0.082}            & \textbf{0.889\footnotesize{$\pm$0.000}}       & 12.27\%                       \\
                                                                          & \emph{RobustECD-SE}  & 0.924\footnotesize{$\pm$0.000}             & \textbf{0.877\footnotesize{$\pm$0.021}}   & 0.923\footnotesize{$\pm$0.009}            & 0.906\footnotesize{$\pm$0.014}             & \textbf{0.915\footnotesize{$\pm$0.018}}   & 0.898\footnotesize{$\pm$0.021}                & \textbf{5.50\%}               
                                                                                                 & \textbf{0.897\footnotesize{$\pm$0.000}}    & \textbf{0.785\footnotesize{$\pm$0.061}}   & 0.881\footnotesize{$\pm$0.018}            & \textbf{0.849\footnotesize{$\pm$0.029}}    & \textbf{0.869\footnotesize{$\pm$0.032}}   & 0.849\footnotesize{$\pm$0.029}                & \textbf{14.52\%}              \\ 
    \hline
    \multirow{5}{*}{Polblogs}                                             & original             & 0.330\footnotesize{$\pm$0.001}             & 0.378\footnotesize{$\pm$0.000}            & 0.318\footnotesize{$\pm$0.000}            & 0.376\footnotesize{$\pm$0.000}             & 0.375\footnotesize{$\pm$0.053}            & 0.458\footnotesize{$\pm$0.067}                & ----                          
                                                                                                 & 0.439\footnotesize{$\pm$0.001}             & 0.528\footnotesize{$\pm$0.000}            & 0.419\footnotesize{$\pm$0.000}            & 0.521\footnotesize{$\pm$0.000}             & 0.515\footnotesize{$\pm$0.105}            & 0.489\footnotesize{$\pm$0.053}                & ----                          \\
                                                                          & WERW-Kpath           & 0.329\footnotesize{$\pm$0.001}             & 0.376\footnotesize{$\pm$0.002}            & 0.316\footnotesize{$\pm$0.001}            & 0.370\footnotesize{$\pm$0.001}             & 0.375\footnotesize{$\pm$0.037}            & 0.453\footnotesize{$\pm$0.025}                & -0.69\%                       
                                                                                                 & 0.437\footnotesize{$\pm$0.003}             & 0.525\footnotesize{$\pm$0.003}            & 0.414\footnotesize{$\pm$0.003}            & 0.515\footnotesize{$\pm$0.002}             & 0.518\footnotesize{$\pm$0.074}            & 0.480\footnotesize{$\pm$0.040}                & -0.77\%                       \\
                                                                          & Edmot                & 0.329\footnotesize{$\pm$0.000}             & 0.378\footnotesize{$\pm$0.000}            & 0.318\footnotesize{$\pm$0.000}            & 0.376\footnotesize{$\pm$0.000}             & 0.376\footnotesize{$\pm$0.053}            & 0.457\footnotesize{$\pm$0.031}                & -0.04\%                       
                                                                                                 & 0.437\footnotesize{$\pm$0.000}             & 0.528\footnotesize{$\pm$0.000}            & 0.419\footnotesize{$\pm$0.000}            & 0.521\footnotesize{$\pm$0.000}             & 0.517\footnotesize{$\pm$0.105}            & 0.486\footnotesize{$\pm$0.049}                & -0.11\%                       \\
                                                                          & \emph{RobustECD-GA}~ & 0.453\footnotesize{$\pm$0.000}             & 0.525\footnotesize{$\pm$0.000}            & 0.504\footnotesize{$\pm$0.000}            & 0.529\footnotesize{$\pm$0.000}             & 0.519\footnotesize{$\pm$0.005}            & 0.381\footnotesize{$\pm$0.002}                & 32.82\%                       
                                                                                                 & 0.472\footnotesize{$\pm$0.000}             & 0.618\footnotesize{$\pm$0.000}            & 0.599\footnotesize{$\pm$0.000}            & 0.622\footnotesize{$\pm$0.000}             & 0.616\footnotesize{$\pm$0.002}            & 0.556\footnotesize{$\pm$0.001}                & 20.04\%                       \\
                                                                          & \emph{RobustECD-SE}  & \textbf{0.517\footnotesize{$\pm$0.007}}    & \textbf{0.551\footnotesize{$\pm$0.006}}   & \textbf{0.556\footnotesize{$\pm$0.009}}   & \textbf{0.551\footnotesize{$\pm$0.005}}    & \textbf{0.529\footnotesize{$\pm$0.007}}   & \textbf{0.499\footnotesize{$\pm$0.006}}       & \textbf{45.64\%}              
                                                                                                 & \textbf{0.619\footnotesize{$\pm$0.005}}    & \textbf{0.642\footnotesize{$\pm$0.004}}   & \textbf{0.643\footnotesize{$\pm$0.006}}   & \textbf{0.644\footnotesize{$\pm$0.004}}    & \textbf{0.628\footnotesize{$\pm$0.005}}   & \textbf{0.569\footnotesize{$\pm$0.005}}       & \textbf{29.66\%}              \\ 
    \hline
    \multirow{4}{*}{\begin{tabular}[c]{@{}l@{}}Amazon\\-sub\end{tabular}} & original             & 0.775\footnotesize{$\pm$0.000}             & 0.592\footnotesize{$\pm$0.000}            & 0.703\footnotesize{$\pm$0.000}            & 0.607\footnotesize{$\pm$0.000}             & 0.760\footnotesize{$\pm$0.001}            & ----                       & ----                          
                                                                                                 & 0.110\footnotesize{$\pm$0.000}             & 0.034\footnotesize{$\pm$0.000}            & 0.048\footnotesize{$\pm$0.000}            & 0.045\footnotesize{$\pm$0.000}             & 0.069\footnotesize{$\pm$0.004}            & ----                       & ----                          \\
                                                                          & WERW-Kpath           & 0.777\footnotesize{$\pm$0.000}             & 0.632\footnotesize{$\pm$0.000}            & \textbf{0.748\footnotesize{$\pm$0.000}}   & 0.633\footnotesize{$\pm$0.000}             & \textbf{0.770\footnotesize{$\pm$0.000}}   & ----                       & 3.80\%                        
                                                                                                 & \textbf{0.112\footnotesize{$\pm$0.000}}    & 0.048\footnotesize{$\pm$0.000}            & 0.052\footnotesize{$\pm$0.000}            & 0.052\footnotesize{$\pm$0.000}             & \textbf{0.076\footnotesize{$\pm$0.000}}   & ----                       & 15.91\%                       \\
                                                                          & Edmot                & 0.775\footnotesize{$\pm$0.000}             & 0.593\footnotesize{$\pm$0.000}            & 0.703\footnotesize{$\pm$0.000}            & 0.607\footnotesize{$\pm$0.000}             & 0.761\footnotesize{$\pm$0.000}            & ----                       & 0.06\%                        
                                                                                                 & 0.110\footnotesize{$\pm$0.000}             & 0.034\footnotesize{$\pm$0.000}            & 0.048\footnotesize{$\pm$0.000}            & 0.045\footnotesize{$\pm$0.000}             & 0.073\footnotesize{$\pm$0.000}            & ----                       & 1.28\%      \\
                                                                          & \emph{RobustECD-SE}  & \textbf{0.779\footnotesize{$\pm$0.000}}    & \textbf{0.663\footnotesize{$\pm$0.009}}   & 0.732\footnotesize{$\pm$0.005}            & \textbf{0.657\footnotesize{$\pm$0.004}}    & 0.768\footnotesize{$\pm$0.001}            & ----                       & \textbf{5.18\%}               
                                                                                                 & 0.107\footnotesize{$\pm$0.000}             & \textbf{0.053\footnotesize{$\pm$0.004}}   & \textbf{0.058\footnotesize{$\pm$0.008}}   & \textbf{0.053\footnotesize{$\pm$0.000}}    & 0.067\footnotesize{$\pm$0.001}            & ----                       & \textbf{18.72\%}              \\ 
    \hline
    \multirow{4}{*}{\begin{tabular}[c]{@{}l@{}}DBLP\\-sub\end{tabular}}   & original             & 0.698\footnotesize{$\pm$0.000}             & 0.348\footnotesize{$\pm$0.000}            & 0.683\footnotesize{$\pm$0.000}            & 0.431\footnotesize{$\pm$0.000}             & 0.436\footnotesize{$\pm$0.000}            & ----                       & ----                          
                                                                                                 & 0.061\footnotesize{$\pm$0.000}             & 0.004\footnotesize{$\pm$0.000}            & 0.004\footnotesize{$\pm$0.000}            & 0.010\footnotesize{$\pm$0.000}             & 0.000\footnotesize{$\pm$0.000}            & ----                       & ----                          \\
                                                                          & WERW-Kpath           & 0.701\footnotesize{$\pm$0.000}             & 0.350\footnotesize{$\pm$0.000}            & \textbf{0.688\footnotesize{$\pm$0.000}}   & 0.438\footnotesize{$\pm$0.000}             & \textbf{0.649\footnotesize{$\pm$0.000}}   & ----                       & 10.44\%                      
                                                                                                 & 0.062\footnotesize{$\pm$0.000}             & 0.004\footnotesize{$\pm$0.000}            & 0.005\footnotesize{$\pm$0.000}            & 0.011\footnotesize{$\pm$0.000}             & \textbf{0.008\footnotesize{$\pm$0.000}}   & ----                       & 6.43\%                        \\
                                                                          & Edmot                & 0.699\footnotesize{$\pm$0.000}             & 0.354\footnotesize{$\pm$0.000}            & 0.683\footnotesize{$\pm$0.000}            & 0.431\footnotesize{$\pm$0.000}             & 0.446\footnotesize{$\pm$0.000}            & ----                       & 0.83\%                        
                                                                                                 & 0.060\footnotesize{$\pm$0.000}             & 0.003\footnotesize{$\pm$0.000}            & 0.004\footnotesize{$\pm$0.000}            & 0.010\footnotesize{$\pm$0.000}             & 0.000\footnotesize{$\pm$0.000}            & ----                       & -2.60\%                       \\
                                                                          & \emph{RobustECD-SE}  & \textbf{0.704\footnotesize{$\pm$0.000}}    & \textbf{0.582\footnotesize{$\pm$0.000}}   & 0.684\footnotesize{$\pm$0.000}            & \textbf{0.598\footnotesize{$\pm$0.000}}    & 0.638\footnotesize{$\pm$0.000}            & ----                       & \textbf{30.66\%}              
                                                                                                 & \textbf{0.064\footnotesize{$\pm$0.000}}    & \textbf{0.007\footnotesize{$\pm$0.000}}   & \textbf{0.011\footnotesize{$\pm$0.000}}   & \textbf{0.019\footnotesize{$\pm$0.000}}    & 0.003\footnotesize{$\pm$0.000}            & ----                       & \textbf{72.08\%}              \\
    \hline\hline
    \end{tabular}}
    \end{table*}

    \begin{table*}
        \renewcommand\arraystretch{1.4}
        \large
        \centering
        \caption{Community detection results in the adversarial networks.}
        \label{tb:adv-effect}
        \resizebox{\textwidth}{!}{%
        \begin{tabular}{l|l|cccccc|c|cccccc|c} 
        \hline\hline
        \multicolumn{1}{c|}{\multirow{3}{*}{Dataset}}                         & \multicolumn{1}{c|}{\multirow{3}{*}{Method}} & \multicolumn{14}{c}{Community Detection}         \\ 
        \cline{3-16}
                                    &                           & \multicolumn{7}{c|}{NMI}            & \multicolumn{7}{c}{ARI}                                                                                                                                                             \\ 
        \cline{3-16}
                                    &                           & \multicolumn{1}{c}{INF} & \multicolumn{1}{c}{FG} & \multicolumn{1}{c}{WT} & \multicolumn{1}{c}{LOU} & \multicolumn{1}{c}{LP} & \multicolumn{1}{c|}{N2VKM} & \multicolumn{1}{c|}{Avg RIMP} & \multicolumn{1}{c}{INF} & \multicolumn{1}{c}{FG} & \multicolumn{1}{c}{WT} & \multicolumn{1}{c}{LOU} & \multicolumn{1}{c}{LP} & \multicolumn{1}{c|}{N2VKM} & \multicolumn{1}{c}{Avg RIMP}  \\ 
        \hline
        \multirow{6}{*}{\begin{tabular}[c]{@{}l@{}}Karate\\(noise)\end{tabular}}     & original                  & 0.699\footnotesize{$\pm$0.000}             & 0.598\footnotesize{$\pm$0.000}            & 0.600\footnotesize{$\pm$0.000}            & 0.587\footnotesize{$\pm$0.000}             & 0.689\footnotesize{$\pm$0.283}            & 0.705\footnotesize{$\pm$0.175}                & 63.90\%                  
                                                                                                                 & 0.702\footnotesize{$\pm$0.000}             & 0.491\footnotesize{$\pm$0.000}            & 0.513\footnotesize{$\pm$0.000}            & 0.462\footnotesize{$\pm$0.000}             & 0.687\footnotesize{$\pm$0.320}            & 0.716\footnotesize{$\pm$0.212}                & 70.66\%                  \\
                                                                                     & Attack                    & 0.000\footnotesize{$\pm$0.000}             & 0.447\footnotesize{$\pm$0.000}            & 0.487\footnotesize{$\pm$0.000}            & 0.250\footnotesize{$\pm$0.000}             & 0.475\footnotesize{$\pm$0.337}            & 0.399\footnotesize{$\pm$0.231}                & ----                     
                                                                                                                 & 0.000\footnotesize{$\pm$0.000}             & 0.361\footnotesize{$\pm$0.000}            & 0.330\footnotesize{$\pm$0.000}            & 0.180\footnotesize{$\pm$0.000}             & 0.498\footnotesize{$\pm$0.354}            & 0.427\footnotesize{$\pm$0.261}                & ----                     \\ 
        \cdashline{3-16}
                                                                                     & WERW-Kpath                & 0.173\footnotesize{$\pm$0.093}             & 0.296\footnotesize{$\pm$0.050}            & 0.444\footnotesize{$\pm$0.082}            & 0.341\footnotesize{$\pm$0.060}             & 0.373\footnotesize{$\pm$0.211}            & 0.384\footnotesize{$\pm$0.198}                & -2.36\%                  
                                                                                                                 & 0.116\footnotesize{$\pm$0.114}             & 0.258\footnotesize{$\pm$0.044}            & 0.359\footnotesize{$\pm$0.133}            & 0.281\footnotesize{$\pm$0.043}             & 0.357\footnotesize{$\pm$0.255}            & 0.401\footnotesize{$\pm$0.238}                & 2.26\%                   \\
                                                                                     & Edmot                     & 0.001\footnotesize{$\pm$0.000}             & 0.447\footnotesize{$\pm$0.000}            & 0.479\footnotesize{$\pm$0.000}            & 0.250\footnotesize{$\pm$0.000}             & 0.418\footnotesize{$\pm$0.345}            & 0.427\footnotesize{$\pm$0.225}                & -1.09\%                  
                                                                                                                 & 0.000\footnotesize{$\pm$0.000}             & 0.361\footnotesize{$\pm$0.000}            & 0.525\footnotesize{$\pm$0.000}            & 0.180\footnotesize{$\pm$0.000}             & 0.446\footnotesize{$\pm$0.367}            & 0.448\footnotesize{$\pm$0.257}                & 8.93\%                   \\
                                                                                     & \emph{RobustECD-GA}       & \textbf{0.720\footnotesize{$\pm$0.132}}    & 0.576\footnotesize{$\pm$0.000}            & \textbf{0.670\footnotesize{$\pm$0.119}}   & 0.352\footnotesize{$\pm$0.128}             & 0.371\footnotesize{$\pm$0.327}            & \textbf{0.836\footnotesize{$\pm$0.183}}       & 44.48\%                  
                                                                                                                 & \textbf{0.768\footnotesize{$\pm$0.122}}    & 0.668\footnotesize{$\pm$0.000}            & \textbf{0.743\footnotesize{$\pm$0.096}}   & 0.367\footnotesize{$\pm$0.131}             & 0.393\footnotesize{$\pm$0.350}            & \textbf{0.882\footnotesize{$\pm$0.152}}       & \textbf{79.39\%}         \\
                                                                                     & \emph{RobustECD-SE}       & 0.425\footnotesize{$\pm$0.336}             & \textbf{0.709\footnotesize{$\pm$0.218}}   & 0.576\footnotesize{$\pm$0.075}            & \textbf{0.484\footnotesize{$\pm$0.173}}    & \textbf{0.828\footnotesize{$\pm$0.197}}   & 0.529\footnotesize{$\pm$0.340}                & \textbf{53.31\%}         
                                                                                                                 & 0.427\footnotesize{$\pm$0.373}             & \textbf{0.720\footnotesize{$\pm$0.252}}   & 0.593\footnotesize{$\pm$0.102}            & \textbf{0.448\footnotesize{$\pm$0.208}}    & \textbf{0.857\footnotesize{$\pm$0.198}}   & 0.552\footnotesize{$\pm$0.352}                & 78.68\%                  \\ 
        \hline
        \multirow{6}{*}{\begin{tabular}[c]{@{}l@{}}Polbooks\\(noise)\end{tabular}}   & original                  & 0.493\footnotesize{$\pm$0.000}             & 0.531\footnotesize{$\pm$0.000}            & 0.559\footnotesize{$\pm$0.000}            & 0.512\footnotesize{$\pm$0.000}             & 0.554\footnotesize{$\pm$0.025}            & 0.556\footnotesize{$\pm$0.017}                & 26.69\%                  
                                                                                                                 & 0.536\footnotesize{$\pm$0.000}             & 0.638\footnotesize{$\pm$0.000}            & 0.681\footnotesize{$\pm$0.000}            & 0.558\footnotesize{$\pm$0.000}             & 0.647\footnotesize{$\pm$0.041}            & 0.662\footnotesize{$\pm$0.009}                & 47.95\%                  \\
                                                                                     & Attack                    & 0.418\footnotesize{$\pm$0.004}             & 0.482\footnotesize{$\pm$0.000}            & 0.393\footnotesize{$\pm$0.000}            & 0.343\footnotesize{$\pm$0.000}             & 0.461\footnotesize{$\pm$0.030}            & 0.462\footnotesize{$\pm$0.012}                & ----                     
                                                                                                                 & 0.459\footnotesize{$\pm$0.010}             & 0.530\footnotesize{$\pm$0.000}            & 0.351\footnotesize{$\pm$0.000}            & 0.252\footnotesize{$\pm$0.000}             & 0.534\footnotesize{$\pm$0.053}            & 0.581\footnotesize{$\pm$0.011}                & ----                     \\ 
        \cdashline{3-16}
                                                                                     & WERW-Kpath                & 0.485\footnotesize{$\pm$0.006}             & 0.560\footnotesize{$\pm$0.019}            & 0.503\footnotesize{$\pm$0.013}            & 0.481\footnotesize{$\pm$0.010}             & 0.552\footnotesize{$\pm$0.032}            & 0.565\footnotesize{$\pm$0.020}                & 23.74\%                  
                                                                                                                 & 0.505\footnotesize{$\pm$0.013}             & 0.641\footnotesize{$\pm$0.028}            & 0.541\footnotesize{$\pm$0.041}            & 0.552\footnotesize{$\pm$0.015}             & 0.649\footnotesize{$\pm$0.054}            & 0.660\footnotesize{$\pm$0.015}                & 39.88\%                  \\
                                                                                     & Edmot                     & 0.490\footnotesize{$\pm$0.000}             & 0.482\footnotesize{$\pm$0.000}            & 0.494\footnotesize{$\pm$0.000}            & 0.367\footnotesize{$\pm$0.000}             & 0.566\footnotesize{$\pm$0.031}            & 0.571\footnotesize{$\pm$0.018}                & 16.05\%                  
                                                                                                                 & 0.533\footnotesize{$\pm$0.000}             & 0.530\footnotesize{$\pm$0.000}            & 0.539\footnotesize{$\pm$0.000}            & 0.344\footnotesize{$\pm$0.000}             & 0.653\footnotesize{$\pm$0.051}            & 0.666\footnotesize{$\pm$0.013}                & 23.85\%                  \\
                                                                                     & \emph{RobustECD-GA}       & 0.559\footnotesize{$\pm$0.000}             & 0.559\footnotesize{$\pm$0.000}            & 0.559\footnotesize{$\pm$0.000}            & 0.559\footnotesize{$\pm$0.000}             & 0.585\footnotesize{$\pm$0.017}            & 0.592\footnotesize{$\pm$0.022}                & 34.99\%                  
                                                                                                                 & 0.646\footnotesize{$\pm$0.000}             & \textbf{0.646\footnotesize{$\pm$0.000}}   & 0.646\footnotesize{$\pm$0.000}            & 0.646\footnotesize{$\pm$0.000}             & \textbf{0.692\footnotesize{$\pm$0.011}}   & 0.658\footnotesize{$\pm$0.015}                & 57.64\%                  \\
                                                                                     & \emph{RobustECD-SE}       & \textbf{0.590\footnotesize{$\pm$0.014}}    & \textbf{0.565\footnotesize{$\pm$0.012}}   & \textbf{0.599\footnotesize{$\pm$0.008}}   & \textbf{0.564\footnotesize{$\pm$0.010}}    & \textbf{0.599\footnotesize{$\pm$0.008}}   & \textbf{0.643\footnotesize{$\pm$0.026}}       & \textbf{40.72\%}         
                                                                                                                 & \textbf{0.656\footnotesize{$\pm$0.015}}    & 0.628\footnotesize{$\pm$0.013}            & \textbf{0.691\footnotesize{$\pm$0.012}}   & \textbf{0.672\footnotesize{$\pm$0.007}}    & 0.665\footnotesize{$\pm$0.009}            & \textbf{0.729\footnotesize{$\pm$0.018}}       & \textbf{62.49\%}         \\ 
        \hline
        \multirow{6}{*}{\begin{tabular}[c]{@{}l@{}}Football\\(noise)\end{tabular}}   & original                  & 0.924\footnotesize{$\pm$0.000}             & 0.698\footnotesize{$\pm$0.000}            & 0.887\footnotesize{$\pm$0.000}            & 0.890\footnotesize{$\pm$0.000}             & 0.888\footnotesize{$\pm$0.037}            & 0.912\footnotesize{$\pm$0.012}                & 11.27\%                  
                                                                                                                 & 0.897\footnotesize{$\pm$0.000}             & 0.474\footnotesize{$\pm$0.000}            & 0.815\footnotesize{$\pm$0.000}            & 0.807\footnotesize{$\pm$0.000}             & 0.784\footnotesize{$\pm$0.103}            & 0.872\footnotesize{$\pm$0.025}                & 52.52\%                  \\
                                                                                     & Attack                    & 0.809\footnotesize{$\pm$0.000}             & 0.658\footnotesize{$\pm$0.000}            & 0.809\footnotesize{$\pm$0.000}            & 0.755\footnotesize{$\pm$0.000}             & 0.800\footnotesize{$\pm$0.051}            & 0.838\footnotesize{$\pm$0.027}                & ----                     
                                                                                                                 & 0.498\footnotesize{$\pm$0.000}             & 0.375\footnotesize{$\pm$0.000}            & 0.498\footnotesize{$\pm$0.000}            & 0.481\footnotesize{$\pm$0.000}             & 0.503\footnotesize{$\pm$0.142}            & 0.719\footnotesize{$\pm$0.055}                & ----                     \\ 
        \cdashline{3-16}
                                                                                     & WERW-Kpath                & 0.795\footnotesize{$\pm$0.003}             & 0.643\footnotesize{$\pm$0.017}            & 0.878\footnotesize{$\pm$0.022}            & 0.812\footnotesize{$\pm$0.025}             & 0.765\footnotesize{$\pm$0.059}            & 0.841\footnotesize{$\pm$0.026}                & 1.34\%                   
                                                                                                                 & 0.486\footnotesize{$\pm$0.001}             & 0.378\footnotesize{$\pm$0.028}            & 0.770\footnotesize{$\pm$0.066}            & 0.623\footnotesize{$\pm$0.055}             & 0.449\footnotesize{$\pm$0.143}            & 0.725\footnotesize{$\pm$0.057}                & 12.10\%                  \\
                                                                                     & Edmot                     & 0.809\footnotesize{$\pm$0.000}             & 0.658\footnotesize{$\pm$0.000}            & 0.809\footnotesize{$\pm$0.000}            & 0.755\footnotesize{$\pm$0.000}             & 0.781\footnotesize{$\pm$0.051}            & 0.835\footnotesize{$\pm$0.027}                & -0.46\%                  
                                                                                                                 & 0.498\footnotesize{$\pm$0.000}             & 0.375\footnotesize{$\pm$0.000}            & 0.498\footnotesize{$\pm$0.000}            & 0.481\footnotesize{$\pm$0.000}             & 0.444\footnotesize{$\pm$0.116}            & 0.714\footnotesize{$\pm$0.061}                & -2.07\%                  \\
                                                                                     & \emph{RobustECD-GA}       & \textbf{0.927\footnotesize{$\pm$0.000}}    & \textbf{0.791\footnotesize{$\pm$0.000}}   & \textbf{0.927\footnotesize{$\pm$0.000}}   & \textbf{0.909\footnotesize{$\pm$0.000}}    & 0.797\footnotesize{$\pm$0.071}            & \textbf{0.870\footnotesize{$\pm$0.015}}       & \textbf{12.20\%}         
                                                                                                                 & \textbf{0.889\footnotesize{$\pm$0.000}}    & \textbf{0.597\footnotesize{$\pm$0.000}}   & \textbf{0.889\footnotesize{$\pm$0.000}}   & \textbf{0.847\footnotesize{$\pm$0.000}}    & 0.430\footnotesize{$\pm$0.211}            & 0.753\footnotesize{$\pm$0.049}                & \textbf{47.09\%}         \\
                                                                                     & \emph{RobustECD-SE}       & 0.809\footnotesize{$\pm$0.000}             & 0.762\footnotesize{$\pm$0.024}            & 0.909\footnotesize{$\pm$0.020}            & 0.886\footnotesize{$\pm$0.014}             & \textbf{0.863\footnotesize{$\pm$0.051}}   & 0.862\footnotesize{$\pm$0.021}                & 9.38\%                   
                                                                                                                 & 0.498\footnotesize{$\pm$0.000}             & 0.488\footnotesize{$\pm$0.056}            & 0.843\footnotesize{$\pm$0.063}            & 0.793\footnotesize{$\pm$0.033}             & \textbf{0.680\footnotesize{$\pm$0.189}}   & \textbf{0.769\footnotesize{$\pm$0.043}}       & 34.40\%                  \\ 
        \hline
        \multirow{6}{*}{\begin{tabular}[c]{@{}l@{}}Polblogs\\(noise)\end{tabular}}  & original                   & 0.330\footnotesize{$\pm$0.001}             & 0.378\footnotesize{$\pm$0.000}            & 0.318\footnotesize{$\pm$0.000}            & 0.376\footnotesize{$\pm$0.000}             & 0.375\footnotesize{$\pm$0.053}            & 0.458\footnotesize{$\pm$0.067}                & 8.60\%                   
                                                                                                                 & 0.439\footnotesize{$\pm$0.001}             & 0.528\footnotesize{$\pm$0.000}            & 0.419\footnotesize{$\pm$0.000}            & 0.521\footnotesize{$\pm$0.000}             & 0.515\footnotesize{$\pm$0.105}            & 0.489\footnotesize{$\pm$0.053}                & 9.42\%                   \\
                                                                                    & Attack                     & 0.303\footnotesize{$\pm$0.001}             & 0.348\footnotesize{$\pm$0.000}            & 0.299\footnotesize{$\pm$0.000}            & 0.336\footnotesize{$\pm$0.000}             & 0.340\footnotesize{$\pm$0.061}            & 0.434\footnotesize{$\pm$0.034}                & ----                     
                                                                                                                 & 0.404\footnotesize{$\pm$0.002}             & 0.490\footnotesize{$\pm$0.000}            & 0.392\footnotesize{$\pm$0.000}            & 0.465\footnotesize{$\pm$0.000}             & 0.453\footnotesize{$\pm$0.122}            & 0.455\footnotesize{$\pm$0.017}                & ----                     \\ 
        \cdashline{3-16} 
                                                                                    & WERW-Kpath                 & 0.302\footnotesize{$\pm$0.007}             & 0.348\footnotesize{$\pm$0.012}            & 0.303\footnotesize{$\pm$0.004}            & 0.318\footnotesize{$\pm$0.003}             & 0.375\footnotesize{$\pm$0.055}            & 0.435\footnotesize{$\pm$0.036}                & 1.03\%                   
                                                                                                                 & 0.401\footnotesize{$\pm$0.005}             & 0.490\footnotesize{$\pm$0.017}            & 0.398\footnotesize{$\pm$0.008}            & 0.359\footnotesize{$\pm$0.007}             & 0.520\footnotesize{$\pm$0.118}            & 0.459\footnotesize{$\pm$0.020}                & -1.02\%                  \\
                                                                                    & Edmot                      & 0.304\footnotesize{$\pm$0.000}             & 0.348\footnotesize{$\pm$0.000}            & 0.299\footnotesize{$\pm$0.000}            & 0.336\footnotesize{$\pm$0.000}             & 0.360\footnotesize{$\pm$0.050}            & 0.431\footnotesize{$\pm$0.037}                & 0.92\%                   
                                                                                                                 & 0.404\footnotesize{$\pm$0.000}             & 0.490\footnotesize{$\pm$0.000}            & 0.392\footnotesize{$\pm$0.000}            & 0.465\footnotesize{$\pm$0.000}             & 0.494\footnotesize{$\pm$0.101}            & 0.454\footnotesize{$\pm$0.018}                & 1.47\%                   \\
                                                                                    & \emph{RobustECD-GA}        & \textbf{0.473\footnotesize{$\pm$0.000}}    & \textbf{0.527\footnotesize{$\pm$0.000}}   & 0.428\footnotesize{$\pm$0.000}            & \textbf{0.532\footnotesize{$\pm$0.000}}    & \textbf{0.530\footnotesize{$\pm$0.023}}   & 0.377\footnotesize{$\pm$0.030}                & 41.96\%                  
                                                                                                                 & 0.535\footnotesize{$\pm$0.000}             & \textbf{0.620\footnotesize{$\pm$0.000}}   & 0.599\footnotesize{$\pm$0.000}            & \textbf{0.625\footnotesize{$\pm$0.000}}    & \textbf{0.625\footnotesize{$\pm$0.031}}   & 0.345\footnotesize{$\pm$0.023}                & 26.66\%                  \\
                                                                                    & \emph{RobustECD-SE}        & 0.444\footnotesize{$\pm$0.007}             & 0.505\footnotesize{$\pm$0.007}            & \textbf{0.558\footnotesize{$\pm$0.005}}   & 0.493\footnotesize{$\pm$0.004}             & 0.500\footnotesize{$\pm$0.005}            & \textbf{0.469\footnotesize{$\pm$0.023}}       & \textbf{46.69\%}         
                                                                                                                 & \textbf{0.562\footnotesize{$\pm$0.005}}    & 0.604\footnotesize{$\pm$0.004}            & \textbf{0.637\footnotesize{$\pm$0.004}}   & 0.601\footnotesize{$\pm$0.003}             & 0.602\footnotesize{$\pm$0.004}            & \textbf{0.548\footnotesize{$\pm$0.011}}       & \textbf{34.58\%}         \\
        \hline\hline
        \end{tabular}}
    \end{table*}

\vspace{-5pt}
\subsection{Evaluation}
% In order to verify the effectiveness of the adversarial enhancement on both the original networks and the networks under adversarial attacks, we respectively use the above-mentioned four methods, including \emph{RobustECD-GA}, \emph{RobustECD-SE}, EdMot and WERW-$\kappa$Path, to show a crosswise comparison, with results reported in TABLE~\ref{ori-effect} , TABLE~\ref{adv-effect-table} and Fig.~\ref{adv-effect}, where the scores are averaged over 20 runs for each method.
We evaluate the benefit of the proposed enhancement strategies, answering the following research questions:
\begin{itemize}
    \item 
    \textbf{RQ1}: Can \emph{RobustECD} improve the performance of community detection in real-world networks combined with existing community detection algorithms?
    \item 
    \textbf{RQ2}: Does \emph{RobustECD} still work when it comes to adversarial networks?
    \item 
    \textbf{RQ3}: How does the selection of various similarity indices in \emph{RobustECD-SE} affect the performance?
    \item 
    \textbf{RQ4}: How does \emph{RobustECD} achieve interpretable enhancement of community detection?
\end{itemize}
\subsubsection{Enhancement in Real Network}
TABLE~\ref{tb:ori-effect} reports the results of the enhancement for six community detection algorithms, from which one can observe that there is a significant boost in detection performance across all six real-world networks.
Firstly, compared with those traditional enhancement algorithms, these detection algorithms combined with the proposed \emph{RobustECD} framework obtain much better results in most cases. The \emph{RobustECD-GA} and \emph{RobustECD-SE} achieves 87.50\% and 88.24\% success rate on enhancing community detection. 
The success rate refers to the percentage of enhanced results which are better than the original results in term of both NMI and ARI.
% \footnote{The success rate refers to the percentage of enhanced results, which are much better than the original results in term of both metrics.}.
These phenomenon provide a positive answer to \textbf{RQ1}, indicating that \emph{RobustECD} can improve the performance of existing community detection algorithms and alleviate the problems of resolution limit and missing data.
Meanwhile, the results in Amazon and DBLP sub-networks also indicates the effectiveness of the \emph{RobustECD-SE} in large-scale networks.

Secondly, we define the relative improvement rate (RIMP) for each metric as follows:
\begin{equation}\label{eq:RIMP}
    \setlength{\abovedisplayskip}{3pt}
    \setlength{\belowdisplayskip}{3pt}
    \mathit{RIMP}=
    \begin{cases}
        (\textit{Met}_\textit{en}-\textit{Met}_\textit{ori})/\textit{Met}_\textit{ori} & \textit{Met}_\textit{ori}>0\\
        \textit{Met}_\textit{en}-\textit{Met}_\textit{ori}                   & \textit{Met}_\textit{ori}=0\\
    \end{cases}
\end{equation}
where $\textit{Met}_\textit{ori}$ and $\textit{Met}_\textit{en}$ refer to the metric of the original and the enhanced results, respectively. Note that we also consider the extreme case that the original metrics may go down to 0 in the adversarial networks.
In TABLE~\ref{tb:ori-effect}, the far-right column of each metrics gives the average relative improvement rate (Avg RIMP) in metric, from which one can see that \emph{RobustECD-GA} and \emph{RobustECD-SE} achieve competitive performance, and significantly outperform baselines.

Thirdly, considered detection algorithms have different performance on the real datasets, but generally obtain more similar community partitions during robust enhancement.
For instance, we use standard deviation to measure the consistency of results of different detection algorithms.
For \emph{RobustECD-SE}, the standard deviations of the original ARI for the six detection algorithms on the six networks are (0.118, 0.059, 0.153, 0.046, 0.030, 0.026), while those of the corresponding enhanced ARI are (0.067, 0.018, 0.039, 0.029, 0.023, 0.025). 
The decrease in standard deviation indicates that \emph{RobustECD} can stable network structure and achieve the consistency of detection performance.
% , suggesting the robustness of \emph{RobustECD} to different detection algorithms.
Moreover, \emph{RobustECD} achieves perfect enhancement on some community detection algorithms applied to small datasets. For instance, when enhancing FG and LOU via \emph{RobustECD-SE} on Karate dataset, both NMI and ARI are equal to 1, suggesting that FG and LOU algorithms can detect community structures completely correctly after enhancement.
\subsubsection{Enhancement in Adversarial Network}
Adversarial attack aims to degrade the performance of algorithms by perturbing the network structure or attacking the computational process. 
In social networks, the adversarial attack on community detection or link prediction probably facilitates to hide the real community structure or sensitive links. 
TABLE~\ref{tb:adv-effect} reports the results of enhancing community detection in adversarial networks, which are generated by slightly modifying the networks structures via certain adversarial attacks.
% In order to address the defensive capability of our adversarial enhancement algorithms against such adversarial attacks, we also design experiments on adversarial networks obtained by slightly modifying the original networks via certain adversarial attacks. 
Note that here we report the community detection results in original networks and adversarial networks as references. 
As we can see, compared with the original results, the performance metrics display a significant decrease in adversarial networks, indicating that adversarial attack has indeed broken the network structure and achieved a community detection deception.
Then during structure enhancement, our algorithms significantly outperform the baselines in all adversarial networks. 
In fact, both \emph{RobustECD-GA} and \emph{RobustECD-SE} help the six community detection algorithms achieve huge improvements on detection performances, which is even better than the results in original networks. 
However, the baselines Edmot and WERW-KPath have mediocre performance and may fail with the increase of the attack strength. 
Such results indicate that our enhancement algorithms could not only help partially or even fully recover the network structures destroyed by adversarial attacks, but also improving the robustness of existing community detection algorithms against adversarial noise, positively answering \textbf{RQ2}.

\subsubsection{Impact of Similarity in RobustECD-SE}\label{sec:sec-lpm-analysis}
\begin{figure}[!t]
    \centering
    \setlength{\abovecaptionskip}{-4pt}
    \includegraphics[width=\linewidth]{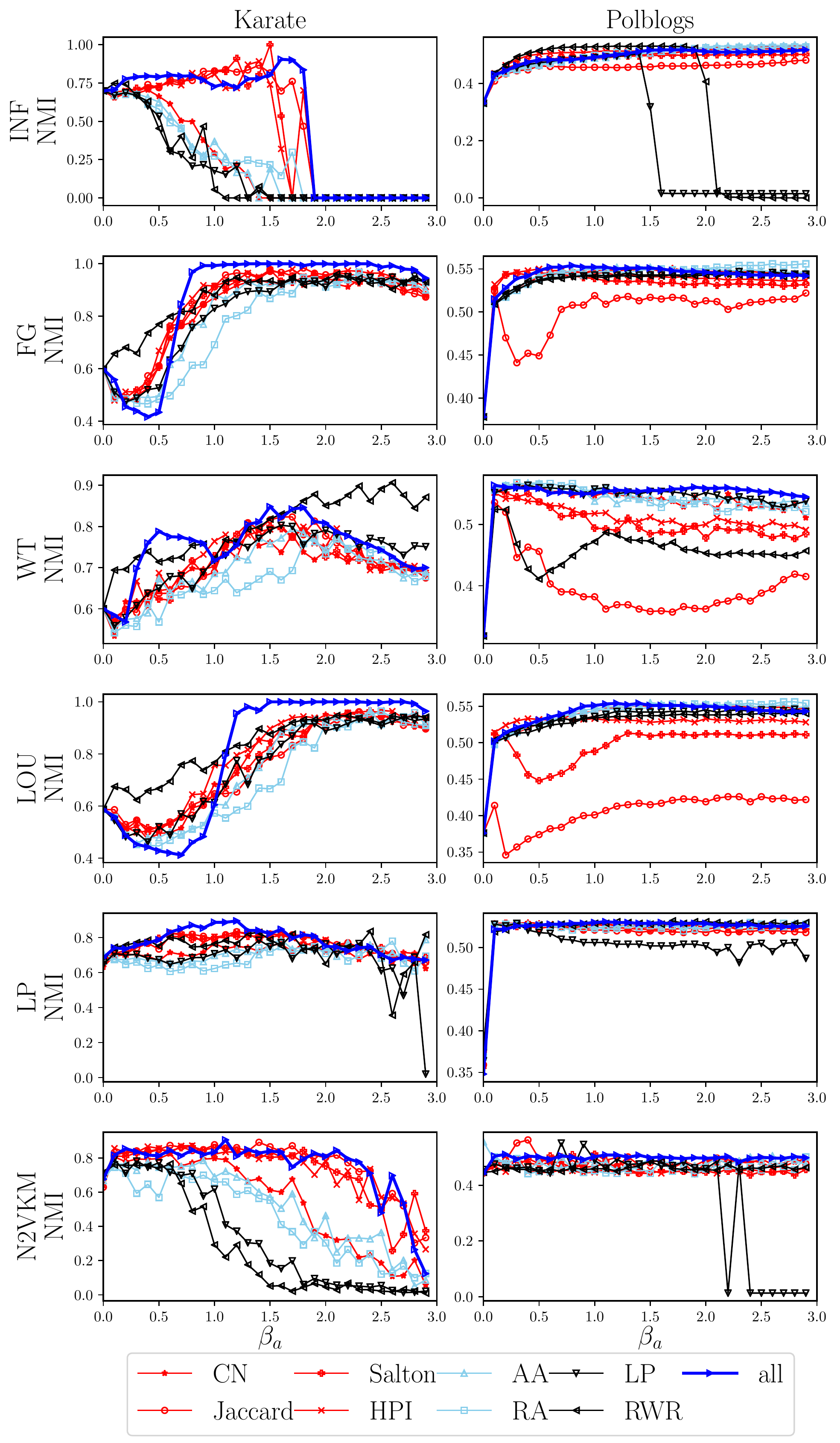}
    \caption{The impact of similarity metrics on the performance of \emph{RobustECD-SE} in term of NMI.}
    \label{fig:lpm-ori}
    \vspace{-10pt}
\end{figure}

Thanks to the outstanding performance of \emph{RobustECD}, we further investigate the impact of the similarity indices in the \emph{RobustECD-SE}.
Fig.~\ref{fig:lpm-ori} shows the results of all similarity indices (\emph{RobustECD-SE(all)}) and single index (\emph{RobustECD-SE(single)}), respectively (see the Appendix~\ref{app:4} and Appendix~\ref{app:si-comb} for more details about the impact of similarity indices).
We first summarized three impact effects:
\begin{itemize}
    \item \textbf{Complementary}: \emph{RobustECD-SE(all)} outperforms all \emph{RobustECD-SE(single)}s, indicating that these single similarity indices are complementary.
    \item \textbf{Redundant}: \emph{RobustECD-SE(single)}s with some indices achieve competitive performance against \emph{RobustECD-SE(all)}, indicating that the other indices in \emph{RobustECD-SE(single)}s that have relative poor performance are redundant.
    \item \textbf{Negative}: \emph{RobustECD-SE(single)}s with some indices outperform \emph{RobustECD-SE(all)}, indicating that the other indices in \emph{RobustECD-SE(single)}s that have relative poor performance are negative.
\end{itemize}

From the comparison results, we observe that \emph{RobustECD-SE(all)} generally outperforms \emph{RobustECD-SE(single)}, and receives more stable results in most cases.
And the impact of single similarity index behaves differently on various networks and various budgets, answering \textbf{RQ3}.
In Karate ($\beta_\textit{a} \in (1.0, 2.0)$), \emph{RobustECD-SE(single)}s with first-order similarity have relatively good performance while those with second-order and high-order similarity have relatively poor performance. 
Since that the scale of Karate is particularly small and first-order similarity are sufficient to capture structure features.
In this case, first-order similarity indices are complementary, second-order and high-order similarity could be redundant or even negative.
In Polblogs ($\beta_\textit{a} \in (0.5, 2.5)$), \emph{RobustECD-SE(single)}s achieve competitive performance against \emph{RobustECD-SE(all)} except for those with Jaccard, Salton and high-order similarity, which turn out to be redundant.
\begin{figure}[!t]
    \centering
    \setlength{\abovecaptionskip}{-4pt}
    \includegraphics[width=\linewidth]{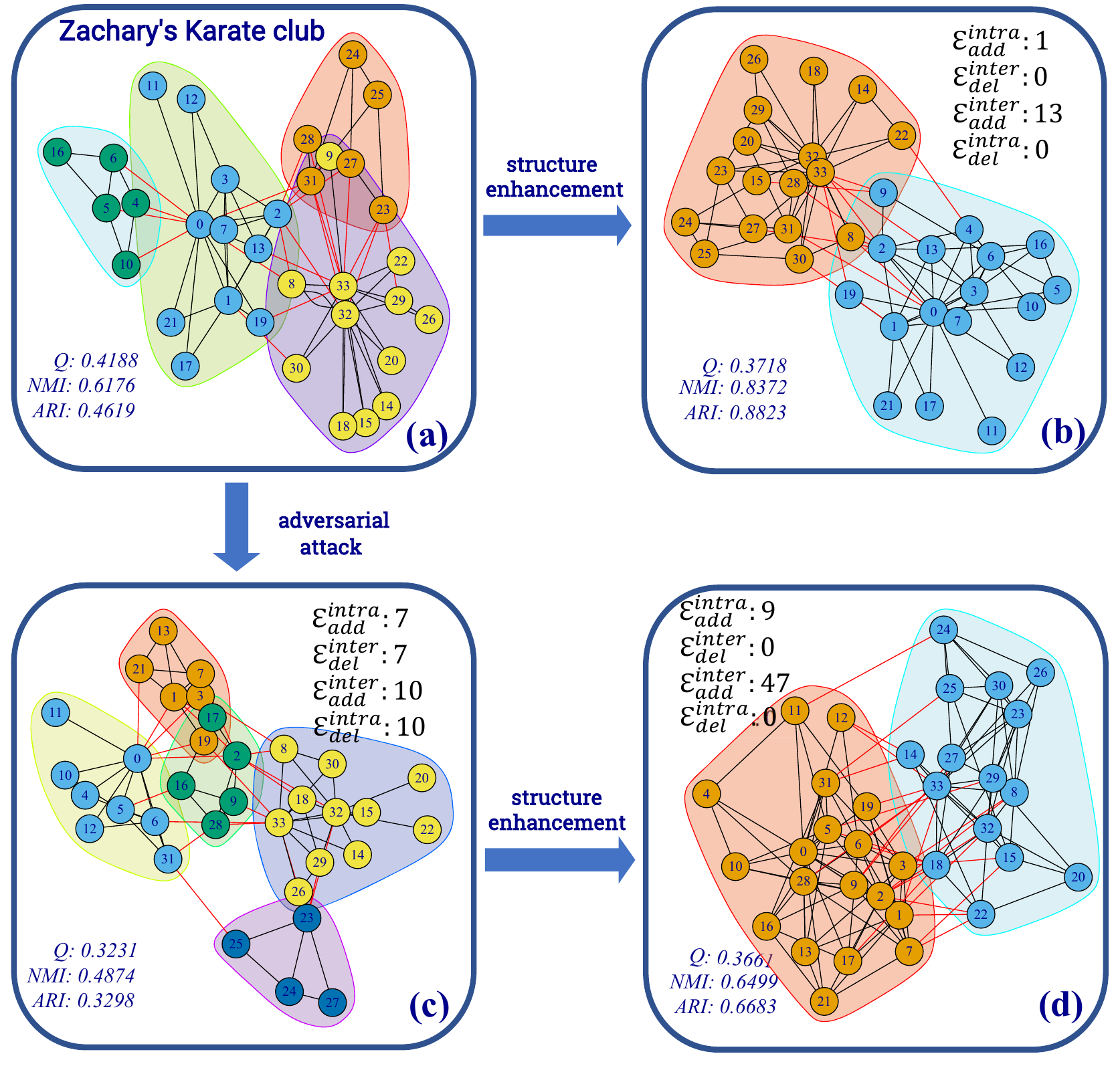}
    \setlength{\abovecaptionskip}{-10pt}
    \caption{Enhancement for LOU in Karate network.}
    \label{fig:karate-vis}\
    \vspace{-18pt}
\end{figure}
\subsubsection{Explanatory Visualization of RobustECD-GA}
%\noindent  {\spaceskip=0.2em\relax \emph{\textbf{What happens with adversarial enhancement via AE-M-GA?}}}
Next, we further investigate how the \emph{RobustECD} optimizes the performances of different detection algorithms.
Since the mechanism of \emph{RobustECD-SE} has been presented in Fig~\ref{fig:vis-puring}, 
we only visualize the \emph{RobustECD-GA} for algorithm LOU in Karate network, as shown in Fig.~\ref{fig:karate-vis}.
%Fig. \ref{karate-vis} shows the adversarial enhancement for Louvain algorithm on karate network. 

The community structure found by LOU in the original network is shown in Fig.~\ref{fig:karate-vis} (a), where there are four communities. 
Since the number of communities found by LOU is more than the ground truth ($\phi_\textit{real}=2$), as mentioned in TABLE~\ref{tb:dataset}, extra inter-community edge addition is available when $\phi_\textit{S} > \phi_\textit{real}$ ($S = \texttt {LOU}$). 
The result of \emph{RobustECD-GA} in the original network is shown in Fig.~\ref{fig:karate-vis} (b), where the enhancement scheme consists of 1 intra-community edge addition and 13 inter-community edge additions, and achieves a significant improvement on community detection, leading to the increase of 35.56\% and 91.02\% in NMI and ARI, respectively. 
Essentially, a large number of inter-community edge additions successfully merge small clusters into larger ones, resulting in more accurate partitions.

% 说明一下fitness设计问题
Notably, the decrease of modularity here (from 0.4188 to 0.3718) can explain why we don't design modularity as fitness function directly. 
By comparing the information in Figs.~\ref{fig:karate-vis} (a) and (b), a community partition with larger modularity does not mean closer to the ground-truth community structure. 
Therefore, if we have knowledge of community information, we can combine modularity with the true number of clusters, to obtain more accurate optimization guidance. 
This has been shown to have excellent performance in enhancing community detection. 
However, when facing unlabeled networks, the fitness function degrades to modularity, i.e., $f=\mathcal{Q}$, so the \emph{RobustECD-GA} may be weakened to a certain extent.

Now, consider the adversarial network obtained by $\mathcal{Q}$-Attack, as shown in Fig.~\ref{fig:karate-vis} (c). 
$\mathcal{Q}$-Attack keeps the number of edges unchanged during community deception and achieves a 22.85\% reduction in modularity with an attack budget of 17. 
As we can see, community structure suffers from structural damage and a new cluster that contains the fringe vertices in the original network is discovered.   
We then deploy structure enhancement to this adversarial network and obtain the enhanced network shown in Fig.~\ref{fig:karate-vis}~(d).
Compared with the community partition in Fig.~\ref{fig:karate-vis}~(c), LOU achieves a better partition, which even surpasses that in the original network shown in Fig.~\ref{fig:karate-vis}~(a). 
Such result suggests that our enhancement algorithms can indeed help the existing community detection algorithms defend against adversarial attacks. 
More interestingly, it seems that such structure enhancement not only repairs the broken network structure caused by adversarial attack, but also further optimizes it to obtain a clearer community structure, answering \textbf{RQ4}. 
\begin{figure*}[!t]
    \centering
    \setlength{\abovecaptionskip}{-2pt}
    \includegraphics[width=\textwidth]{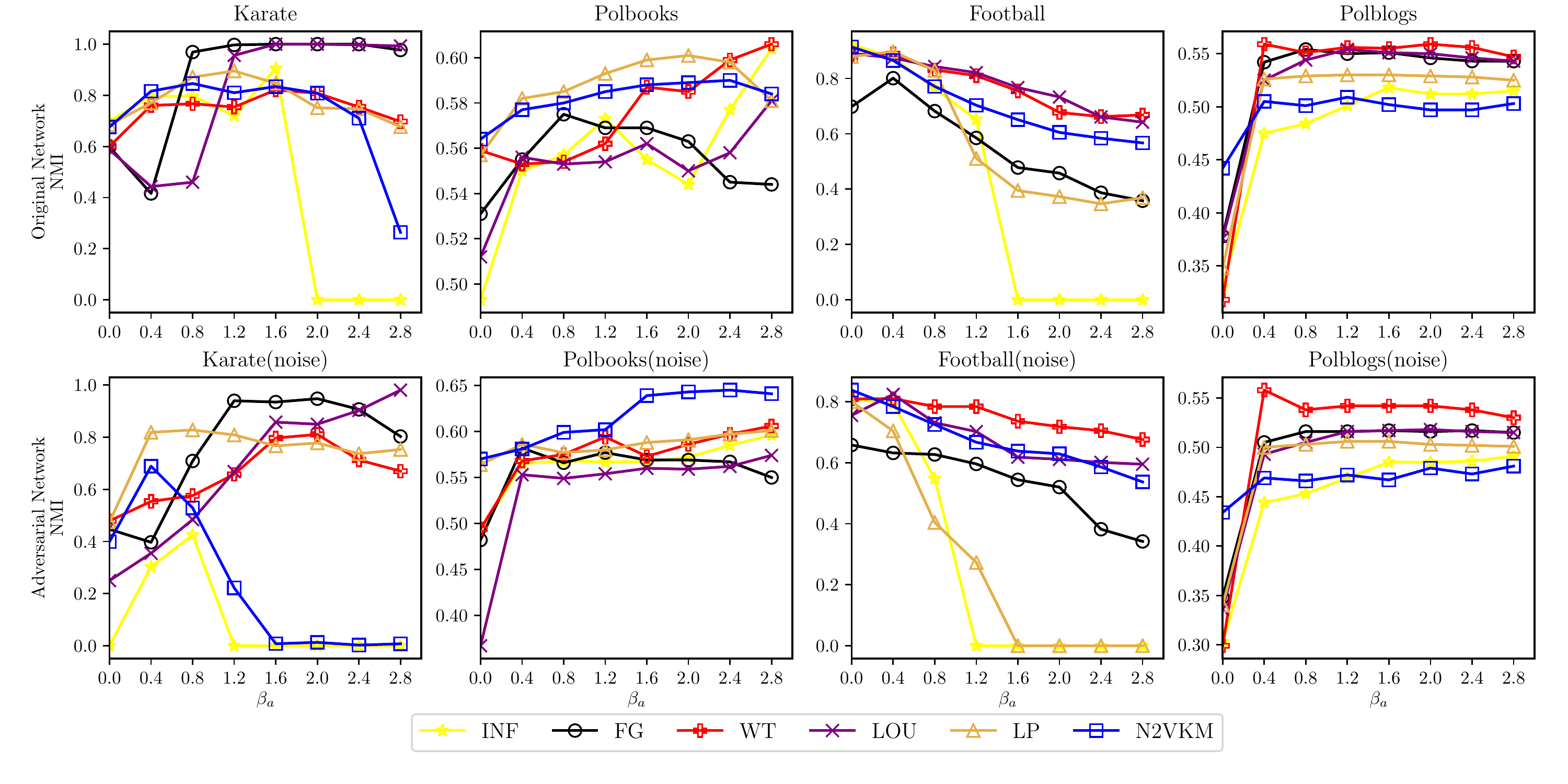}
    \caption{The impact of modification budget $\beta_\textit{a}$ on the performance of \emph{RobustECD-SE}.}
    \label{fig:el-para-impact}
    \vspace{-15pt}
\end{figure*}

\begin{figure}[!t]
    \centering
    \setlength{\abovecaptionskip}{-2pt}
    \includegraphics[width=\linewidth]{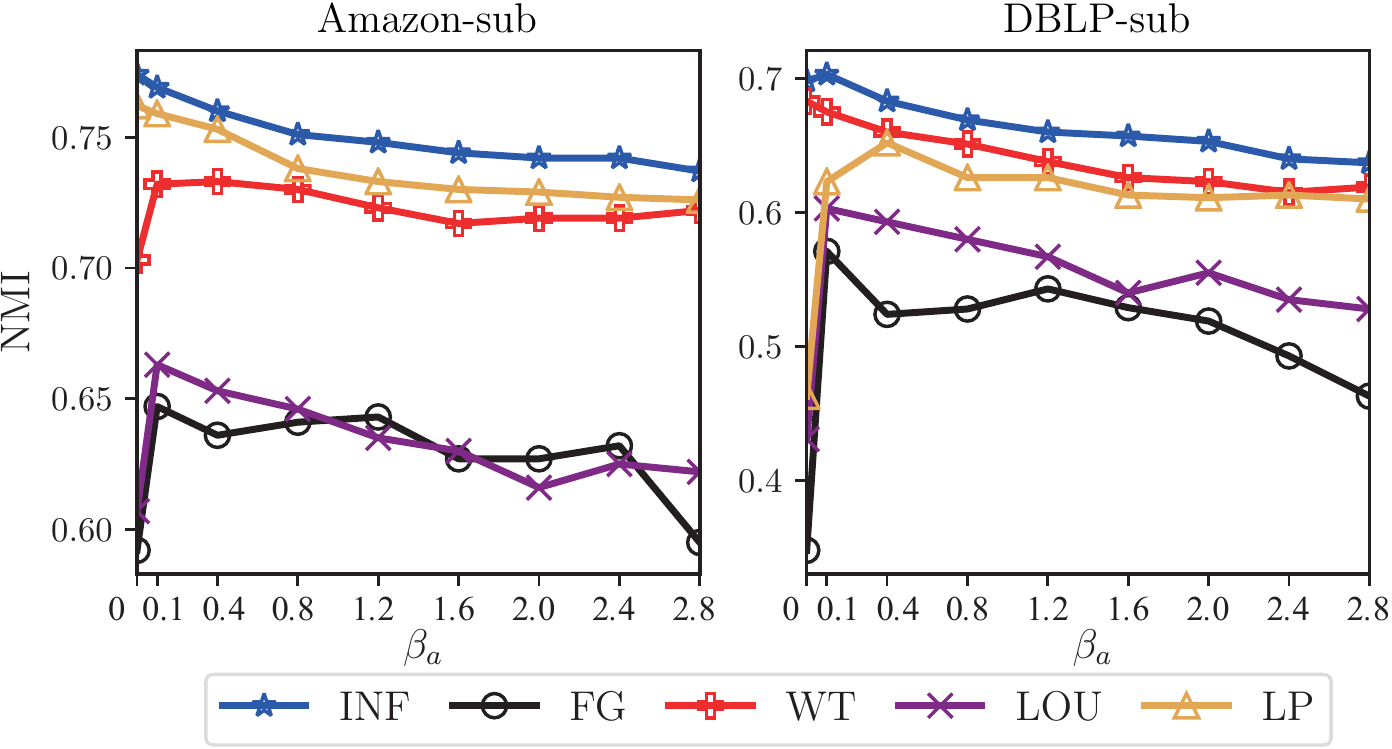}
    \caption{The impact of modification budgets $\beta_\textit{a}$ on the performance of \emph{RobustECD-SE} for large-scale networks.}
    \label{fig:large-para-impact}
    \vspace{-20pt}
\end{figure}
% \subsubsection{Impact of modification budget in \emph{RobustECD-SE}}\label{analysis}
\subsubsection{Parameter Sensitivity}
In this subsection, we discuss the impact of key parameters on the performance of \emph{RobustECD-SE} (see the Appendix~\ref{app:3} for the sensitivity analysis in \emph{RobustECD-GA}).
The metrics with a budget of 0 corresponds to the original results.

% Firstly, we present the evaluate results of \emph{RobustECD-SE} on both original and noise datasets.
% Due to space limit and similarity between NMI and ARI, Fig.~\ref{fig:el-para-impact} only show the results in term of NMI \textcolor{blue}{(see the Appendix for more details about ARI)}.
Firstly, we present the evaluate results of \emph{RobustECD-SE} in Fig.~\ref{fig:el-para-impact}, from which one can see that such impact behaves differently on various networks.
Specifically, Polblogs has a low average NMI equal to 0.373, and the performance of \emph{RobustECD-SE} is relatively stable with the increase of budget; Polbooks has an average NMI equal to 0.534, and its performance curve is messy but basically goes up; for Karate with the average NMI equal to 0.646, when the budget is relatively large, FG and LOU performs well while INF and N2VKM suffers from negative enhancement; Football has an average NMI up to 0.869, the performance of \emph{RobustECD-SE} drops steadily with the increase of budget.
Essentially, the impact of modification budget on the performance of \emph{RobustECD-SE} is influenced by the network structure. That is, for networks with weak community structures and low average performance metrics, like Polblogs, most community detection algorithms have huge spaces to be enhanced. But, for those networks with strong community structures like Football, it is difficult for \emph{RobustECD-SE} to further enhance the community detection, i.e., adding or deleting more links may even weaken the stable community structure, leading to the degradation of performance.

% Next, Fig.~\ref{fig:ga-para-impact} shows the evaluate results of \emph{RobustECD-GA}, from which one can see that \emph{RobustECD-GA} performs well in small-scale networks.
% In general, \emph{RobustECD-GA} is not strictly sensitive to different parameter settings in most cases.
% As a reasonable explanation, \emph{RobustECD-GA} is based on evolutionary computation, which is capable of finding the optimal modification scheme in a small range of solution space.
% And with the increase of budget, the solution space becomes larger and \emph{RobustECD-GA} is prone to fall into local optimum.
% Moreover, with the increase of network scale, like Polblogs, \emph{RobustECD-GA} performs poorly.
% Note that the range of budget in \emph{RobustECD-GA} is much smaller than that in \emph{RobustECD-SE}, since that the candidate sets of edge modification are conditioned.

Finally, Fig.~\ref{fig:large-para-impact} shows the evaluate results of \emph{RobustECD-SE} in large-scale networks.
As we can see, \emph{RobustECD-SE} still works in large-scale networks in most cases, and the performance drops slowly as the budget increase.

\begin{table}[!t]
    \renewcommand\arraystretch{1.2}
    \centering
    \setlength{\abovecaptionskip}{0pt}
    \caption{The average running time ($s$) of the four enhancement algorithms. The test is performed on LOU with the same experimental setup.}
    \label{tb:time}
    \resizebox{\linewidth}{!}{%
    \begin{tabular}{l|rrrrrr} 
        \hline\hline
        \diagbox{Method}{Time (s)}{Dataset} & \multicolumn{1}{c}{Karate} & \multicolumn{1}{c}{Polbooks} & \multicolumn{1}{c}{Football} & \multicolumn{1}{c}{Polblogs} & \multicolumn{1}{c}{Amazon} & \multicolumn{1}{c}{DBLP}  \\ 
        \hline
        Edmot                     & 0.01                       & 0.03                         & 0.05                         & 1.37                         & 2.50                       & 24.10                     \\
        WERW-Kpath                & 2.10                       & 11.40                        & 25.77                        & 1200.00                      & 5000.00                    & 68900.00                  \\
        \emph{RobustECD-GA}              & 47.40                      & 136.00                       & 170.00                       & 13900.00                     & ------                     & ------                    \\
        \emph{RobustECD-SE}              & 0.12                       & 0.24                         & 0.38                         & 17.10                        & 357.00                     & 5440.00                   \\
        \hline\hline
        \end{tabular}
    }
    \vspace{-10pt}
\end{table}
    
% analysis \emph{RobustECD-GA} fitness and time complex
\subsubsection{Computational Complexity Analysis}
In order to compare the efficiency of our \emph{RobustECD}, we roughly estimated their time complexity as follows.
\begin{itemize}
    \item The most computationally expensive part of \emph{RobustECD-GA} is the calculation of fitness, which consists of modularity $\mathcal{Q}$ and the number of communities $\phi_{\textit{S}}$, so an extra community detection is necessary before the calculation of fitness.
    Besides, selection, crossover and mutation are consisted of sampling and edge operations, and have a cost of $\mathcal{O}(|\mathcal{E}|)$, where $|\mathcal{E}|$ is the number of edges in the original network.
    So \emph{RobustECD-GA} runs in time $\mathcal{O}(\phi_\textit{p} \cdot \mathcal{T}_\textit{ga} \cdot \max(|\textit{S}|,|\mathcal{E}| ))$, where $|\textit{S}|$ is the time complexity of the target community detection algorithm.
    % For evolutionary algorithms, a fitness function is used to evaluate the quality of the results during iteration, which directly affects the optimization effectiveness and convergence speeds of the algorithms. 
    % In \emph{RobustECD-GA}, the fitness function consists of modularity $\mathcal{Q}$ and the number of communities $\phi_{\textit{S}}$, which are obtained from the community structure, so an extra community detection is necessary before the calculation of fitness. 
    % \item \emph{RobustECD-SE} runs in time $\mathcal{O}(m\cdot \beta_\textit{a} + |\textit{S}| + |\mathcal{H}|)$, where $|\mathcal{H}|$ is the time complexity of similarity metric. 
    % Note that $\mathcal{O(|H|)}$ is the maximum time complexity among selected similarity metrics.
    % The weighted random sampling without replacement has a time complexity of $m \cdot \beta_\textit{a}$. During the process of partition ensemble, which runs in time $\mathcal{O}(n^2)$, the selection of a threshold has a cost of $\mathcal{O}(|\mathcal{E}_{co}|)$, where $\mathcal{E}_{co}$ is the edge set of the co-occurrence network, for which the maximum possible number of edges is $n(n-1)/2$. 
    \item The most computationally expensive part of \emph{RobustECD-SE} is the threshold selection,
    which has a cost of $\mathcal{O}(|\mathcal{E}_\textit{co}|)$, where $|\mathcal{E}_\textit{co}|$ is the number of edges in the co-occurrence network. $\mathcal{G}_\textit{co}$ can be much denser than the original network and have up to $n(n-1)/2$ edges. 
    So the time complexity of \emph{RobustECD-SE} is no more than $\mathcal{O}(n^2)$.
\end{itemize}
Moreover, we evaluate the efficiency of \emph{RobustECD} by directly comparing the running time with baselines. 
The average running time (in seconds) of the four algorithms are presented in TABLE~\ref{tb:time}. 
As we can see, although the \emph{RobustECD-GA} performs well on small-scale networks, it is limited by the optimization mode and does not scale well on large networks. 
Instead, \emph{RobustECD-SE} has a relatively small time complexity and scales well on large networks.

\section{Conclusion} \label{sec:conclusion}
In this paper, we proposed to enhance network structure to improve the performance of existing community detection algorithms.
In particular, we put forward two structure enhancement algorithms, namely \emph{RobustECD-GA} and \emph{RobustECD-SE}, taking both robustness and generalization into account. Extensive experimental results demonstrate the superiority of our methods in helping six common community detection algorithms achieve significant performance improvements for both real-world networks and adversarial networks, and further solve the resolution limit in modularity optimization and achieve consensus partitions.
We believe this could be a fruitful avenue of future research that address more complex situations like overlapping community in dynamic networks.
% Our findings inspire more ideas for future works. For instance, the current selected community detection algorithms mainly focus on non-overlapping community detection, therefore extending robust enhancement to overlapping case for community detection, could be interesting topics for further studies.
% Moreover, we designed a fitness function with the number of clusters in \emph{RobustECD-GA} and the process of threshold selection in \emph{RobustECD-SE}, which can help the existing community detection algorithms solve the resolution limit in modularity optimization and achieve consensus partitions. Although there is a restriction of time complexity in \emph{RobustECD-GA},  \emph{RobustECD-SE} is effective and scales well on large networks.
% Finally, our findings inspire more ideas for future works. For instance, the current selected community detection algorithms mainly focus on nonoverlapping community detection, therefore extending adversarial enhancement to overlapping case for community detection, could be interesting topics for further studies.

% \appendices
% \section{Impact of similarity in RobustECD-SE}
% TABLE

%\section{123}

\ifCLASSOPTIONcompsoc
  \section*{Acknowledgments}
\else
  \section*{Acknowledgment}
\fi
The authors would like to thank all the members in the IVSN Research Group, Zhejiang University of Technology for the valuable discussions about the ideas and technical details presented in this paper. This work was partially supported by the National Natural Science Foundation of China under Grant 61973273, by the Zhejiang Provincial Natural Science Foundation of China under Grant LR19F030001, and by the Hong Kong Research Grants Council under the GRF Grant CityU11200317.

%\newpage



\bibliographystyle{IEEEtran}
\bibliography{MyBibliography}


%   H.~Kopka and P.~W. Daly, \emph{A Guide to \LaTeX}, 3rd~ed.\hskip 1em plus

% \end{thebibliography}

% \vspace{-45pt}
\begin{IEEEbiography}[{\includegraphics[width=1in,height=1.25in,clip,keepaspectratio]{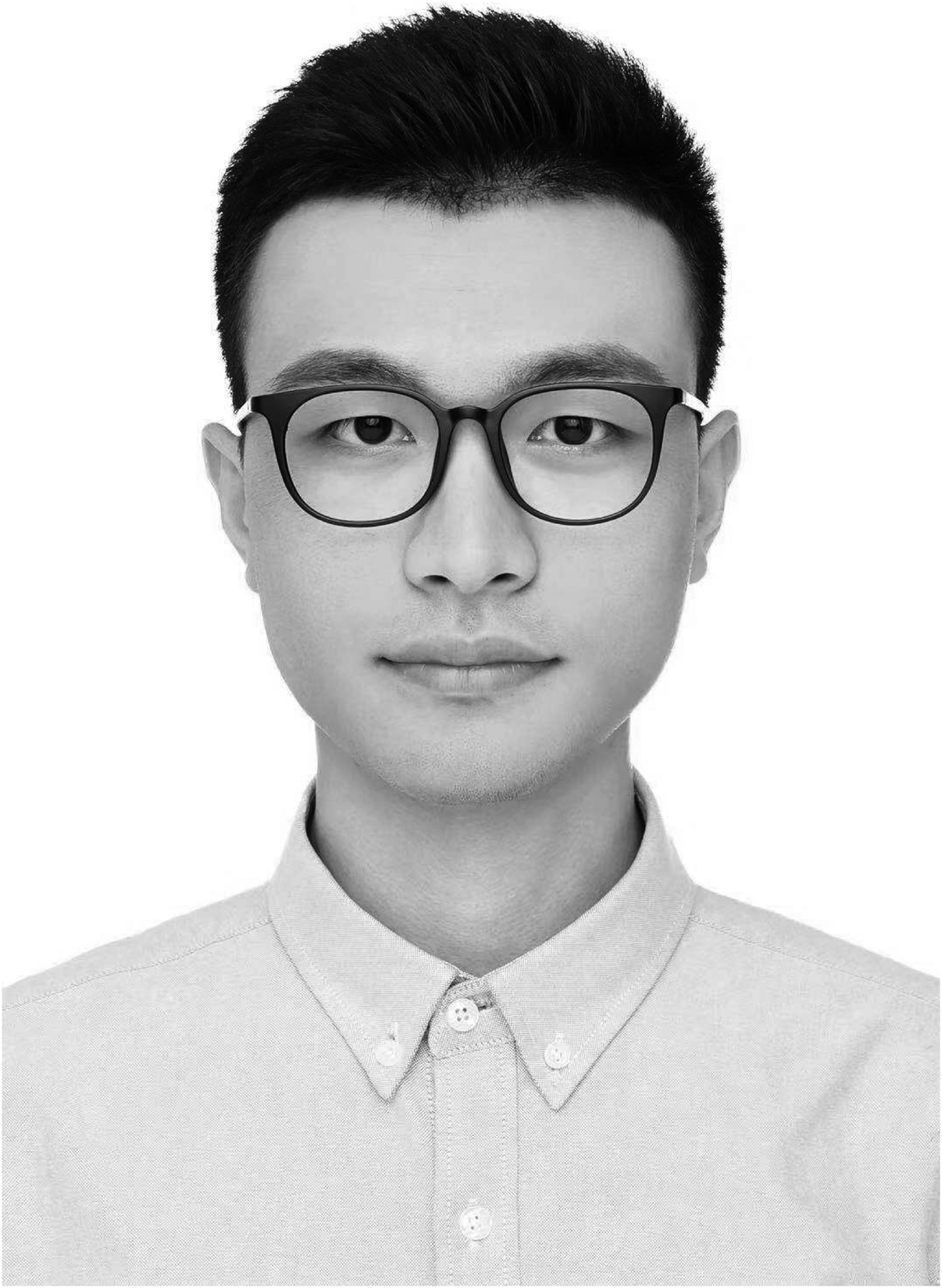}}]{Jiajun Zhou}
received the BS degree in automation from the Zhejiang University of Technology, Hangzhou, China, in 2018, where he is currently pursuing the MS degree in control theory and engineering with the College of Information Engineering.
His current research interests include graph mining and deep learning, especially for network security.
\end{IEEEbiography}
\vspace{-35pt}
\begin{IEEEbiography}[{\includegraphics[width=1in,height=1.25in,clip,keepaspectratio]{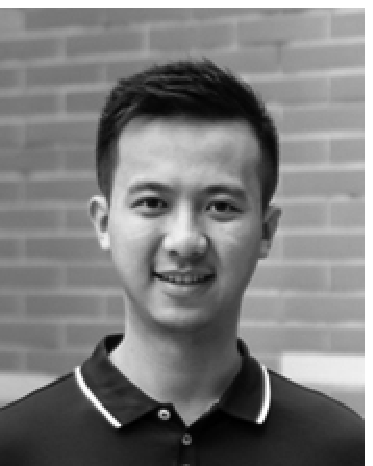}}]{Zhi Chen} received the BS and MS degrees in EECS from UC Berkeley, in 2019 and 2020, respectively. He is currently working toward the PhD degree in computer science with the University of Illinois, Urbana-Champaign. At UC Berkeley, he was a research assistant with the Center for Long-Term Cybersecurity from 2019 to 2020, and with BAIR Lab in 2018. His research interests include security and machine learning. 
\end{IEEEbiography}
% \vspace{-35pt}
\begin{IEEEbiography}[{\includegraphics[width=1in,height=1.25in,clip,keepaspectratio]{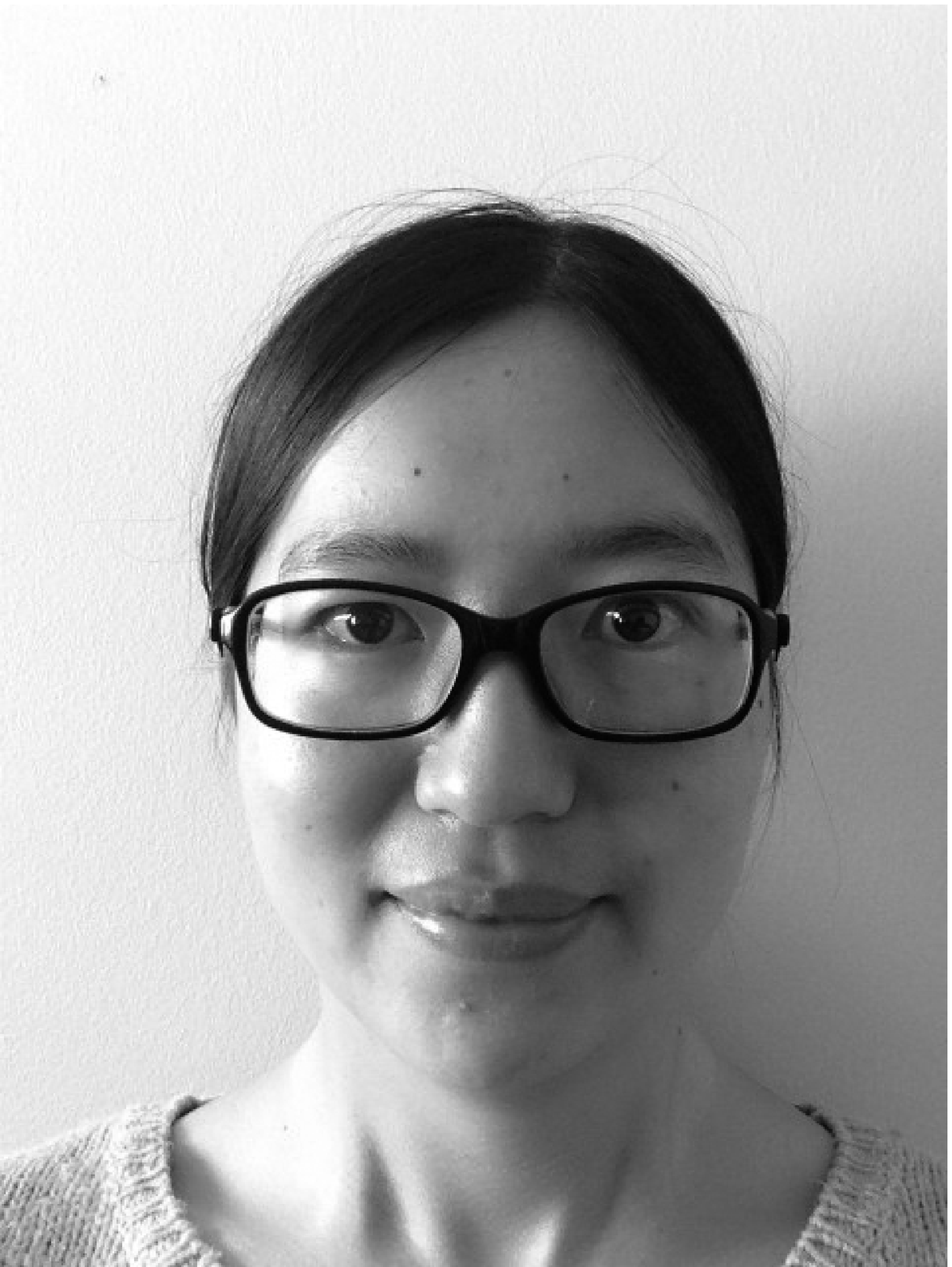}}]{Min Du}
received the PhD degree from the School of Computing, University of Utah in 2018,
after completing the bachelor’s degree and the master's degree from Beihang University. She is currently a Postdoctoral scholar in EECS Department, UC Berkeley. Her research interests include big data analytics and machine learning security.
\end{IEEEbiography}
\vspace{-35pt}
\begin{IEEEbiography}[{\includegraphics[width=1in,height=1.25in,clip,keepaspectratio]{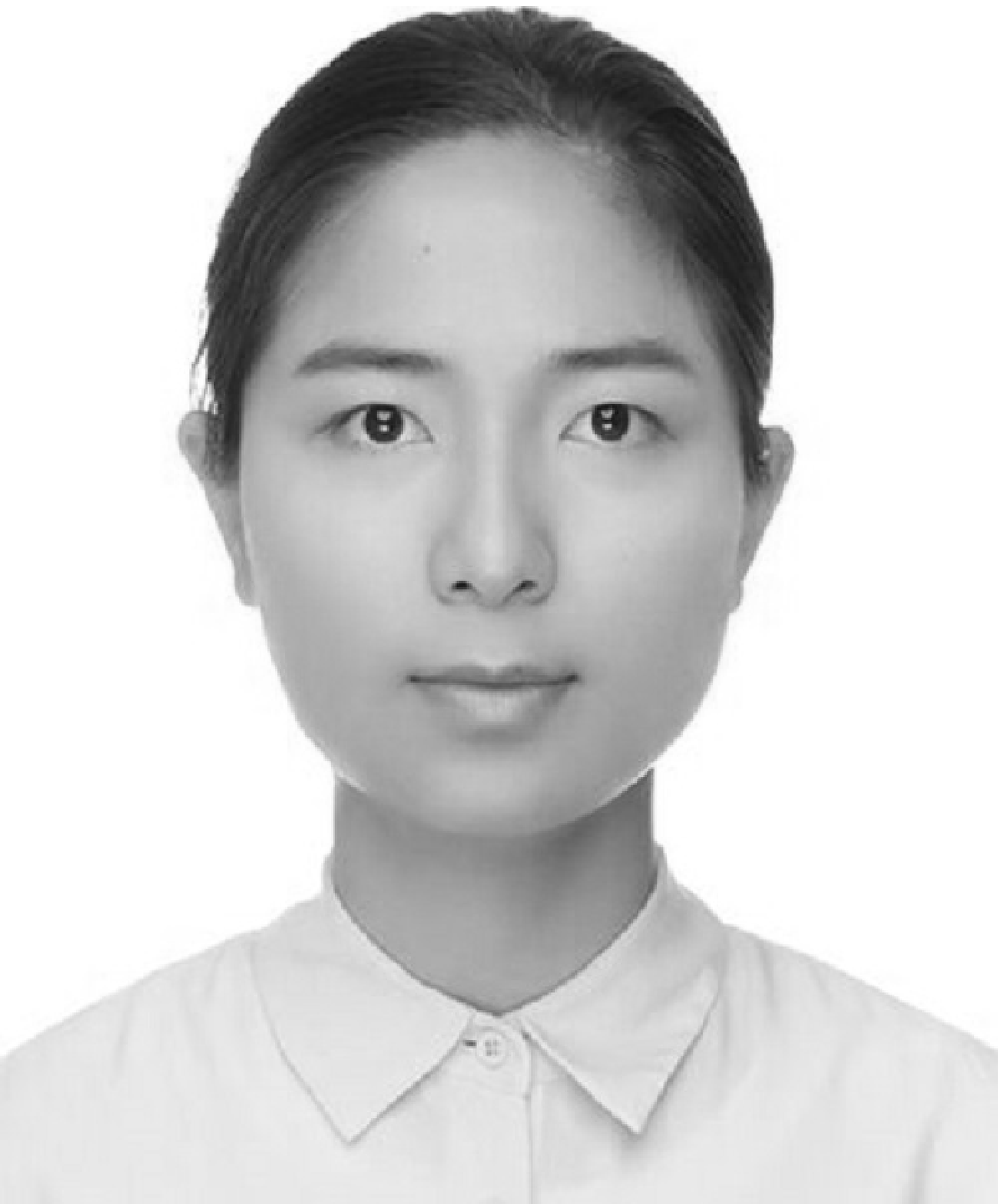}}]{Lihong Chen}
received the BS degree from Zhejiang University of Technology, Hangzhou, China, in 2018. She is currently pursuing the MS degree at College of Information Engineering, Zhejiang University of Technology, Hangzhou, China. Her current research interests include social network analysis, evolutionary computing and deep learning.
\end{IEEEbiography}
\vspace{-35pt}
\begin{IEEEbiography}[{\includegraphics[width=1in,height=1.25in,clip,keepaspectratio]{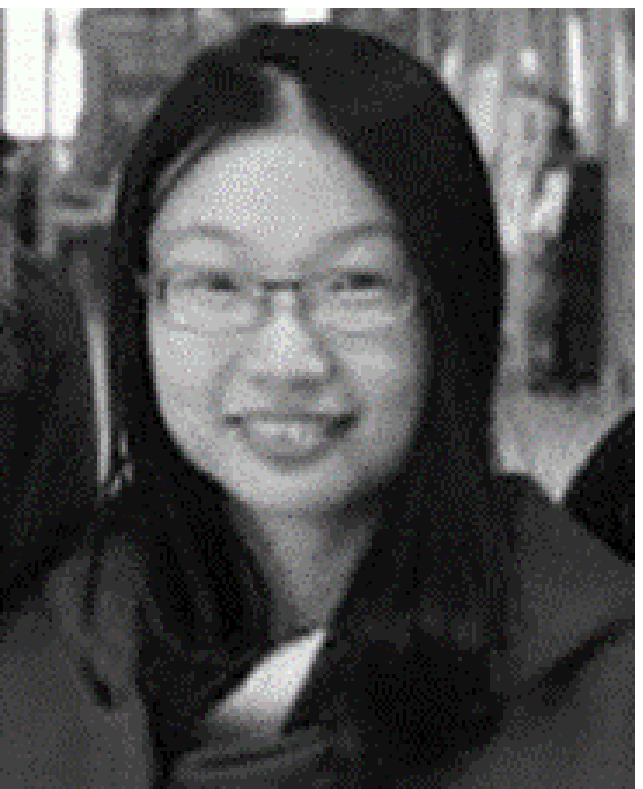}}]{Shanqing Yu} received the MS and PhD degrees from the Graduate School of Information, Production and Systems, Waseda University,
Japan, and the School of Computer Engineering, in 2008 and 2011 respectively. She is currently a lecturer at the College of Information Engineering, Zhejiang University of Technology. Her research interests cover intelligent computation, data mining and intelligent transport systems.
\end{IEEEbiography}
\vspace{-35pt}
\begin{IEEEbiography}[{\includegraphics[width=1in,height=1.25in,clip,keepaspectratio]{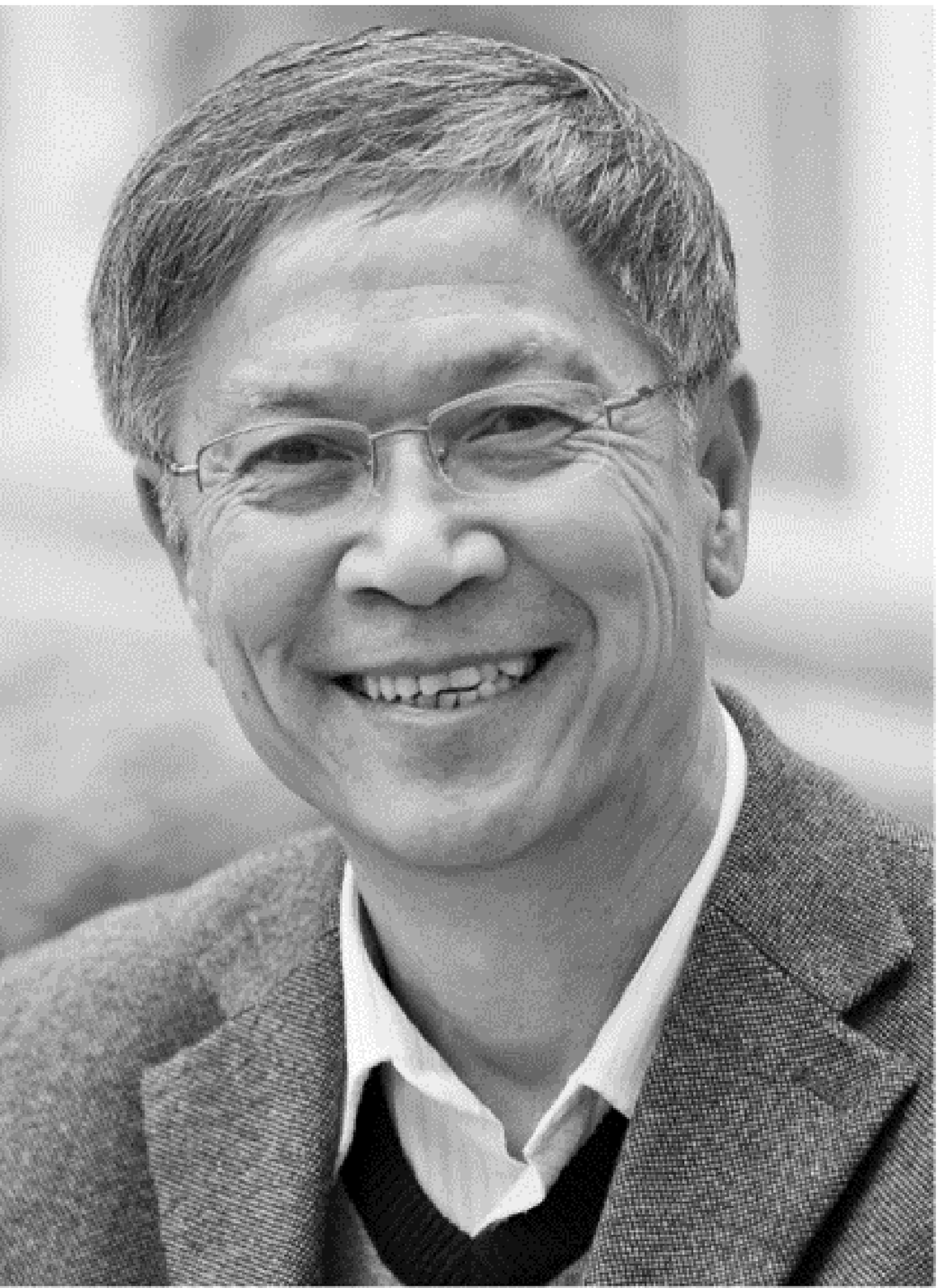}}]{Guanrong Chen}(M’89-SM’92-F’97) 
received the MSc degree in computer science from Sun Yat-sen University, Guangzhou, China in 1981, and the PhD degree in applied mathematics from Texas A\&M University, College Station, Texas, in 1987. 
He has been a chair professor and the founding director of the Centre for Chaos and Complex Networks at the City University of Hong Kong since 2000. 
Prior to that, he was a tenured full professor with the University of Houston, Texas. He was awarded the 2011 Euler Gold Medal, Russia, and conferred a Honorary Doctorate by the Saint Petersburg State University, Russia in 2011 and by the University of Le Havre, Normandy, France in 2014. 
He is a member of the Academy of Europe and a fellow of The World Academy of Sciences, and is a Highly Cited Researcher in Engineering as well as in Mathematics according to Thomson Reuters.
\end{IEEEbiography}
\vspace{-35pt}
\begin{IEEEbiography}[{\includegraphics[width=1in,height=1.25in,clip,keepaspectratio]{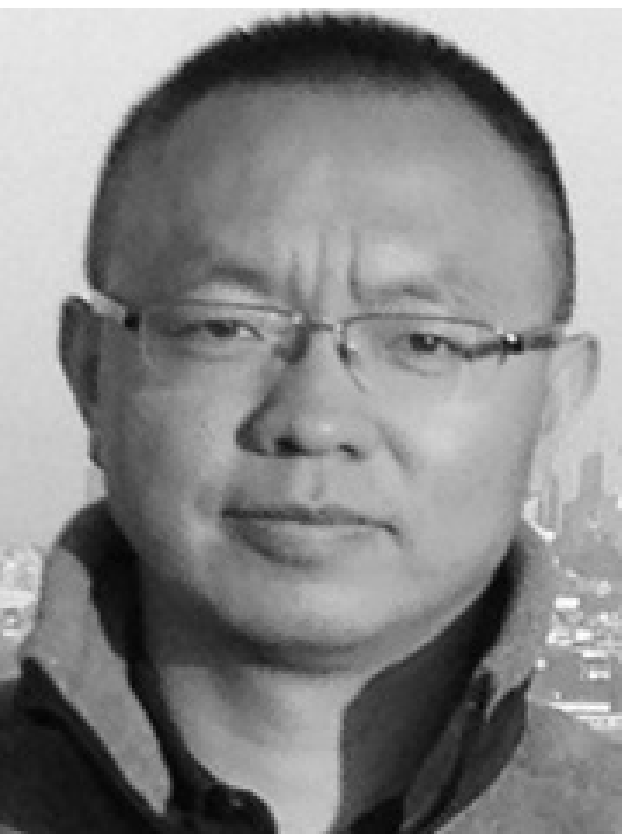}}]{Qi Xuan}(M'18) received the BS and PhD degrees in control theory and engineering from Zhejiang University, Hangzhou, China, in 2003 and 2008, respectively. He was a Post-Doctoral Researcher with the Department of Information Science and Electronic Engineering, Zhejiang University, from 2008 to 2010, and a Research Assistant with the Department of Electronic Engineering, City University of Hong Kong, Hong Kong, in 2010 and 2017. From 2012 to 2014, he was a Post-Doctoral Fellow with the Department of Computer Science, University of California at Davis, CA, USA. He is a member of the IEEE and is currently a Professor with the Institute of Cyberspace Security, College of Information Engineering, Zhejiang University of Technology, Hangzhou, China. His current research interests include network science, graph data mining, cyberspace security, machine learning, and computer vision.
\end{IEEEbiography}


%\newpage


%\vfill


\clearpage

\appendices
    \section{}

    \subsection{Discussion on edge deletion in RobustECD-SE}\label{app:1}
    % \textbf{Discussion on edge addition/deletion in RobustECD-SE}
    \noindent Similar to \emph{RobustECD-GA}, these two parameters ($\beta_\textit{a}$, $\beta_\textit{d}$) are also available in \emph{RobustECD-SE}.
    We extensively study the effect of budget parameters in \emph{RobustECD-SE} to support our decision mentioned in Sec.~\ref{sec:RobustECD-SE-nr}, i.e., we only consider edge addition in \emph{RobustECD-SE} and neglect edge deletion.

\begin{figure}[!htb]
        \centering
        \includegraphics[width=\linewidth]{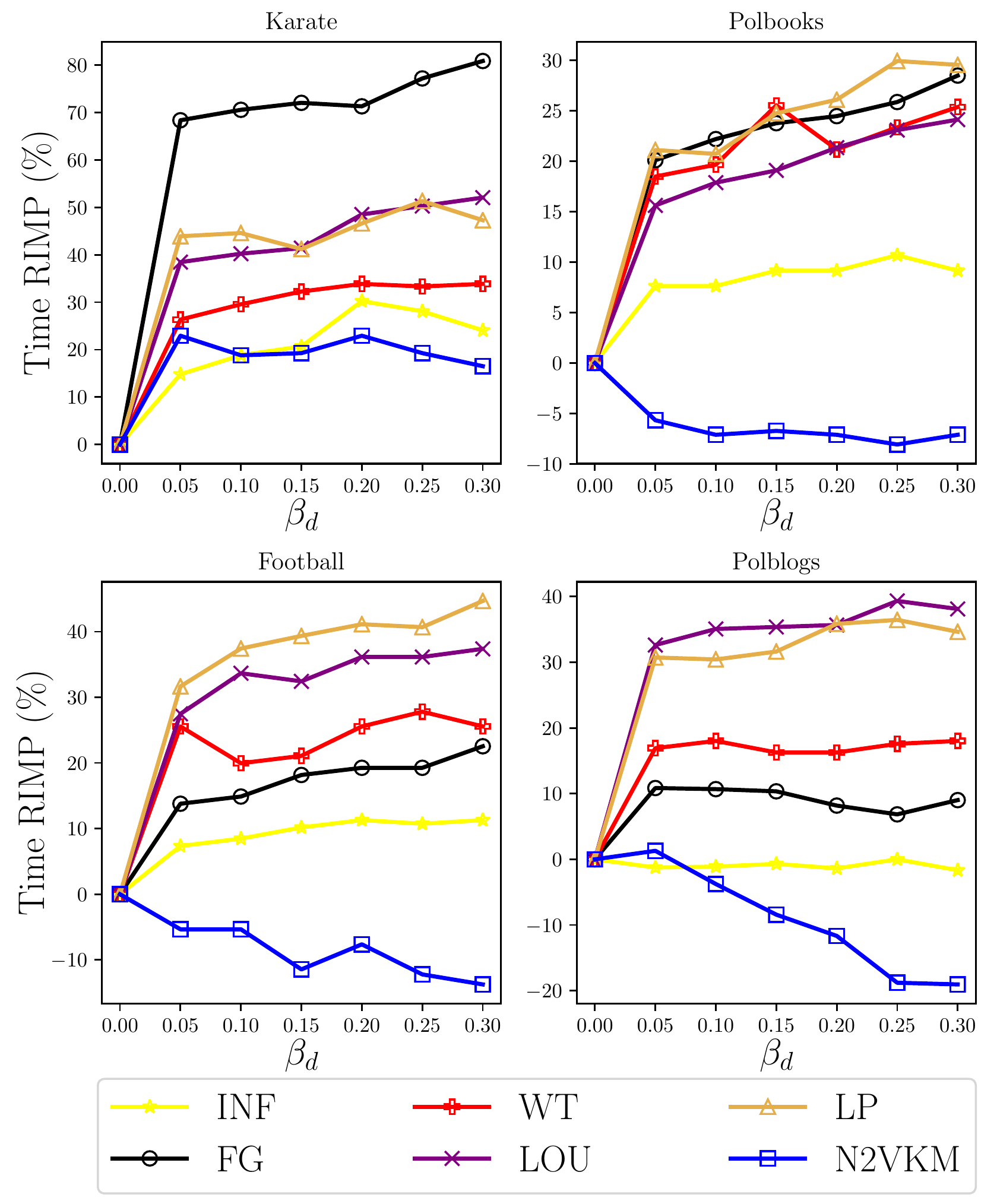}
        \caption{Extra time consumption of edge deletion in \emph{RobustECD-SE}.}
        \label{fig:app:ad-time}
    \end{figure}

    Fig.~\ref{fig:app:ad-time} show the relative improvement rate of time in \emph{RobustECD-SE} involving edge deletion.
    From the comparison results, one can observe that edge deletion brings up extra time consumption in most cases.

\begin{figure*}[!htb]
        \centering
        \includegraphics[width=\textwidth]{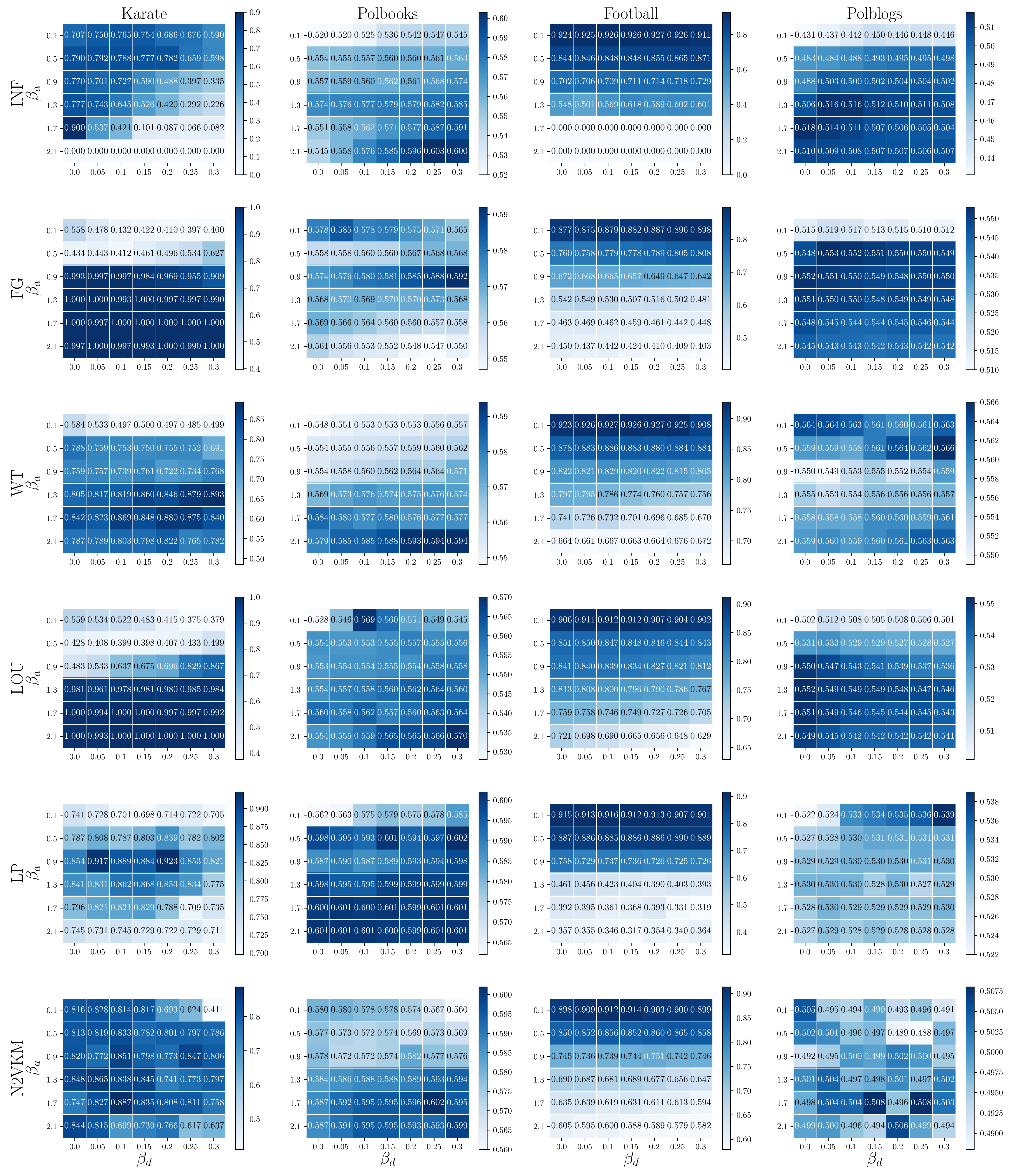}
        \caption{The impact of both $\beta_\textit{a}$ and $\beta_\textit{d}$ in \emph{RobustECD-SE} in term of NMI.}
        \label{fig:app:ad}
    \end{figure*}

    Moreover, the impact of edge addition/deletion is shown in Fig.~\ref{fig:app:ad}, from which one can observe that \emph{RobustECD-SE} with extra edge deletion has the same or worse effect as that with only edge addition in most cases.
    
    \subsection{Adversarial networks details}\label{app:2}
    Adversarial perturbation is treated as malicious noise, which can be generated via adversarial attack.
    We consider the following two community deception methods to deploy adversarial attack:
    \begin{itemize}
        \item {\textbf {$\mathcal{Q}$-Attack}~\cite{chen2019ga}.} 
        It is an evolutionary attack strategy based on genetic algorithms, in which the modularity is used to design the fitness function. This strategy deploys attack via negligible network rewiring, which doesn't change the degree of vertices, and achieves the state-of-the-art attack effect.
        \item {\textbf {$\mathcal{D}_{m}$-Deception via Modularity}~\cite{fionda2017community}.} 
        $\mathcal{D}_{m}$ is a community deception algorithm based on modularity, which can hide a target community via intra-community edge deletion and inter-community edge addition.
    \end{itemize}
    We generate adversarial networks for all small benchmark networks.
    Details of attack setup are shown in TABLE~\ref{app:tb:adv-detail}.
    \begin{table}
        \renewcommand\arraystretch{1.2}
        \centering
        \caption{Details of attack setup.}
        \label{app:tb:adv-detail}
        \resizebox{\linewidth}{!}{%
        \begin{tabular}{lccc} 
        \hline\hline
        \multirow{2}{*}{Adversarial network} & \multirow{2}{*}{Attack method} & \multicolumn{2}{c}{Parameter}  \\ 
        \cline{3-4}
                             &               & $\mathcal{S}$   &$\beta$ \\ 
        \hline
        Karate(noise)                         & $\mathcal{Q}$-Attack      & LOU  & 5      \\
        Polbooks(noise)                        & $\mathcal{Q}$-Attack      & LOU  & 20      \\
        Football(noise)                        & $\mathcal{D}_{m}$         & WT   & 100     \\
        Polblogs(noise)                        & $\mathcal{Q}$-Attack      & LOU  & 100     \\
        \hline\hline
        \end{tabular}
        }
    \end{table}
    
    % For two large-scale networks, we don't deploy the attack since that the noise has been introduced during sub-network extraction.
    For two large-scale networks, we generate networks with missing data as follows:
    \begin{enumerate}
        \item Select a certain number ($x$) of vertices as community seeds via weighted random sampling, in which the sampling weights are associate with vertex degree;
        \item Extract $h$-hop ego-networks of these seed vertices, and remove those vertices that have multiple community labels;
        \item Aggregate these ego-networks to form a connected subgraphs.
    \end{enumerate}
    Details of parameter setup are reported in TABLE~\ref{app:tb:se-detail}.
    \begin{table}
        % \tiny
        \renewcommand\arraystretch{1.2}
        \centering
        \caption{Details of parameter setup in subgraph extraction.}
        \label{app:tb:se-detail}
        \resizebox{\linewidth}{!}{%
        \begin{tabular}{lccc} 
        \hline\hline
        \multirow{2}{*}{Subgraph of real network} & \multirow{2}{*}{missing data} & \multicolumn{2}{c}{Parameter}  \\ 
        \cline{3-4}
                            &       & $x$  & $h$       \\ 
        \hline
        Amazon-sub       & $\surd$        & 15   & 4         \\
        DBLP-sub         & $\surd$        & 10   & 3        \\
        \hline\hline
        \end{tabular}
        }
    \end{table}
    \begin{figure*}[!t]
        \centering
        \includegraphics[width=\textwidth]{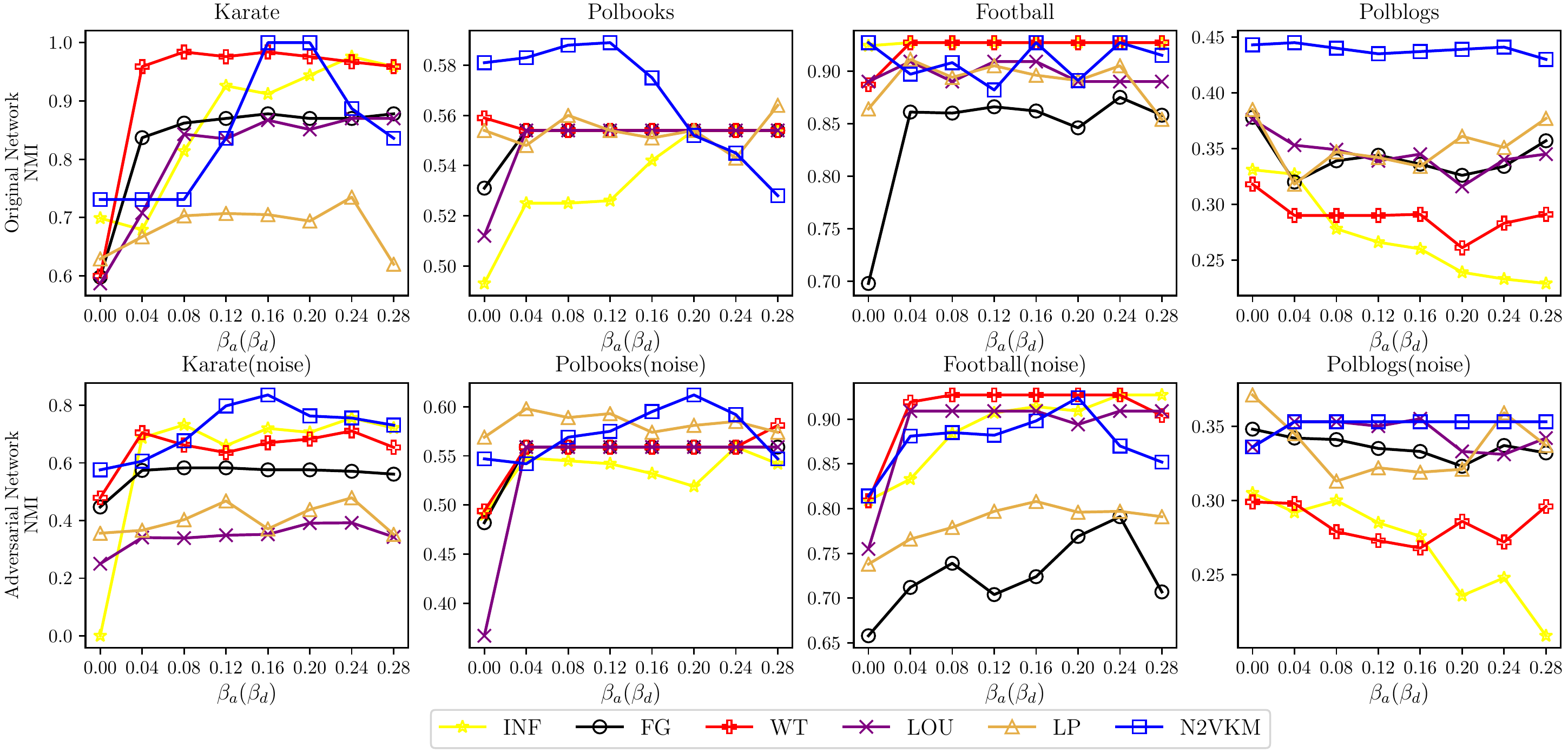}
        \caption{The impact of modification budgets $\beta_\textit{a}$ and $\beta_\textit{d}$ on the performance of \emph{RobustECD-GA}.}
        \label{fig:app:ga-para}
        % \vspace{-10pt}
    \end{figure*}
    \begin{figure*}[!t]
        \centering
        \includegraphics[width=\textwidth]{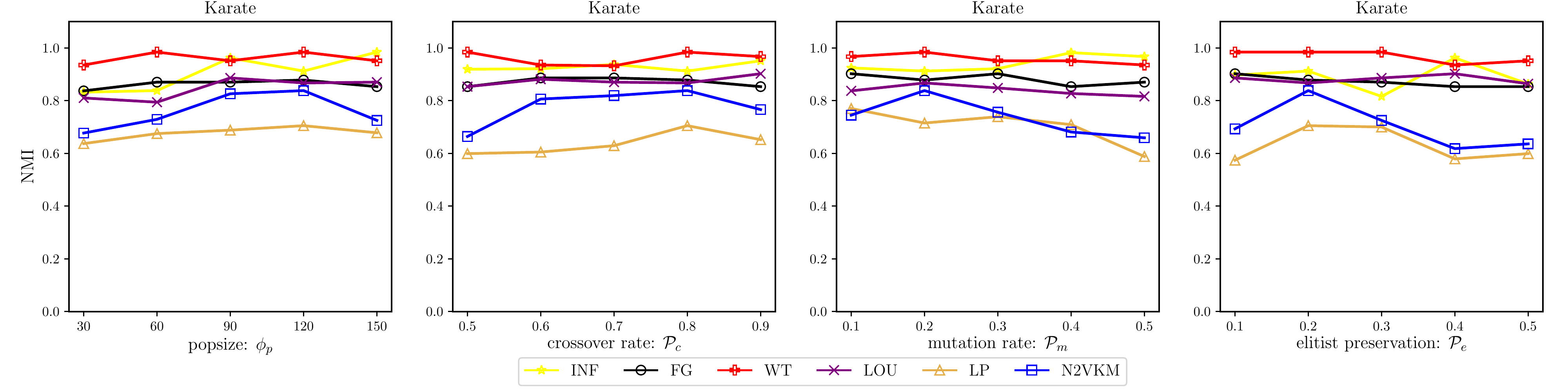}
        \caption{The impact of GA paratemeters on the performance of \emph{RobustECD-GA}.}
        \label{fig:app:ga-para-combin}
        % \vspace{-10pt}
    \end{figure*}

    \subsection{Parameter sensitivity of RobustECD-GA}\label{app:3}
    In this subsection, we discuss the impact of key parameters on the performance of \emph{RobustECD-GA}.
    Fig.~\ref{fig:app:ga-para} shows the parameter sensitivity of \emph{RobustECD-GA}, from which one can see that \emph{RobustECD-GA} performs well in small-scale networks like Karate, Polbooks and Football.
    In general, \emph{RobustECD-GA} is not strictly sensitive to different parameter settings in most cases, since that the sensitivity curves are smooth over the interval $[0.04, 0.28]$.
    Moreover, with the increase of network scale, like Polblogs, \emph{RobustECD-GA} performs poorly.
    As a reasonable explanation, \emph{RobustECD-GA} is based on evolutionary computation, which is capable of finding the optimal modification scheme in a small range of solution space.
    And with the increase of budget, the solution space becomes larger and \emph{RobustECD-GA} is prone to fall into local optimum, leading to a poor performance.

    In addition, we also conduct experiments to show the sensitivity with respect to the parameters in genetic algorithm, as shown in Fig.~\ref{fig:app:ga-para-combin}.
    As we can see, \emph{RobustECD-GA} is roughly not sensitive to different parameter settings in most cases. 
    Note that the curves of LP and N2VKM have relatively obvious fluctuation, since that both of the two community detection algorithms have greater randomness.

    \subsection{Vertex similarity details}\label{sec: detail-si}
    In this subsection, we briefly summarize the definition of similarity indices used in this paper.
    \begin{itemize}
        \item Common Neighbors (CN). It is defined as: 
        \begin{equation}
            \mathcal{H}_{ij}=|\Gamma_i \cap \Gamma_j|,
        \end{equation}
        where $\Gamma_i$ denotes the set of 1-hop neighbors of vertex $v_i$ and $|\cdot|$ is the cardinality of the set. It's easy to calculate CN by the adjacency matrix $\mathcal{A}$, i.e., $\mathcal{H}_{ij}=(\mathcal{A}^2)_{ij}$.
        \item Salton Index. It is defined as:
            \begin{equation}
                \mathcal{H}_{ij}=\frac{|\Gamma_i \cap \Gamma_j|}{\sqrt{k_i \times k_j}},
            \end{equation}
            where $k_i$ denotes the degree of vertex $v_i$.
        \item Jaccard Index. It is defined as:
            \begin{equation}
                \mathcal{H}_{ij}=\frac{|\Gamma_i \cap \Gamma_j|}{|\Gamma_i \cup \Gamma_j|}.
            \end{equation}
        \item Hub Promoted Index (HPI). It is defined as:
            \begin{equation}
                \mathcal{H}_{ij}=\frac{|\Gamma_i \cap \Gamma_j|}{\min \{k_i, k_j\}}.
            \end{equation}
            Under this measure, the links adjacent to hubs are likely to be assigned higher scores since the denominator is determined by the lower degree.
        \item Adamic–Adar Index (AA). It is defined as:
            \begin{equation}
                \mathcal{H}_{ij}=\sum_{z \in \Gamma_i \cap \Gamma_j} \frac{1}{\log k_z},
            \end{equation}
            where $z$ is the common neighbor of $v_i$ and $v_j$. The main assumption of this index is that the common neighbors of smaller degrees contribute more to the similarity.
        \item Resource Allocation Index (RA). It is defined as:
            \begin{equation}
                \mathcal{H}_{ij}=\sum_{z \in \Gamma_i \cap \Gamma_j} \frac{1}{k(z)}.
            \end{equation}
            RA index is similar to AA, but punish more heavily on their common neighbors of high-degree.
        \item Local Path Index (LP). It is defined as:
            \begin{equation}
                \mathcal{H}_{ij}=(\mathcal{A}^{2})_{ij}+\alpha (\mathcal{A}^{3})_{ij},
            \end{equation}
            where $\alpha$ is a free parameter. LP considers the contribution of third-order neighbors on the basis of Common neighbors (CN), and degenerates to CN when $\alpha=0$.
        \item RandomWalk with Restart (RWR). 
            Consider a random walker starting from vertex $v_i$, who will iteratively moves to a random neighbor with probability $c$ and return to vertex $v_i$ with probability $1-c$. Denote by $q_{ij}$ the probability this random walker locates at vertex $v_j$ in the steady state, we have
            \begin{equation}
                \boldsymbol{q}_{i}(t+1)=c \boldsymbol{P}^{\mathrm{T}} \boldsymbol{q}_{i}(t)+(1-c) \boldsymbol{e}_{i},
            \end{equation}
            where $\boldsymbol{P}$ is the transition matrix with $\boldsymbol{P}_{ij}=1/k_i$ if $v_i$ and $v_j$ are connected, and $\boldsymbol{P}_{ij}=0$ otherwise.
            The solution is straightforward, as
            \begin{equation}
                \boldsymbol{q}_{i}=(1-c)\left(I-c \boldsymbol{P}^{\mathrm{T}}\right)^{-1} \boldsymbol{e}_{i},
            \end{equation}
            The RWR index is thus defined as
            \begin{equation}
                \mathcal{H}_{ij}^{\mathrm{RWR}}=q_{ij}+q_{ji},
            \end{equation}
            where $q_{ij}$ is the $j$th element of the vector $\boldsymbol{q}_{i}$.
        
    \end{itemize}

    \subsection{Impact of single similarity index in RobustECD-SE}\label{app:4}
\begin{figure*}[!t]
        \centering
        \includegraphics[width=\textwidth]{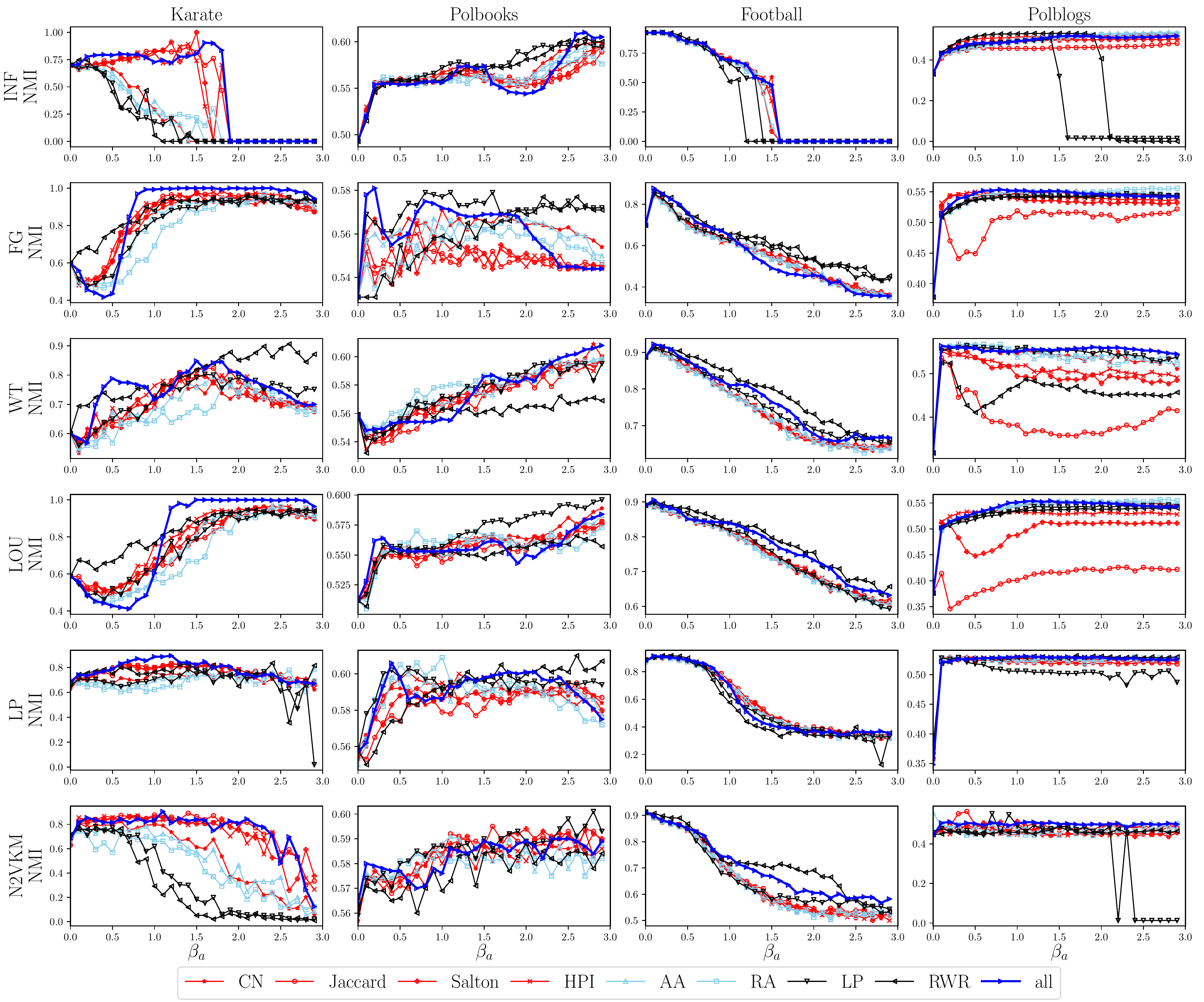}
        \caption{The impact of single similarity in \emph{RobustECD-SE} for real networks.}
        \label{fig:app:ori-lpm}
        % \vspace{-10pt}
    \end{figure*}

\begin{figure*}[!t]
        \centering
        \includegraphics[width=\textwidth]{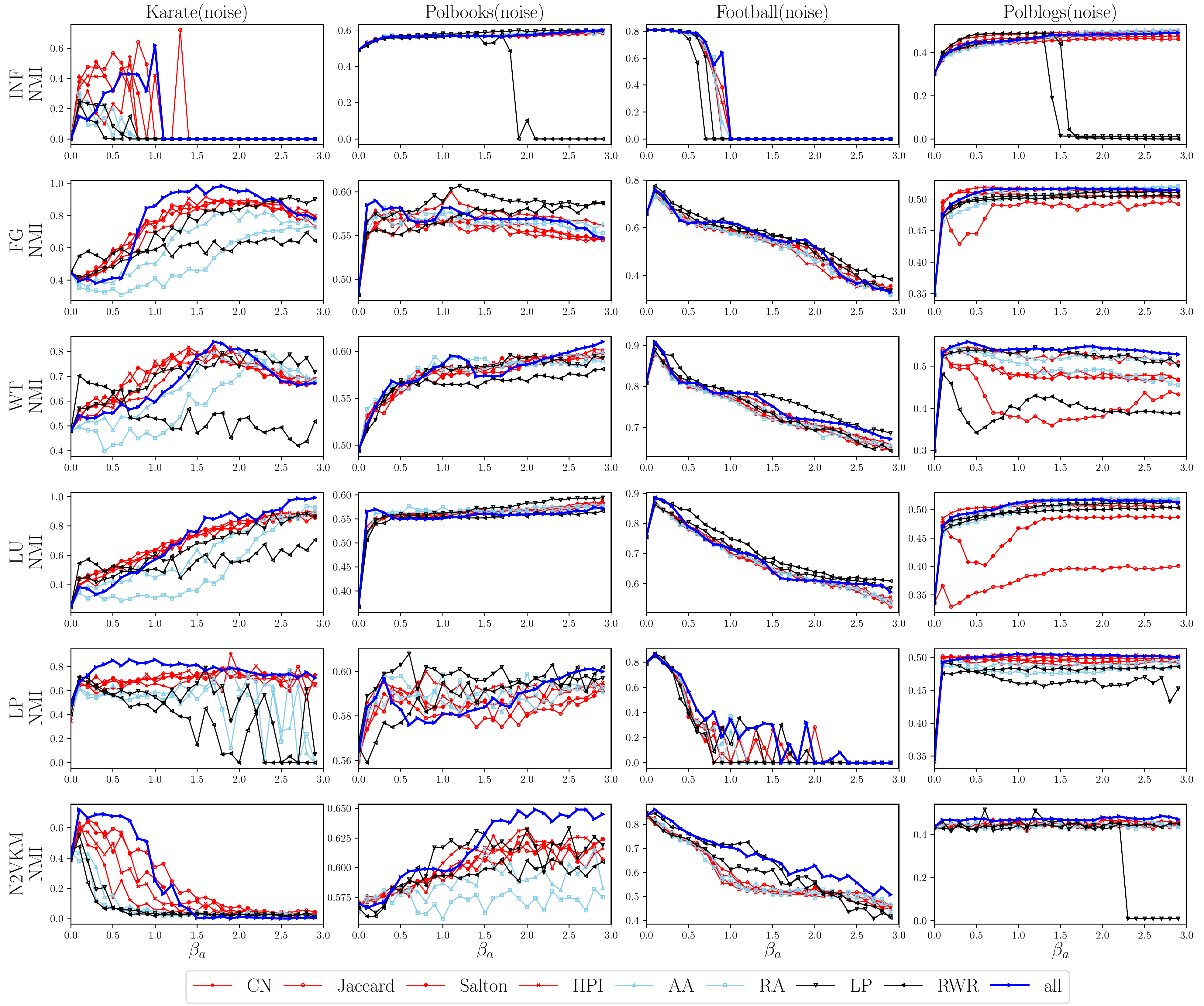}
        \caption{The impact of single similarity in \emph{RobustECD-SE} for adversarial networks.}
        \label{fig:app:adv-lpm}
        % \vspace{-10pt}
    \end{figure*}

    Fig.~\ref{fig:app:ori-lpm} and Fig.~\ref{fig:app:adv-lpm} show the results of all similarity indices (\emph{RobustECD-SE(all)}) and single index (\emph{RobustECD-SE(single)}) in real networks and adversarial networks, respectively.
    In Karate and Karate(noise) ($\beta_\textit{a} \in (1.0, 2.0)$), \emph{RobustECD-SE(single)}s with first-order similarity have relatively good performance while those with second-order and high-order similarity have relatively poor performance. Since that the scale of Karate is particularly small and first-order similarity are sufficient to capture structure features, so first-order similarity indices are complementary, second-order and high-order similarity are redundant or even negative in some cases.
    For Polbooks and Polbooks(noise), \emph{RobustECD-SE(single)} with LP similarity receives the best performance in most cases. The specific definition of Local Path (LP) is $\mathcal{H}=\mathcal{A}^2+\alpha\mathcal{A}^3$, where $\alpha$ is adjustable parameter. LP considers the contribution of third-order neighbors on the basis of Common neighbors (CN). So other similarity indices are redundant or even negative.
    For Football and Football(noise) ($\beta_\textit{a} \in (0.0, 0.5)$), all \emph{RobustECD-SE(single)}s have similarity results and achieve competitive performance against \emph{RobustECD-SE(all)}. Since that Football network has a strong community structures, and single similarity index is sufficient to capture structure features.
    For polblogs and Polblogs(noise), \emph{RobustECD-SE(single)}s achieve competitive performance against \emph{RobustECD-SE(all)} except for those with Jaccard, Salton and high-order similarity, which turn out to be redundant.

    \subsection{Impact of combination of similarity indices in RobustECD-SE}\label{app:si-comb}
    Combined with the conclusions in Appendix~\ref{app:4}, we further explore how to design proper combinations of similarity indices.
    Table~\ref{app:tb:se-si-combin-ori} and Table~\ref{app:tb:se-si-combin-adv} report the results of several special combinations of similarity indices, from which we come to the following conclusions:
    \begin{itemize}
        \item \emph{RobustECD-SE(combination:4)} obtains competitive results compared to \emph{RobustECD-SE(all)}, i.e., after pruning half of the indices according to the category of similarity, \emph{RobustECD-SE} still performs robustly.
        \item For very small networks like Karate, the combinations of first-order similarity indices (i.e., \emph{RobustECD-SE(combination:1)}) are sufficient to achieve excellent performance while the combinations of higher-order similarity indices is counterproductive.
        \item For those networks with strong community structures like Football, different combinations obtain relatively consistent performances, and even a single similarity index is sufficient.
              
    \end{itemize}

\begin{table*}
    \renewcommand\arraystretch{1.2}
    \large
    \centering
    \caption{Community detection results under different combinations of similarity indices in the real networks.}
    \label{app:tb:se-si-combin-ori}
    \resizebox{\textwidth}{!}{%
    % \caption{Community detection results under different combinations of similarity indices in the original networks.}
    \begin{tabular}{l|l|cccccccc|cccccc|c} 
        \hline\hline
        \multicolumn{1}{c|}{\multirow{3}{*}{Dataset}} & \multicolumn{1}{c|}{\multirow{3}{*}{Method}} & \multicolumn{8}{c|}{Similarity indices}                                                                                                                                                                             & \multicolumn{7}{c}{Community Detection}                                                                                                                                                 \\ 
        \cline{3-17}
        \multicolumn{1}{c|}{}     & \multicolumn{1}{c|}{}    & \multicolumn{4}{c|}{1}     & \multicolumn{2}{c|}{2}      & \multicolumn{2}{c|}{3}        & \multicolumn{7}{c}{NMI}              \\ 
        \cline{3-17}
        \multicolumn{1}{c|}{}                         & \multicolumn{1}{c|}{}                        & \multicolumn{1}{c}{CN} & \multicolumn{1}{c}{Jaccard} & \multicolumn{1}{c}{Salton} & \multicolumn{1}{c|}{HPI} & \multicolumn{1}{c}{AA} & \multicolumn{1}{c|}{RA} & \multicolumn{1}{c}{LP} & \multicolumn{1}{c|}{RWR} & \multicolumn{1}{c}{INF} & \multicolumn{1}{c}{FG} & \multicolumn{1}{c}{WT} & \multicolumn{1}{c}{LOU} & \multicolumn{1}{c}{LP} & \multicolumn{1}{c|}{N2VKM} & \multicolumn{1}{c}{Avg RIMP}  \\ 
        \hline
        \multirow{7}{*}{Karate}             & Original & \multicolumn{8}{l|}{}                                                                                      &         0.699\footnotesize{$\pm$0.000}  &         0.598\footnotesize{$\pm$0.000}  &         0.600\footnotesize{$\pm$0.000}  &         0.587\footnotesize{$\pm$0.000}  &         0.689\footnotesize{$\pm$0.283}  &         0.705\footnotesize{$\pm$0.175}  & ----  \\
                                            & \emph{RobustECD-SE(all)        }     & $\surd$ & $\surd$ & $\surd$ & $\surd$ & $\surd$ & $\surd$ & $\surd$ & $\surd$  & \textbf{0.825\footnotesize{$\pm$0.220}} & \textbf{1.000\footnotesize{$\pm$0.000}} &         0.821\footnotesize{$\pm$0.088}  & \textbf{1.000\footnotesize{$\pm$0.000}} & \textbf{0.847\footnotesize{$\pm$0.136}} & \textbf{0.834\footnotesize{$\pm$0.114}} & \textbf{38.95\%} \\
                                            & \emph{RobustECD-SE(combination:1)}   & $\surd$ & $\surd$ & $\surd$ & $\surd$ &         &         &         &          & \textbf{0.874\footnotesize{$\pm$0.241}} & \textbf{0.997\footnotesize{$\pm$0.023}} & \textbf{0.825\footnotesize{$\pm$0.099}} &         0.933\footnotesize{$\pm$0.138}  & \textbf{0.839\footnotesize{$\pm$0.144}} & \textbf{0.855\footnotesize{$\pm$0.109}} & \textbf{38.54\%} \\
                                            & \emph{RobustECD-SE(combination:2)}   &         &         &         &         & $\surd$ & $\surd$ &         &          &         0.379\footnotesize{$\pm$0.218}  &         0.974\footnotesize{$\pm$0.060}  &         0.767\footnotesize{$\pm$0.120}  &         0.913\footnotesize{$\pm$0.152}  &         0.771\footnotesize{$\pm$0.131}  &         0.571\footnotesize{$\pm$0.263}  & 15.56\%   \\
                                            & \emph{RobustECD-SE(combination:3)}   &         &         &         &         &         &         & $\surd$ & $\surd$  &         0.341\footnotesize{$\pm$0.191}  &         0.942\footnotesize{$\pm$0.099}  &         0.784\footnotesize{$\pm$0.144}  &         0.899\footnotesize{$\pm$0.119}  &         0.729\footnotesize{$\pm$0.125}  &         0.112\footnotesize{$\pm$0.190}  & 1.97\%    \\
                                            & \emph{RobustECD-SE(combination:4)}   &         &         & $\surd$ & $\surd$ &         & $\surd$ &         & $\surd$  &         0.730\footnotesize{$\pm$0.227}  & \textbf{0.997\footnotesize{$\pm$0.023}} & \textbf{0.855\footnotesize{$\pm$0.119}} & \textbf{0.975\footnotesize{$\pm$0.080}} &         0.809\footnotesize{$\pm$0.140}  &         0.829\footnotesize{$\pm$0.111}  & 35.79\%  \\
                                            & \emph{RobustECD-SE(single)     }     &         &         & $\surd$ &         &         &         &         &          & \textbf{0.825\footnotesize{$\pm$0.255}} &         0.967\footnotesize{$\pm$0.065}  &         0.797\footnotesize{$\pm$0.120}  &         0.821\footnotesize{$\pm$0.152}  &         0.799\footnotesize{$\pm$0.136}  &         0.833\footnotesize{$\pm$0.119}  & 31.09\%  \\ 
        \hline
        \multirow{7}{*}{Polbooks}           & Original & \multicolumn{8}{l|}{}                                                                                      &         0.493\footnotesize{$\pm$0.000}  &         0.531\footnotesize{$\pm$0.000}  &         0.559\footnotesize{$\pm$0.000}  &         0.512\footnotesize{$\pm$0.000}  &         0.554\footnotesize{$\pm$0.025}  &         0.556\footnotesize{$\pm$0.017}  & ----   \\
                                            & \emph{RobustECD-SE(all)        }     & $\surd$ & $\surd$ & $\surd$ & $\surd$ & $\surd$ & $\surd$ & $\surd$ & $\surd$  & \textbf{0.574\footnotesize{$\pm$0.014}} & \textbf{0.569\footnotesize{$\pm$0.001}} & \textbf{0.586\footnotesize{$\pm$0.017}} &         0.560\footnotesize{$\pm$0.011}  & \textbf{0.598\footnotesize{$\pm$0.009}} & \textbf{0.589\footnotesize{$\pm$0.009}} & \textbf{8.61\%}\\
                                            & \emph{RobustECD-SE(combination:1)}   & $\surd$ & $\surd$ & $\surd$ & $\surd$ &         &         &         &          &         0.561\footnotesize{$\pm$0.014}  &         0.558\footnotesize{$\pm$0.014}  &         0.581\footnotesize{$\pm$0.016}  &         0.556\footnotesize{$\pm$0.009}  &         0.592\footnotesize{$\pm$0.014}  &         0.588\footnotesize{$\pm$0.007}  & 7.34\%   \\
                                            & \emph{RobustECD-SE(combination:2)}   &         &         &         &         & $\surd$ & $\surd$ &         &          &         0.562\footnotesize{$\pm$0.018}  &         0.568\footnotesize{$\pm$0.007}  &         0.584\footnotesize{$\pm$0.021}  &         0.561\footnotesize{$\pm$0.018}  &         0.594\footnotesize{$\pm$0.013}  &         0.582\footnotesize{$\pm$0.021}  & 7.82\%  \\
                                            & \emph{RobustECD-SE(combination:3)}   &         &         &         &         &         &         & $\surd$ & $\surd$  & \textbf{0.577\footnotesize{$\pm$0.016}} & \textbf{0.576\footnotesize{$\pm$0.014}} &         0.579\footnotesize{$\pm$0.015}  & \textbf{0.571\footnotesize{$\pm$0.013}} & \textbf{0.604\footnotesize{$\pm$0.017}} &         0.589\footnotesize{$\pm$0.024}  & \textbf{9.26\%}  \\
                                            & \emph{RobustECD-SE(combination:4)}   &         &         & $\surd$ & $\surd$ &         & $\surd$ &         & $\surd$  &         0.572\footnotesize{$\pm$0.011}  &         0.566\footnotesize{$\pm$0.010}  & \textbf{0.588\footnotesize{$\pm$0.022}} & \textbf{0.563\footnotesize{$\pm$0.012}} &         0.591\footnotesize{$\pm$0.015}  & \textbf{0.591\footnotesize{$\pm$0.014}} & 8.46\%  \\
                                            & \emph{RobustECD-SE(single)     }     &         &         & $\surd$ &         &         &         &         &          &         0.562\footnotesize{$\pm$0.016}  &         0.551\footnotesize{$\pm$0.025}  &         0.575\footnotesize{$\pm$0.018}  &         0.560\footnotesize{$\pm$0.019}  &         0.586\footnotesize{$\pm$0.020}  &         0.589\footnotesize{$\pm$0.013}  & 6.95\%  \\ 
        \hline
        \multirow{7}{*}{Football}           & Original  & \multicolumn{8}{l|}{}                                                                                     &         0.924\footnotesize{$\pm$0.000}  &         0.698\footnotesize{$\pm$0.000}  &         0.887\footnotesize{$\pm$0.000}  &         0.890\footnotesize{$\pm$0.000}  &         0.888\footnotesize{$\pm$0.037}  &         0.912\footnotesize{$\pm$0.012}  & ----  \\
                                            & \emph{RobustECD-SE(all)        }     & $\surd$ & $\surd$ & $\surd$ & $\surd$ & $\surd$ & $\surd$ & $\surd$ & $\surd$  & \textbf{0.924\footnotesize{$\pm$0.000}} & \textbf{0.877\footnotesize{$\pm$0.021}} & \textbf{0.923\footnotesize{$\pm$0.009}} & \textbf{0.906\footnotesize{$\pm$0.014}} & \textbf{0.915\footnotesize{$\pm$0.018}} &         0.898\footnotesize{$\pm$0.021}  & \textbf{5.50\%}  \\
                                            & \emph{RobustECD-SE(combination:1)}   & $\surd$ & $\surd$ & $\surd$ & $\surd$ &         &         &         &          & \textbf{0.925\footnotesize{$\pm$0.001}} & \textbf{0.869\footnotesize{$\pm$0.023}} & \textbf{0.920\footnotesize{$\pm$0.009}} &         0.902\footnotesize{$\pm$0.015}  &         0.908\footnotesize{$\pm$0.019}  &         0.894\footnotesize{$\pm$0.017}  & 4.99\%  \\
                                            & \emph{RobustECD-SE(combination:2)}   &         &         &         &         & $\surd$ & $\surd$ &         &          &         0.924\footnotesize{$\pm$0.001}  &         0.860\footnotesize{$\pm$0.026}  &         0.918\footnotesize{$\pm$0.012}  &         0.898\footnotesize{$\pm$0.019}  &         0.913\footnotesize{$\pm$0.016}  &         0.890\footnotesize{$\pm$0.017}  & 4.67\%  \\
                                            & \emph{RobustECD-SE(combination:3)}   &         &         &         &         &         &         & $\surd$ & $\surd$  &         0.924\footnotesize{$\pm$0.001}  &         0.857\footnotesize{$\pm$0.028}  &         0.916\footnotesize{$\pm$0.014}  & \textbf{0.908\footnotesize{$\pm$0.019}} &         0.910\footnotesize{$\pm$0.015}  & \textbf{0.899\footnotesize{$\pm$0.017}} & 4.85\%  \\
                                            & \emph{RobustECD-SE(combination:4)}   &         &         & $\surd$ & $\surd$ &         & $\surd$ &         & $\surd$  &         0.924\footnotesize{$\pm$0.001}  &         0.867\footnotesize{$\pm$0.021}  &         0.918\footnotesize{$\pm$0.009}  &         0.903\footnotesize{$\pm$0.017}  & \textbf{0.914\footnotesize{$\pm$0.019}} &         0.892\footnotesize{$\pm$0.021}  & 4.98\%  \\
                                            & \emph{RobustECD-SE(single)     }     &         &         &         &         &         &         &         & $\surd$  &         0.924\footnotesize{$\pm$0.001}  &         0.864\footnotesize{$\pm$0.030}  &         0.912\footnotesize{$\pm$0.015}  &         0.903\footnotesize{$\pm$0.019}  &         0.911\footnotesize{$\pm$0.014}  & \textbf{0.908\footnotesize{$\pm$0.017}} & \textbf{5.04\%} \\ 
        \hline
        \multirow{7}{*}{Polblogs}           & Original  & \multicolumn{8}{l|}{}                                                                                     &         0.330\footnotesize{$\pm$0.001}  &         0.378\footnotesize{$\pm$0.000}  &         0.318\footnotesize{$\pm$0.000}  &         0.376\footnotesize{$\pm$0.000}  &         0.375\footnotesize{$\pm$0.053}  &         0.458\footnotesize{$\pm$0.067}  & ----  \\
                                            & \emph{RobustECD-SE(all)        }     & $\surd$ & $\surd$ & $\surd$ & $\surd$ & $\surd$ & $\surd$ & $\surd$ & $\surd$  &         0.517\footnotesize{$\pm$0.007}  & \textbf{0.551\footnotesize{$\pm$0.006}} &         0.556\footnotesize{$\pm$0.009}  & \textbf{0.551\footnotesize{$\pm$0.005}} &         0.529\footnotesize{$\pm$0.007}  & \textbf{0.499\footnotesize{$\pm$0.006}} & \textbf{45.64\% }\\
                                            & \emph{RobustECD-SE(combination:1)}   & $\surd$ & $\surd$ & $\surd$ & $\surd$ &         &         &         &          &         0.501\footnotesize{$\pm$0.006}  &         0.544\footnotesize{$\pm$0.005}  &         0.540\footnotesize{$\pm$0.007}  &         0.518\footnotesize{$\pm$0.020}  &         0.522\footnotesize{$\pm$0.004}  &         0.496\footnotesize{$\pm$0.003}  & 41.80\%  \\
                                            & \emph{RobustECD-SE(combination:2)}   &         &         &         &         & $\surd$ & $\surd$ &         &          &         0.514\footnotesize{$\pm$0.011}  & \textbf{0.555\footnotesize{$\pm$0.009}} & \textbf{0.559\footnotesize{$\pm$0.009}} & \textbf{0.555\footnotesize{$\pm$0.008}} & \textbf{0.531\footnotesize{$\pm$0.007}} &         0.486\footnotesize{$\pm$0.012}  & 45.61\%  \\
                                            & \emph{RobustECD-SE(combination:3)}   &         &         &         &         &         &         & $\surd$ & $\surd$  & \textbf{0.525\footnotesize{$\pm$0.006}} &         0.548\footnotesize{$\pm$0.007}  & \textbf{0.560\footnotesize{$\pm$0.011}} &         0.549\footnotesize{$\pm$0.006}  & \textbf{0.532\footnotesize{$\pm$0.007}} & \textbf{0.498\footnotesize{$\pm$0.007}} & \textbf{46.13\% }\\
                                            & \emph{RobustECD-SE(combination:4)}   &         &         & $\surd$ & $\surd$ &         & $\surd$ &         & $\surd$  & \textbf{0.523\footnotesize{$\pm$0.008}} &         0.548\footnotesize{$\pm$0.009}  &         0.534\footnotesize{$\pm$0.012}  &         0.546\footnotesize{$\pm$0.008}  &         0.526\footnotesize{$\pm$0.007}  &         0.453\footnotesize{$\pm$0.005}  & 42.63\%  \\
                                            & \emph{RobustECD-SE(single)     }     &         &         &         &         &         & $\surd$ &         &          &         0.521\footnotesize{$\pm$0.010}  &         0.547\footnotesize{$\pm$0.009}  &         0.537\footnotesize{$\pm$0.038}  &         0.549\footnotesize{$\pm$0.008}  &         0.521\footnotesize{$\pm$0.013}  &         0.457\footnotesize{$\pm$0.007}  & 42.70\%  \\
        \hline\hline
    \end{tabular}}
\end{table*}

\begin{table*}
    \renewcommand\arraystretch{1.2}
    \large
    \centering
    \caption{Community detection results under different combinations of similarity indices in the adversarial networks.}
    \label{app:tb:se-si-combin-adv}
    \resizebox{\textwidth}{!}{%
    \begin{tabular}{l|l|cccccccc|cccccc|c} 
    \hline\hline
    \multirow{3}{*}{Dataset}  & \multicolumn{1}{c|}{\multirow{3}{*}{Method}}              & \multicolumn{8}{c|}{Similarity indices}               & \multicolumn{7}{c}{Community Detection}                                                                                                                   \\ 
    \cline{3-17}
                              &                                     & \multicolumn{4}{c|}{1}             & \multicolumn{2}{c|}{2} & \multicolumn{2}{c|}{3}  & \multicolumn{7}{c}{NMI}                                                                                                                                     \\ 
    \cline{3-17}
                              &                                     & CN & Jaccard & Salton & HPI & AA & RA & LP & RWR           & INF & FG & WT & LOU                  & LP                   & N2VKM                & Avg RIMP          \\ 
    \hline
    \multirow{8}{*}{\begin{tabular}[c]{@{}l@{}}Karate\\(noise)\end{tabular}}                 
                                            & Original  & \multicolumn{8}{l|}{\multirow{2}{*}{}}                                                                    &         0.699\footnotesize{$\pm$0.000}  &         0.598\footnotesize{$\pm$0.000}  &         0.600\footnotesize{$\pm$0.000}  &         0.587\footnotesize{$\pm$0.000}  &         0.689\footnotesize{$\pm$0.283}  &         0.705\footnotesize{$\pm$0.175}  & 63.90\%           \\
                                            & Attack    & \multicolumn{8}{l|}{}                                                                                     &         0.000\footnotesize{$\pm$0.000}  &         0.447\footnotesize{$\pm$0.000}  &         0.487\footnotesize{$\pm$0.000}  &         0.250\footnotesize{$\pm$0.000}  &         0.475\footnotesize{$\pm$0.337}  &         0.399\footnotesize{$\pm$0.231}  & ----              \\ 
    \cdashline{2-17}        
                                            & \emph{RobustECD-SE(all)        }     & $\surd$ & $\surd$ & $\surd$ & $\surd$ & $\surd$ & $\surd$ & $\surd$ & $\surd$  &         0.425\footnotesize{$\pm$0.336}  & \textbf{0.709\footnotesize{$\pm$0.218}} &         0.576\footnotesize{$\pm$0.075}  &         0.484\footnotesize{$\pm$0.173}  & \textbf{0.828\footnotesize{$\pm$0.197}} & \textbf{0.529\footnotesize{$\pm$0.340}} & 53.31\%           \\
                                            & \emph{RobustECD-SE(combination:1)}   & $\surd$ & $\surd$ & $\surd$ & $\surd$ &         &         &         &          &         0.594\footnotesize{$\pm$0.212}  & \textbf{0.844\footnotesize{$\pm$0.179}} & \textbf{0.702\footnotesize{$\pm$0.151}} & \textbf{0.598\footnotesize{$\pm$0.121}} & \textbf{0.828\footnotesize{$\pm$0.106}} & \textbf{0.744\footnotesize{$\pm$0.165}} & \textbf{82.06\%}  \\
                                            & \emph{RobustECD-SE(combination:2)}   &         &         &         &         & $\surd$ & $\surd$ &         &          &         0.052\footnotesize{$\pm$0.083}  &         0.343\footnotesize{$\pm$0.142}  &         0.542\footnotesize{$\pm$0.128}  &         0.324\footnotesize{$\pm$0.134}  &         0.573\footnotesize{$\pm$0.160}  &         0.063\footnotesize{$\pm$0.123}  & -6.79\%           \\
                                            & \emph{RobustECD-SE(combination:3)}   &         &         &         &         &         &         & $\surd$ & $\surd$  &         0.000\footnotesize{$\pm$0.000}  &         0.532\footnotesize{$\pm$0.173}  &         0.600\footnotesize{$\pm$0.140}  &         0.465\footnotesize{$\pm$0.144}  &         0.621\footnotesize{$\pm$0.137}  &         0.039\footnotesize{$\pm$0.112}  & 11.46\%           \\
                                            & \emph{RobustECD-SE(combination:4)}   &         &         & $\surd$ & $\surd$ &         & $\surd$ &         & $\surd$  & \textbf{0.608\footnotesize{$\pm$0.266}} &         0.617\footnotesize{$\pm$0.259}  &         0.594\footnotesize{$\pm$0.119}  &         0.411\footnotesize{$\pm$0.170}  &         0.781\footnotesize{$\pm$0.181}  &         0.264\footnotesize{$\pm$0.308}  & 35.96\%           \\
                                            & \emph{RobustECD-SE(single)     }     &         &         & $\surd$ &         &         &         &         &          & \textbf{0.628\footnotesize{$\pm$0.319}} &         0.657\footnotesize{$\pm$0.618}  & \textbf{0.707\footnotesize{$\pm$0.173}} & \textbf{0.579\footnotesize{$\pm$0.155}} &         0.689\footnotesize{$\pm$0.155}  &         0.445\footnotesize{$\pm$0.301}  & \textbf{57.19}\%           \\ 
    \hline
    \multirow{8}{*}{\begin{tabular}[c]{@{}l@{}}Polbooks\\(noise)\end{tabular}}               
                                            & Original  & \multicolumn{8}{l|}{\multirow{2}{*}{}}                                                                    &         0.493\footnotesize{$\pm$0.000}  &         0.531\footnotesize{$\pm$0.000}  &         0.559\footnotesize{$\pm$0.000}  &         0.512\footnotesize{$\pm$0.000}  &         0.554\footnotesize{$\pm$0.025}  &         0.556\footnotesize{$\pm$0.017}  & 26.69\%           \\
                                            & Attack    & \multicolumn{8}{l|}{}                                                                                     &         0.418\footnotesize{$\pm$0.004}  &         0.482\footnotesize{$\pm$0.000}  &         0.393\footnotesize{$\pm$0.000}  &         0.343\footnotesize{$\pm$0.000}  &         0.461\footnotesize{$\pm$0.030}  &         0.462\footnotesize{$\pm$0.012}  & ----              \\ 
    \cdashline{2-17}
                                            & \emph{RobustECD-SE(all)        }     & $\surd$ & $\surd$ & $\surd$ & $\surd$ & $\surd$ & $\surd$ & $\surd$ & $\surd$  & \textbf{0.590\footnotesize{$\pm$0.014}} & \textbf{0.565\footnotesize{$\pm$0.012}} & \textbf{0.599\footnotesize{$\pm$0.008}} &         0.564\footnotesize{$\pm$0.010}  & \textbf{0.599\footnotesize{$\pm$0.008}} & \textbf{0.643\footnotesize{$\pm$0.026}} & \textbf{40.72\%}  \\
                                            & \emph{RobustECD-SE(combination:1)}   & $\surd$ & $\surd$ & $\surd$ & $\surd$ &         &         &         &          &         0.587\footnotesize{$\pm$0.016}  &         0.545\footnotesize{$\pm$0.005}  & \textbf{0.599\footnotesize{$\pm$0.012}} &         0.562\footnotesize{$\pm$0.014}  &         0.593\footnotesize{$\pm$0.014}  &         0.635\footnotesize{$\pm$0.027}  & 39.31\%           \\
                                            & \emph{RobustECD-SE(combination:2)}   &         &         &         &         & $\surd$ & $\surd$ &         &          & \textbf{0.590\footnotesize{$\pm$0.016}} &         0.560\footnotesize{$\pm$0.015}  &         0.593\footnotesize{$\pm$0.008}  &         0.576\footnotesize{$\pm$0.016}  &         0.593\footnotesize{$\pm$0.016}  &         0.595\footnotesize{$\pm$0.062}  & 38.93\%           \\
                                            & \emph{RobustECD-SE(combination:3)}   &         &         &         &         &         &         & $\surd$ & $\surd$  & \textbf{0.594\footnotesize{$\pm$0.013}} & \textbf{0.586\footnotesize{$\pm$0.018}} &         0.590\footnotesize{$\pm$0.018}  & \textbf{0.578\footnotesize{$\pm$0.018}} & \textbf{0.599\footnotesize{$\pm$0.010}} &         0.632\footnotesize{$\pm$0.037}  & \textbf{41.51\%}  \\
                                            & \emph{RobustECD-SE(combination:4)}   &         &         & $\surd$ & $\surd$ &         & $\surd$ &         & $\surd$  &         0.587\footnotesize{$\pm$0.015}  &         0.554\footnotesize{$\pm$0.013}  &         0.591\footnotesize{$\pm$0.008}  &         0.565\footnotesize{$\pm$0.016}  &         0.594\footnotesize{$\pm$0.015}  & \textbf{0.638\footnotesize{$\pm$0.034}} & 39.57\%           \\
                                            & \emph{RobustECD-SE(single)     }     &         &         & $\surd$ &         &         &         &         &          &         0.580\footnotesize{$\pm$0.022}  &         0.548\footnotesize{$\pm$0.009}  &         0.592\footnotesize{$\pm$0.016}  & \textbf{0.578\footnotesize{$\pm$0.022}} &         0.586\footnotesize{$\pm$0.019}  &         0.615\footnotesize{$\pm$0.035}  & 38.64\%           \\ 
    \hline
    \multirow{8}{*}{\begin{tabular}[c]{@{}l@{}}Football\\(noise)\end{tabular}}               
                                            & Original  & \multicolumn{8}{l|}{\multirow{2}{*}{}}                                                           &         0.924\footnotesize{$\pm$0.000}  &         0.698\footnotesize{$\pm$0.000}  &         0.887\footnotesize{$\pm$0.000}  &         0.890\footnotesize{$\pm$0.000}  &         0.888\footnotesize{$\pm$0.037}  &         0.912\footnotesize{$\pm$0.012}  & 11.27\%           \\
                                            & Attack    & \multicolumn{8}{l|}{}                                                                            &         0.809\footnotesize{$\pm$0.000}  &         0.658\footnotesize{$\pm$0.000}  &         0.809\footnotesize{$\pm$0.000}  &         0.755\footnotesize{$\pm$0.000}  &         0.800\footnotesize{$\pm$0.051}  &         0.838\footnotesize{$\pm$0.027}  & ----              \\ 
    \cdashline{2-17}
                                            & \emph{RobustECD-SE(all)        }     & $\surd$ & $\surd$ & $\surd$ & $\surd$ & $\surd$ & $\surd$ & $\surd$ & $\surd$  &         0.809\footnotesize{$\pm$0.000}  &         0.762\footnotesize{$\pm$0.024}  & \textbf{0.909\footnotesize{$\pm$0.020}} & \textbf{0.886\footnotesize{$\pm$0.014}} & \textbf{0.863\footnotesize{$\pm$0.051}} & \textbf{0.862\footnotesize{$\pm$0.021}} & \textbf{9.38\%}   \\
                                            & \emph{RobustECD-SE(combination:1)}   & $\surd$ & $\surd$ & $\surd$ & $\surd$ &         &         &         &          &         0.809\footnotesize{$\pm$0.001}  &         0.759\footnotesize{$\pm$0.039}  &         0.906\footnotesize{$\pm$0.021}  &         0.881\footnotesize{$\pm$0.018}  &         0.843\footnotesize{$\pm$0.049}  &         0.857\footnotesize{$\pm$0.023}  & 8.61\%            \\
                                            & \emph{RobustECD-SE(combination:2)}   &         &         &         &         & $\surd$ & $\surd$ &         &          &         0.809\footnotesize{$\pm$0.004}  &         0.750\footnotesize{$\pm$0.036}  &         0.898\footnotesize{$\pm$0.027}  &         0.873\footnotesize{$\pm$0.031}  &         0.841\footnotesize{$\pm$0.049}  &         0.839\footnotesize{$\pm$0.030}  & 7.64\%            \\
                                            & \emph{RobustECD-SE(combination:3)}   &         &         &         &         &         &         & $\surd$ & $\surd$  &         0.809\footnotesize{$\pm$0.000}  & \textbf{0.768\footnotesize{$\pm$0.035}} &         0.905\footnotesize{$\pm$0.021}  & \textbf{0.887\footnotesize{$\pm$0.019}} &         0.849\footnotesize{$\pm$0.045}  &         0.852\footnotesize{$\pm$0.023}  & 8.98\%            \\
                                            & \emph{RobustECD-SE(combination:4)}   &         &         & $\surd$ & $\surd$ &         & $\surd$ &         & $\surd$  &         0.809\footnotesize{$\pm$0.001}  & \textbf{0.768\footnotesize{$\pm$0.034}} & \textbf{0.911\footnotesize{$\pm$0.014}} &         0.883\footnotesize{$\pm$0.020}  &         0.857\footnotesize{$\pm$0.049}  & \textbf{0.860\footnotesize{$\pm$0.023}} & \textbf{9.34\%}   \\
                                            & \emph{RobustECD-SE(single)     }     &         &         &         &         &         &         &         & $\surd$  &         0.809\footnotesize{$\pm$0.000}  &         0.767\footnotesize{$\pm$0.048}  &         0.906\footnotesize{$\pm$0.021}  &         0.877\footnotesize{$\pm$0.027}  & \textbf{0.863\footnotesize{$\pm$0.041}} & \textbf{0.860\footnotesize{$\pm$0.031}} & 9.20\%            \\ 
    \hline
    \multirow{8}{*}{\begin{tabular}[c]{@{}l@{}}Polblogs\\(noise)\end{tabular}}               
                                            & Original & \multicolumn{8}{l|}{\multirow{2}{*}{}}                                                            &         0.330\footnotesize{$\pm$0.001}  &         0.378\footnotesize{$\pm$0.000}  &         0.318\footnotesize{$\pm$0.000}  &         0.376\footnotesize{$\pm$0.000}  &         0.375\footnotesize{$\pm$0.053}  &         0.458\footnotesize{$\pm$0.067}  & 8.60\%            \\
                                            & Attack   & \multicolumn{8}{l|}{}                                                                             &         0.303\footnotesize{$\pm$0.001}  &         0.348\footnotesize{$\pm$0.000}  &         0.299\footnotesize{$\pm$0.000}  &         0.336\footnotesize{$\pm$0.000}  &         0.340\footnotesize{$\pm$0.061}  &         0.434\footnotesize{$\pm$0.034}  & ----              \\ 
    \cdashline{2-17}
                                            & \emph{RobustECD-SE(all)        }     & $\surd$ & $\surd$ & $\surd$ & $\surd$ & $\surd$ & $\surd$ & $\surd$ & $\surd$  &         0.444\footnotesize{$\pm$0.007}  &         0.505\footnotesize{$\pm$0.007}  & \textbf{0.558\footnotesize{$\pm$0.005}} &         0.493\footnotesize{$\pm$0.004}  & \textbf{0.500\footnotesize{$\pm$0.005}} & \textbf{0.469\footnotesize{$\pm$0.023}} & \textbf{46.69\%}    \\
                                            & \emph{RobustECD-SE(combination:1)}   & $\surd$ & $\surd$ & $\surd$ & $\surd$ &         &         &         &          &         0.444\footnotesize{$\pm$0.010}  & \textbf{0.513\footnotesize{$\pm$0.007}} &         0.539\footnotesize{$\pm$0.009}  & \textbf{0.495\footnotesize{$\pm$0.007}} &         0.499\footnotesize{$\pm$0.005}  & \textbf{0.470\footnotesize{$\pm$0.003}} & 46.10\%    \\
                                            & \emph{RobustECD-SE(combination:2)}   &         &         &         &         & $\surd$ & $\surd$ &         &          &         0.426\footnotesize{$\pm$0.009}  &         0.488\footnotesize{$\pm$0.007}  &         0.541\footnotesize{$\pm$0.011}  &         0.479\footnotesize{$\pm$0.007}  &         0.492\footnotesize{$\pm$0.006}  &         0.439\footnotesize{$\pm$0.010}  & 41.70\%    \\
                                            & \emph{RobustECD-SE(combination:3)}   &         &         &         &         &         &         & $\surd$ & $\surd$  & \textbf{0.453\footnotesize{$\pm$0.009}} &         0.499\footnotesize{$\pm$0.007}  & \textbf{0.545\footnotesize{$\pm$0.009}} &         0.487\footnotesize{$\pm$0.007}  &         0.496\footnotesize{$\pm$0.006}  &         0.450\footnotesize{$\pm$0.009}  & 44.95\%    \\
                                            & \emph{RobustECD-SE(combination:4)}   &         &         & $\surd$ & $\surd$ &         & $\surd$ &         & $\surd$  & \textbf{0.458\footnotesize{$\pm$0.009}} & \textbf{0.511\footnotesize{$\pm$0.007}} &         0.536\footnotesize{$\pm$0.009}  & \textbf{0.497\footnotesize{$\pm$0.007}} & \textbf{0.501\footnotesize{$\pm$0.006}} &         0.453\footnotesize{$\pm$0.006}  & \textbf{46.15\%}    \\
                                            & \emph{RobustECD-SE(single)     }     &         &         &         &         &         & $\surd$ &         &          &         0.419\footnotesize{$\pm$0.009}  &         0.484\footnotesize{$\pm$0.007}  &         0.529\footnotesize{$\pm$0.032}  &         0.478\footnotesize{$\pm$0.008}  &         0.488\footnotesize{$\pm$0.010}  &         0.431\footnotesize{$\pm$0.009}  & 39.90\%    \\
    \hline\hline
    \end{tabular}}
    \end{table*}

\end{document}